\documentclass[%
 aip,
 jcp,
 reprint,%
]{revtex4-1}

\usepackage{graphicx}
\usepackage{dcolumn}
\usepackage{bm}

\usepackage{amsmath}
\usepackage{amssymb}
\usepackage{wasysym}

\usepackage[utf8]{inputenc}
\usepackage[T1]{fontenc}
\usepackage{mathptmx}
\usepackage{etoolbox}

\makeatletter
\def\@email#1#2{%
 \endgroup
 \patchcmd{\titleblock@produce}
  {\frontmatter@RRAPformat}
  {\frontmatter@RRAPformat{\produce@RRAP{*#1\href{mailto:#2}{#2}}}\frontmatter@RRAPformat}
  {}{}
}%
\makeatother
\begin{document}


\title{Kinetics of Carbon Condensation in Detonation of High Explosives: First-Order Phase Transition Theory Perspective}
\author{Apoorva Purohit}
\altaffiliation{Present Address: Center for Integrative Chemical Biology and Drug Discovery, UNC Eshelman School of Pharmacy, 301 Pharmacy Lane, University of North Carolina, Chapel Hill, NC 27599, USA}
\affiliation{Theoretical Division, Los Alamos National Laboratory, Los Alamos, NM 87545, USA}

\author{Kirill A. Velizhanin}
 \email{kirill@lanl.gov}
\affiliation{Theoretical Division, Los Alamos National Laboratory, Los Alamos, NM 87545, USA}

\date{\today}

\begin{abstract}
The kinetics of carbon condensation, or carbon clustering, in detonation of carbon-rich high explosives is modeled by solving a system of rate equations for concentrations of carbon particles. Unlike previous efforts, the rate equations account not only for the aggregation of particles, but also for their fragmentation in a thermodynamically consistent manner. Numerical simulations are performed, yielding the distribution of particle concentrations as a function of time. In addition to that, analytical expressions are obtained for all the distinct steps and regimes of the condensation kinetics, which facilitates the analysis of the numerical results and allows one to study the sensitivity of the kinetic behavior to the variation of system parameters. The latter is important because the numerical values of many parameters are not reliably known at present. The theory of the kinetics of first-order phase transitions is found adequate to describe the general kinetic trends of carbon condensation, as described by the rate equations. Such physical phenomena and processes as the coagulation, nucleation, growth, and Ostwald ripening are observed and their dependence on various system parameters is studied and reported. It is believed that the present work will become useful when analyzing the present and future results for the kinetics of carbon condensation, obtained from experiments or atomistic simulations. 
\end{abstract}

\maketitle


\section{Introduction\label{sec:Introduction}}

In the detonation of the so-called \emph{ideal} high explosives (HEs),
e.g., cyclotetramethylene-tetranitramine (HMX) or pentaerythritol
tetranitrate (PETN), HE molecules undergo rapid shock-driven decomposition
into a dense supercritical fluid of small molecules such as ${\rm H_{2}O}$,
${\rm N_{2}}$, ${\rm CO}$, and ${\rm CO_{2}}$.\citep{Fickett-2000-Detonation,Mader-2007-Explosives}
This decomposition consists of very fast chemical reactions and the associated heat release that occur on the time scale of, e.g., $\sim10\text{-}20\,{\rm ns}$
and $\lesssim5\,{\rm ns}$ for HMX and PETN, respectively.\citep{Loboiko-2000-716,Tarver-2006-1026}
The decomposition time is $\sim1$-$10$ ns, depending on how this time is defined, for the HMX-based PBX 9501 HE.\citep{Aslam-2020-030001}
However, the heat release becomes significantly slower in the detonation
of carbon-rich (or oxygen-deficient) HEs such as trinitrotoluene (TNT) or triaminotrinitrobenzene (TATB). What is assumed to happen is an
HE molecule, exemplified by TATB (${\rm C_{6}H_{6}N_{6}O_{6}}$, $258\,{\rm g/mol}$),
decomposes in detonation as \citep{Menikoff-2012-1140}
\begin{equation}
{\rm C_{6}H_{6}N_{6}O_{6}}\rightarrow\begin{cases}
3{\rm H_{2}O}+3{\rm N_{2}}+1.5{\rm CO_{2}}+4.5{\rm C,} & \text{}\\
3{\rm H_{2}O}+3{\rm N_{2}}+3{\rm CO}+3{\rm C}, & \text{}
\end{cases}\label{eq:HE_decomp}
\end{equation}
where the upper net decomposition reaction is energy favored, and
the lower one is entropy-favored at high temperatures. In addition
to small gaseous molecules, this decomposition thus also produces
the \emph{excess} carbon (the last product terms in the reactions)
that ultimately condenses into nanometer-sized carbon particles (e.g.,
nanodiamonds).\citep{Ornellas-1982-52821,Shaw-1987-2080,Titov-1989-372,Mochalin-2012-11,Danilenko-2017-93}
The mass fraction of the excess carbon, relative to the total initial
mass of HE, is $f_{C}=0.14\text{-}0.21$ in Eq. (\ref{eq:HE_decomp}).
Thermochemical calculations with the Los Alamos National Laboratory (LANL)
thermochemical code Magpie\citep{Ticknor-2020-030033} produce $f_{C}\approx0.19$
for bulk carbon in the form of diamond\citep{Velizhanin-2020-070051}
at the temperature and pressure of $T=2500\,{\rm K}$ and $P=30\,{\rm GPa}$,
respectively. Experimentally, the amount of recovered excess carbon
ranges from $f_{C}=0.04$ to $0.18$.\citep{Ornellas-1982-52821,Titov-1989-372,Mochalin-2012-11,Danilenko-2017-93}

Kinetically, the net decomposition reaction, Eq. (\ref{eq:HE_decomp}),
can be separated into two stages, the first being the initial decomposition
into gaseous molecules and small (few atoms) carbon-rich fragments.\citep{Tarver-1997-4845}
This first stage - fast reaction zone - is comparable in duration to the
detonation-driven decomposition of the ideal HEs. The second stage
is the slow diffusion-limited\citep{Chevrot-2012-084506} exothermic
condensation of carbon fragments into bigger ones, the so-called carbon
clustering.\citep{Shaw-1987-2080} That the two stages are present
could be deduced from a peculiar dependence of the steady-state detonation
velocity on the diameter of the detonated HE cylinder charge for TATB-based HE PBX-9502.\citep{Campbell-1984-183,Jackson-2015-224} The durations of
the fast and slow stages in PBX-9502 were obtained as $22$ and $280\,{\rm ns}$,
respectively, in the gas gun experiments.\citep{Dattelbaum-2014-396}
The slow energy release due to the second stage, which can amount up to $\sim 1\text{-}2\,{\rm kJ}$ per gram of HE,\citep{Menikoff-2012-1140} requires special treatment
and extra care, compared to e.g., HMX or PETN, when predicting the
HE performance.\citep{Shaw-1987-2080,Handley-2018-011303} More specifically,
reactive burn models,\citep{Handley-2018-011303} which are required
to perform hydrodynamic simulations of detonation at Los Alamos National
Laboratory and elsewhere, need to introduce and carefully parameterize
the slow second stage for the detonation simulations to be accurate.\citep{Wescott-2005-053514,Menikoff-2012-1140,Aslam-2018-145901}

The experimental data on carbon condensation could be grouped into two
categories. The first one concerns the analysis of carbon particles,
recovered in the form of soot after the detonation event. Diamond
and graphite-like particles recovered in this way turn out to be almost
spherical with the diameter of $\sim2\text{-}6$ nm, which corresponds
to $\sim3\times10^{3}$ - $3\times10^{4}$ carbon atoms, assuming
the graphite or diamond density.\citep{Greiner-1988-440,Titov-1989-372,Mochalin-2012-11,Bagge-Hansen-2015-245902,Kashkarov-2016-012072,Danilenko-2017-93,Watkins-2017-23129,Huber-2018-289,Hammons-2021-5286}
The second category - the data on the \emph{kinetics} of carbon clustering
- is rather scarce to date due to experimental challenges arising
from the need to observe the formation of nanometer-sized particles
with nanosecond time resolution in an extreme environment ($P\sim30\,{\rm GPa}$,
$T\sim3000\,{\rm K}$). Nevertheless, recent small-angle x-ray scattering (SAXS) experiments demonstrate
the sub-microsecond to few microsecond carbon clustering times and
the actual kinetics of the aggregation of carbon particles in detonation.\citep{Ten-2010-387,Ten-2014-369,Bagge-Hansen-2015-245902,Rubtsov-2016-012071,Watkins-2017-23129,Watkins-2018-821,Huber-2018-114,Hammons-2021-5286}
However, the convoluted nature of the experimental data does not presently
allow for the direct parameterization of reactive burn models, so
the development of the accurate theory of carbon condensation is required.

The openly published research on the theory of carbon clustering started
in 1987 with the seminal paper by Shaw and Johnson,\citep{Shaw-1987-2080}
where they applied the Smoluchowski coagulation theory,\citep{Smoluchowski-1916-585,Chandrasekhar-1943-1}
represented by a set of rate equations, to estimate the kinetics of heat release in the process of carbon condensation in detonation.
The classified developments in USA and Russia have probably started
well before that.\citep{Danilenko-2004-595} The Shaw-Johnson approach,
with various modifications, has been used to model the kinetics
of carbon condensation over the years,\citep{Ershov-1990-102, Ershov-1991-231,Ershov-1993-99,Kupershtokh-1996-393,Ree-1998-265,Viecelli-1999-237,Viecelli-2002-11352,Bastea-2012-214106,Chevrot-2012-084506,Watkins-2017-23129}
resulting in some degree of agreement with experiments.\citep{Bastea-2012-214106,Watkins-2017-23129}
Notably, the fragmentation of carbon particles, neglected in the original
Shaw-Johnson treatment, was accounted for in Ref.~\onlinecite{Viecelli-2002-11352}.
That fragmentation, however, was not included in the thermodynamically
consistent manner where the fragmentation and coagulation rate constants
would be related through equilibrium constants, which themselves depend
on thermodynamic properties of carbon particles. The thermodynamics
of carbon particles in detonation was considered in Refs.~\onlinecite{Viecelli-2001-2730,Bastea-2017-42151},
but was not applied to kinetics of carbon clustering.

Another string of theoretical research into the kinetics of carbon
clustering has started only recently, thanks to the development of efficient and accurate computation chemistry tools. The atomistic classical molecular
dynamics and density functional tight binding simulations, including those capable of directly simulating
carbon condensation in detonation, have produced promising results
only very recently,\citep{Chevrot-2009-3392,Zhang-2009-10619,Chevrot-2012-084506,Armstrong-2020-353,Lindsey-2020-054103,Lindsey-2021-164115} even though the initial crude attempts date back to the early 1990s.\citep{Ershov-1993-99,Kupershtokh-1996-393}
Of special interest are recent simulations by Lindsey,\citep{Armstrong-2020-353,Lindsey-2020-054103}
where the carbon particles were demonstrated to increase in size via the exchange
of very small carbon clusters - the so-called Ostwald ripening.\citep{Lifshitz-1961-35,Kahlweit-1975-1,Krapivsky-2010-Kinetic}
Unlike coagulation, the thermodynamically consistent description of
particle fragmentation is essential for the accurate treatment of the
Ostwald ripening. Therefore, the rate equations-based approaches,
developed to date, cannot be applied to analyze these recent atomistic
results. On the other hand, the atomistic approaches to carbon condensation
are expensive computationally, and are also still in the process of
being developed and calibrated. We thus believe that the analysis
of their results would greatly benefit from the development of a more
``coarse-grained'' rate equations-based framework where carbon clustering
is described in a thermodynamically consistent manner. In this paper, exactly
this latter task is performed. In particular, the rate equations with
thermodynamically consistent aggregation and fragmentation rates are
introduced and discussed in Sec. \ref{sec:Kinetic_model}. The energetics
of carbon particles, needed to properly relate the aggregation and
fragmentation rate constants, is discussed in Sec. \ref{sec:Energetics}.
The numerical results for a specific parameterization of energetics
of carbon particles are presented in Sec. \ref{sec:Numerical_Results},
and the various kinetic steps of carbon condensation along with the
sensitivity of the kinetic behavior on various system parameters are
discussed in Secs. \ref{sec:Coagulation}-\ref{sec:Coarsening}. The
more general discussions are provided in Sec. \ref{sec:Discussion},
and Sec. \ref{sec:Conclusion} concludes. The physics of carbon condensation,
when analyzed from the standpoint of the rate equations in this work,
was found to be the one generally described by the theory of kinetics
of first-order phase transitions,\citep{Binder-1976-343,Penrose-1978-Kinetics,Binder-1987-783,Slezov-2009-Kinetics,Krapivsky-2010-Kinetic}
hence the title of this paper. This paper is sufficiently self-contained
in that we describe all the kinetic steps of the condensation, which
is rarely done in a single paper in the literature. Furthermore, for the
two steps, the growth in Sec. \ref{sec:Growth} and Ostwald ripening
in Sec. \ref{subsec:Ostwald-Ripening}, we go beyond the so-called
monomer-only approximation. It can thus be expected that the present
paper can be of pedagogical value.

The final word of caution is that we found that the details of the carbon condensation kinetics could sometimes be very sensitive to even slight variations
of particle energetics, as well as the temperature and the concentration
of the excess carbon. Unfortunately, the energetics of carbon particles
is itself known only approximately. This is why it was a considerable
effort on our side to not so much rely on a specific parameterization,
but instead (i) show the entire palette of the realizable kinetic
behaviors, and (ii) analyze the conditions at which transitions between
different kinetic regimes occur when parameters change. Accordingly, we believe
that this work will be useful, as a framework and not as a source
of specific numerical values, when analyzing the results of experimental
studies and atomistic simulations towards the understanding of carbon
condensation in detonation.

\section{Kinetic Modeling of Carbon Condensation\label{sec:Kinetic_model}}

The time evolution of concentration $c_{n}$ of particles made of
$n$ carbon atoms is given by the kinetic model in Eq.~(\ref{cnplusminus})
that includes the rates of aggregation and fragmentation of particles
\begin{equation}
\frac{dc_{n}}{dt}=\left[\frac{dc_{n}}{dt}\right]_{+}+\left[\frac{dc_{n}}{dt}\right]_{-}.\label{cnplusminus}
\end{equation}
Assuming only binary collisions, the contribution of the aggregation
is \citep{Friedlander-2000-Smoke,Krapivsky-2010-Kinetic,Seinfeld-2016-Atmospheric}
\begin{equation}
\left[\frac{dc_{n}}{dt}\right]_{+}=\sum_{m=1}^{n-1}\frac{1}{2}K_{n-m,m}^{+}c_{n-m}c_{m}-\sum_{m=1}^{N-n}K_{n,m}^{+}c_{n}c_{m},\label{cnplus}
\end{equation}
where $K_{n,m}^{+}$ is the rate constant of the aggregation process
$(n)+(m)\rightarrow(n+m)$, in which two carbon particles of sizes
$m$ and $n$ aggregate into a single particle of size $n+m$. The
term ``particle size'' in this work denotes the number of carbon
atoms in the particle. The maximum allowed cluster size is $N$. Physically,
$N$ should be set to infinity, but it is finite in numerical simulations.

The contribution of the fragmentation to Eq.~(\ref{cnplusminus})
is
\begin{equation}
\left[\frac{dc_{n}}{dt}\right]_{-}=-\sum_{m=1}^{n-1}\frac{1}{2}K_{n-m,m}^{-}c_{n}+\sum_{m=1}^{N-n}K_{n,m}^{-}c_{n+m},\label{cnminus}
\end{equation}
where $K_{n,m}^{-}$ is the rate constant of the fragmentation process
$(n+m)\rightarrow(n)+(m)$. The factors of $1/2$ in the first rhs terms of Eqs.~(\ref{cnplus}) and (\ref{cnminus}) compensate for
over-counting when physically the same processes are counted twice
in summations. This over-counting and some related issues are discussed
in App.~\ref{app:over_count}. It can be demonstrated that Eqs.~(\ref{cnplus})
and (\ref{cnminus}) preserve the total concentration of carbon in
the system (number of carbon atoms per unit volume) defined as
\begin{equation}
c_{tot}=M_{1},\label{eq:ctot}
\end{equation}
where 
\begin{equation}
M_{p}=\sum_{n=1}^{N}n^{p}c_{n}\label{eq:moment_cn}
\end{equation}
 is the $p^{{\rm th}}$ moment of the distribution of concentrations.

The reaction of aggregation is assumed diffusion-limited, which is
in line with earlier molecular dynamics results,\citep{Chevrot-2012-084506}
with the rate constant given by\citep{Smoluchowski-1916-585,Chandrasekhar-1943-1,Shaw-1987-2080,Chevrot-2012-084506}
\begin{equation}
K_{n,m}^{+}=4\pi R_{nm}D_{nm},\label{eq:Kplus_diff_lim}
\end{equation}
where $D_{nm}=D_{n}+D_{m}$ is the relative diffusion coefficient.
Assuming spherical particles, the diffusion coefficient of the particle
of size $n$ is given by the Stokes-Einstein relation \citep{Friedlander-2000-Smoke}
$D_{n}=\frac{k_{B}T}{6\pi R_{n}\eta}$, where $R_{n}$ is the radius of the particle of size $n$, $k_{B}$ is the Boltzmann constant, $T$ is the temperature,
and $\eta$ is the dynamic viscosity. The radius can be estimated
by 
\begin{equation}
R_{n}=(3n/(4\pi n_{C}))^{1/3},\label{eq:Rn_nC}
\end{equation}
where $n_{C}$ is the atomic density of carbon ($[n_{C}]={\rm 1/m^{3}}$)
in carbon particles.

The distance of coalescence, $R_{nm}$ in Eq. (\ref{eq:Kplus_diff_lim}),
is the distance between the centers of particles $(n)$ and $(m)$
at which the particles are in contact. Accordingly, it is set to be
$R_{nm}=R_{n}+R_{m}$; see however the historical perspective in App.
\ref{app:Rnm}. Substituting this expression for the distance of coalescence,
and that for the relative diffusion constant, to Eq.~(\ref{eq:Kplus_diff_lim})
produces
\begin{equation}
K_{n,m}^{+}=k_{0}\left(n^{1/3}+m^{1/3}\right)\left(n^{-1/3}+m^{-1/3}\right),\label{eq:Knm_k0}
\end{equation}
where $k_{0}=\frac{2k_{B}T}{3\eta}$. For example, $k_{0}=2.76\times10^{-12}\,{\rm m^{3}/s}$
for $T=3000\,{\rm K}$ and $\eta=10^{-3}\,{\rm kg/(m\cdot s)}$, see
parameters in Tab. \ref{tab:std_parms} below. 

\subsection{Thermodynamics of Detonation Products\label{subsec:Products_Thermo}}

One of the chief assumptions in this work is that the chemical state
of the system, described by the rate equations above, is a mixture of a supercritical fluid of chemically
neutral small molecules such as ${\rm H_{2}O}$, ${\rm N_{2}}$, ${\rm CO_{2}}$,
${\rm CO}$ etc. and carbon particles undergoing condensation. This
can only be assumed if there is a sufficient time scale separation
between the initial decomposition of the HE molecules and carbon condensation.
For example, Refs.~\onlinecite{Menikoff-2012-1140,Dattelbaum-2014-396}
suggest $\sim10$-$20$ ns and $\sim300$ ns for the durations
of the two processes, respectively, in detonating TATB-based HE PBX-9502,
so the assumption is accurate. An extensive volume of such a system
at a given pressure and temperature is
\begin{equation}
V(P,T)=V_{f}(P,T)+\sum_{n}N_{n}v_{n}(P,T),\label{eq:V_Vf_vn}
\end{equation}
where $V_{f}$ denotes the fluid volume, and $N_{n}$ is the number
of carbon particles of size $n$, each occupying volume $v_{n}$ in
the fluid. This single-particle volume is not necessarily exactly
proportional to the number of carbon atoms $n$ in the particle\footnote{We, however, assumed this proportionality for simplicity when deriving Eq.~(\ref{eq:Knm_k0})} due to, for example, steric interactions of the fluid with the surface
of the particle, or the formation of the ligand layer on the surface
of the carbon particle. This can lead to a non-trivial dependence
of the total volume on carbon redistribution in particles of different
sizes even though the total amount of carbon, $\sum_{n}nN_{n}$, is
preserved.

An extensive Helmholtz free energy for the system is given by 
\begin{gather}
A(P,T)=A_{f}(P,T)+\sum_{n}N_{n}f_{n}^{(ntr)}(P,T)\nonumber \\
+\beta^{-1}\sum_{n}N_{n}k_{B}T\ln\left(\frac{N_{n}}{eV(P,T)}\Lambda_{C}^{3}n^{-3/2}\right),\label{eq:A_PT_ext}
\end{gather}
where the first rhs term is the Helmholtz free energy of the fluid, and $f_{n}^{(ntr)}$ in the second term is the non-translational (i.e.,
vibrational, rotational, electronic, and due to carbon particle-fluid
interaction) Helmholtz free energy of the carbon particle of size
$n$. The third rhs term describes the translational Helmholtz
free energy\citep{McQuarrie-2000-StatMech} of the carbon particles,
where the thermal de Broglie wavelength for a single carbon atom of
mass $m_{C}\approx2.0\times10^{-26}\,{\rm kg}$ is given by
\begin{equation}
\Lambda_{C}=\left(\frac{2\pi\hbar^{2}\beta}{m_{C}}\right)^{1/2},
\end{equation}
where $\beta^{-1}=k_{B}T$. The constant $e=2.71828...$ in the third rhs term of Eq.~\eqref{eq:A_PT_ext} is the base of the natural logarithm.
Gibbs free energy is $G(P,T)=A(P,T)+PV(P,T)$,
and the chemical potential of particles of size $n$ is (omitting
explicit $P,T$ arguments)
\begin{gather}
\mu_{n}=\left(\frac{\partial G}{\partial N_{n}}\right)=f_{n}^{(ntr)}+\beta^{-1}\ln\left(c_{n}\Lambda_{C}^{3}n^{-3/2}\right)\nonumber \\
+\left(P-\beta^{-1}\sum_{m}c_{m}\right)v_{n},\label{eq:mun_fntr}
\end{gather}
where the partial derivative with respect to $N_{n}$ is evaluated
at constant pressure, temperature and $\{N_{m}|m\neq n\}$. Concentrations
of carbon particles are defined as $c_{n}=N_{n}/V$. At near-detonation
conditions, the actual pressure in the system, $P$, is much greater
than the ``ideal'' pressure of carbon particles, $\beta^{-1}\sum_{m}c_{m}$,
so the latter one is neglected. The chemical potential then becomes
\begin{equation}
\mu_{n}=\mu_{n}^{\circ}+\beta^{-1}\ln(c_{n}/c^{\circ}),\label{eq:mun_muno}
\end{equation}
where
\begin{equation}
\mu_{n}^{\circ}=\mu_{n}^{(ntr)}+\beta^{-1}\ln\left(c^{\circ}\Lambda_{C}^{3}n^{-3/2}\right)
\end{equation}
is the standard chemical potential, and $c^{\circ}$ is some arbitrarily
chosen standard concentration. The non-translational part of the standard
chemical potential is defined as $\mu_{n}^{(ntr)}=f_{n}^{(ntr)}+Pv_{n}$.
For large particles where the contribution of the surface is relatively small, one has $\mu^{(ntr)}\approx n\mu_{b}$, where $\mu_{b}$ is
the chemical potential of a carbon atom in bulk carbon. We introduce
the surface chemical potential $\mu_{s,n}$ via
\begin{equation}
\mu_{n}^{(ntr)}=n\mu_{b}+\mu_{s,n},\label{eq:mu_ntr_mub_musn}
\end{equation}
and, similarly, $f_{n}^{(ntr)}=nf_{b}+f_{s,n}$ and $v_{n}=nv_{b}+v_{s,n}$,
so that $\lim_{n\rightarrow\infty}n^{-1}\left|x_{s,n}/x_{b}\right|=0$,
where $x=v$, $f$ or $\mu$. Since $\mu_{b}$ is defined up to a
constant, we choose $\mu_{b}=0$ at the pressure and temperature of
interest. The final result is that 
\begin{equation}
\mu_{n}^{\circ}=\mu_{s,n}+\beta^{-1}\ln\left(c^{\circ}\Lambda_{C}^{3}n^{-3/2}\right).\label{eq:mu_n_o_mu_s_n}
\end{equation}

At equilibrium, the aggregation and fragmentation rates are the same
due to the detailed balance, so that $K_{n,m}^{+}c_{n}^{(eq)}c_{m}^{(eq)}=K_{n,m}^{-}c_{n+m}^{(eq)}$.
Furthermore, the chemical potential of carbon, $\mu$, can be introduced
so that $\mu_{n}=n\mu$. Eq. (\ref{eq:mun_muno}) then produces the
equilibrium concentration of particles as
\begin{equation}
c_{n}^{(eq)}(\mu)=c^{\circ}e^{-\beta(\mu_{n}^{\circ}-n\mu)},\label{eq:cn_eq}
\end{equation}
 which finally leads to a relationship between the coagulation and
fragmentation rate constants as
\begin{equation}
K_{n,m}^{-}=c^{\circ}e^{\beta\left(\mu_{n+m}^{\circ}-\mu_{n}^{\circ}-\mu_{m}^{\circ}\right)}K_{n,m}^{+}.\label{eq:Kminus_Kplus}
\end{equation}

\section{Energetics of Carbon Particles\label{sec:Energetics}}

Eq. (\ref{eq:Kminus_Kplus}) makes it possible to calculate the rate
constants of fragmentation $K_{n,m}^{-}$ from the diffusion-limited
rate constants of aggregation, Eq. (\ref{eq:Kplus_diff_lim}), provided that the standard chemical potentials $\mu_{n}^{\circ}$ are known. The
second rhs term of Eq. (\ref{eq:mu_n_o_mu_s_n}) is evaluated trivially,
and so this section focuses on how to estimate $\mu_{s,n}$. Experimental
studies \citep{Greiner-1988-440,Titov-1989-372,Ten-2009-102,Watkins-2017-23129}
have validated that the large carbon particles ($>1000$ atoms) obtained
from post-detonation products are almost spherical in shape. For such
particles, the deviation of the non-translational standard chemical
potential $\mu_{n}^{(ntr)}$ from $n\mu_{b}$ in Eq. (\ref{eq:mu_ntr_mub_musn})
is expected to be proportional to the surface area of the cluster,\citep{Machlin-2007-Aspects,Kelton-2010-Nucleation}
\begin{equation}
\mu_{s,n}=\mu_{s}n^{2/3},\label{eq:musn_mus}
\end{equation}
which is the so-called capillarity approximation. Here, the surface
chemical potential coefficient $\mu_{s}$ does not depend on the particle
size, as long as the size is large. The contribution of free rotations
of the particle as the whole would produce a different dependence
of $\mu_{s,n}$ on $n$, but these rotations are expected to be hindered due to the
interaction between the surface of the particle and the surrounding
fluid. The rotations thus become vibrations and the total number of
vibrational modes due to the interaction of the particle surface with
the fluid is proportional to the surface area of the particle, so
the $\propto n^{2/3}$ scaling stands. The order of magnitude of possible
$\mu_{s}$ values can be illustrated by results of Viecelli \emph{et
al.,}\citep{Viecelli-2001-2730} where the values of $\mu_{s}=3.04$,
$1.73$ and $0.048\,{\rm eV}$ were adopted for particles
made of diamond, graphite, and liquid carbon, respectively. For comparison,
the excess internal energy of $3.47n^{2/3}\,{\rm eV}$ was
adopted in Ref.~\onlinecite{Shaw-1987-2080}.

\begin{figure*}
\includegraphics[width=0.4\paperwidth]{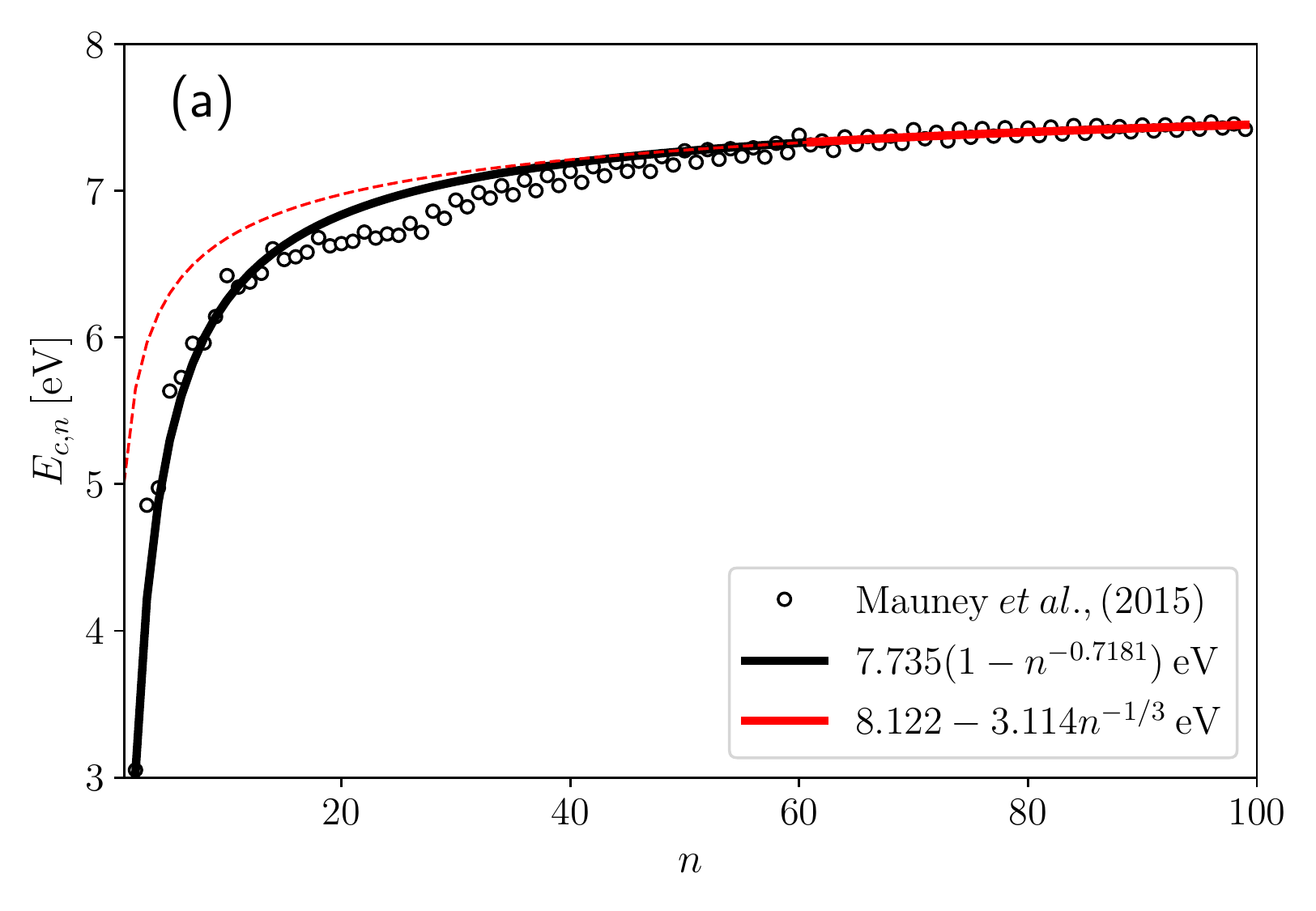}
\includegraphics[width=0.4\paperwidth]{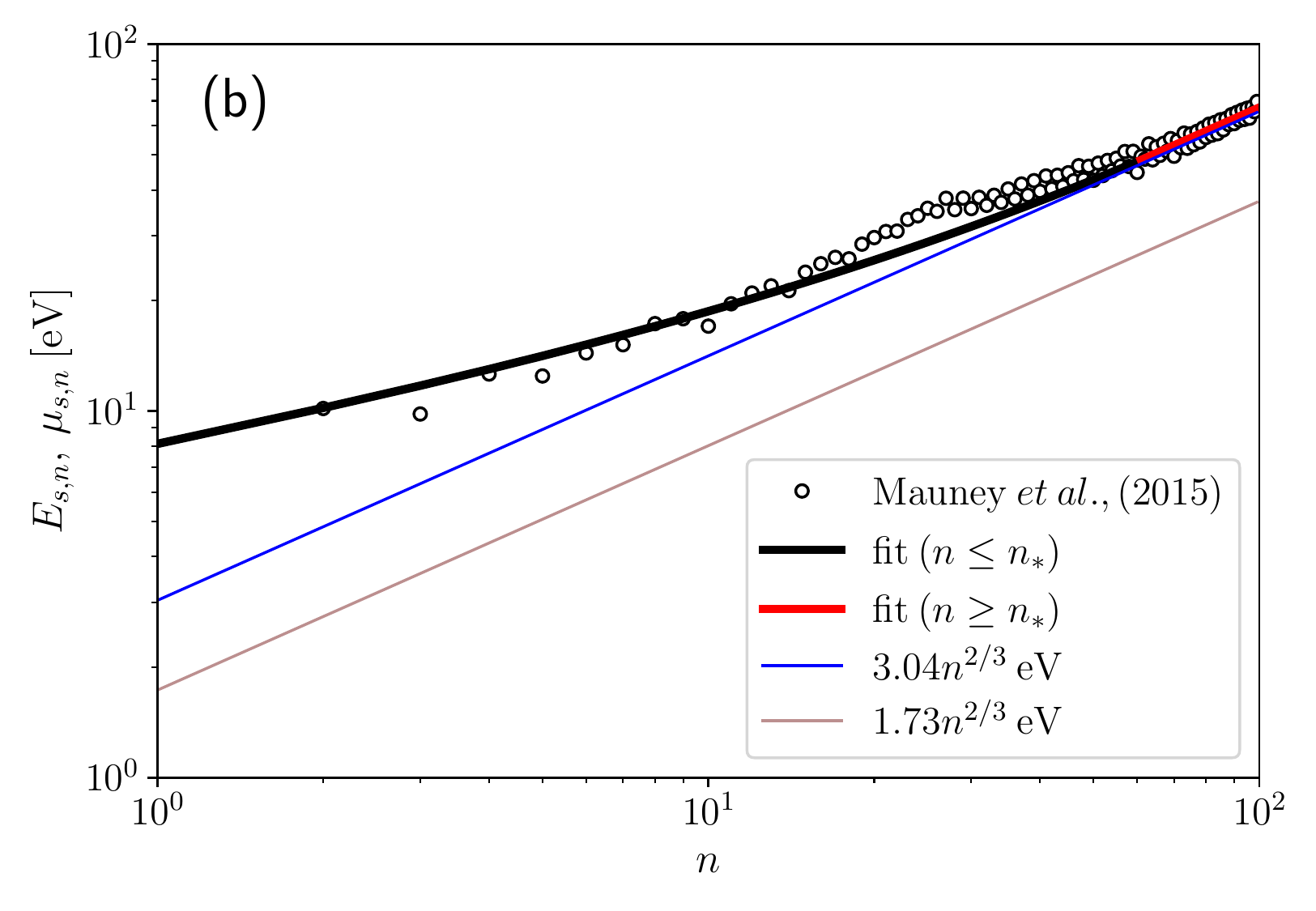}
\caption{\label{fig:Ec_vs_n}(a) Cohesive energy $E_{c,n}$ as a function of
carbon particle size. (b) Surface cold energy and surface chemical
potential as a function of carbon particle size.}
\end{figure*}

For small particles (clusters), the capillarity approximation breaks
down\citep{Kelton-2010-Nucleation} as it is impossible to distinguish
between the core and the surface of the particle any longer. Since the values
of $\mu_{s,n}$ are not directly available from the literature, we illustrate
the capillarity approximation and its break down for small particles
by analyzing the so-called cohesive energy of carbon particles. Theoretical
studies on energetics of carbon particles have attracted attention
from both physicists and chemists, who have applied \emph{ab initio}
techniques, as well as empirical interatomic potential-based methods, to calculate the cohesive energies of carbon particles as a function
of particle size at zero temperature.\citep{Tomanek-1991-2331,Eggen-1994-3029,Kosimov-2008-235433,Yu-2009-184708,Mauney-2015-30}
The cohesive energy $E_{c,n}$ for a particle of size $n$ is typically
defined in the literature as
\begin{equation}
E_{c,n}=-\left(E_{n}-nE_{1}\right)/n.\label{eq:Ecn}
\end{equation}
Here, $E_{n}$ is the total {\it cold energy} of the particle of size
$n$, defined as the zero-temperature internal energy of an isolated particle
in its optimal geometry, without accounting for the zero-point energy
of vibrational modes. Accordingly, $E_{1}$ stands for the ground-state
energy of an isolated carbon atom. The results of the \emph{ab initio}
calculations for the cohesive energies of carbon particles with sizes
$2\leq n\leq100$ by Mauney \emph{et al.\citep{Mauney-2015-30}} are
shown by black circles in Fig. \ref{fig:Ec_vs_n}(a).
Similar to Eq. (\ref{eq:mu_ntr_mub_musn}), we split the cold energy
of the particle into the bulk and surface parts as
\begin{equation}
E_{n}=nE_{b}+E_{s,n},\label{eq:En_Eb_Esn}
\end{equation}
where $E_{s,n}=E_{s}n^{2/3}$ for large particles. Using these expressions,
Eq. (\ref{eq:Ecn}) becomes $E_{c,n}=E_{c,\infty}-E_{s}n^{-1/3}$,
where 
\begin{equation}
E_{c,\infty}=E_{1}-E_{b}\label{eq:Ecinf_E1_Eb}
\end{equation}
is the cohesive energy for bulk carbon. We perform the least-squares fitting of the cohesive energy by Mauney \emph{et al.\citep{Mauney-2015-30}}
(black circles in Fig. \ref{fig:Ec_vs_n}(a)) with a piecewise analytic
function
\begin{equation}
E_{c,n}=\begin{cases}
A\left(1-n^{-\gamma}\right), & n\leq n'\\
E_{c,\infty}-E_{s}n^{-1/3}, & n\geq n'
\end{cases}\label{eq:Ecn_fit_form}
\end{equation}
where $E_{c,\infty}$, $E_{s}$ and $\gamma$ are independent fitting
parameters and $A$ is constrained by the requirement that the function
is continuous at $n=n'$, which is set to $n'=60$. The functional
form for the small particles is chosen to produce $E_{c,1}=0$, in
accordance with Eq. (\ref{eq:Ecn}). The fitting produces (rounded
to four significant digits) $E_{c,\infty}=8.122\,{\rm eV}$, $E_{s}=3.114\,{\rm eV}$,
$\gamma=0.7181$ and $A=7.735\,{\rm eV}$, and the two pieces of the fitting function are plotted by the black and red solid lines
in Fig. \ref{fig:Ec_vs_n}(a). The obtained value of $E_{c,\infty}$
is in agreement with $\sim8\,{\rm eV}$ reported in many previous studies.\citep{Dappe-2006-235124,Shin-2014-114702}
The red dashed line corresponds to the large-particle functional form
in Eq. (\ref{eq:Ecn_fit_form}) plotted at $n\leq n'$. Its disagreement
with the numerical results demonstrates that the energetics of small
particles cannot be accurately modelled by using the capillarity approximation,
$E_{s,n}=E_{s}n^{2/3}$.

Substituting Eqs. (\ref{eq:En_Eb_Esn}) and (\ref{eq:Ecinf_E1_Eb})
into Eq. (\ref{eq:Ecn}), we obtain the surface cold energy as
\begin{equation}
E_{s,n}=n\left(E_{c,\infty}-E_{c,n}\right).\label{eq:Esn}
\end{equation}
Substituting the numerical cohesive energies by Mauney \emph{et al.\citep{Mauney-2015-30},}
together with $E_{c,\infty}=8.122\,{\rm eV}$ obtained from the fitting,
into this expression produces black circles in Fig. \ref{fig:Ec_vs_n}(b).
The actual fitting results, produced by combining Eqs. (\ref{eq:Ecn_fit_form})
and (\ref{eq:Esn}), are represented by the thick black and red lines.
The thin blue and brown lines represent $\mu_{s,n}=\mu_{s}n^{2/3}$, with,
respectively, $\mu_{s}=3.04\,{\rm eV}$ and $\mu_{s}=1.73\,{\rm eV}$
- the values for diamond and graphite, respectively, adopted from the work of Viecelli \emph{et al}.\citep{Viecelli-2001-2730} Even though the
surface cold energy coefficient $E_{s}$ and surface chemical potential
coefficient $\mu_{s}$ cannot be directly compared, it is interesting
to note that the value of the latter for diamond, $\mu_{s}=3.04\,{\rm eV}$,\citep{Viecelli-2001-2730}
is numerically very similar to $E_{s}=3.114\,{\rm eV}$ obtained here,
resulting in a good agreement between the thick red and thin blue
lines in Fig. \ref{fig:Ec_vs_n}(b). Nevertheless, the fitting results
disagree with $\mu_{s,n}=3.04n^{2/3}\,{\rm eV}$ for small particles
due to the break down of the capillarity approximation.

To evaluate $\mu_{s,n}$ from $E_{s,n}$ obtained here, we need to
further account for the surface contributions to (i) the Helmholtz
free energy of vibrational motion, $f_{s,n}^{(v)}$, and (ii) the
particle volume, $v_{s,n}$. To perform a very rough order-of-magnitude
estimation of these values for a large particle, we assume the following:
\begin{enumerate}
\item The packing of carbon atoms in the outmost one-atom-thick layer of
a particle is less dense than the rest of the particle.
\item The variations of values of specific (per unit mass) thermodynamic
variables between the particle surface and the core are order-of-magnitude
comparable to those in the phase transition between graphite and diamond,
respectively. The equation of state for these two carbon phases is
adopted from Ref.~\onlinecite{Velizhanin-2020-070051}, and the calculations are performed using the LANL thermochemical code Magpie\citep{Ticknor-2020-030033} at $P=30\,{\rm GPa}$, $T=3000\,{\rm K}$.
\end{enumerate}
The number of carbon atoms in the one-atom-thick outmost layer can
be estimated as
\begin{equation}
n_{s}=4\pi R_n^{2}/d^{2}=4\pi\left(\frac{3}{4\pi}\right)^{2/3}n^{2/3}\approx4.8n^{2/3},
\end{equation}
where $R_n$ is the radius of particle of size $n$. The size of the particle is given
by $n=\frac{4}{3}\pi R_n^{3}n_{C}$, where $n_{C}$ is defined right
after Eq. (\ref{eq:Rn_nC}). The characteristic distance between carbon
atoms in the particle is $d=n_{C}^{-1/3}$. The difference between
the vibrational contributions to the Helmholtz free energy of graphite
($gr$) and diamond ($dim$) is
\begin{equation}
F_{gr}^{(v)}-F_{dim}^{(v)}\approx-8.1\times10^{5}\,{\rm J/kg},
\end{equation}
and so the contribution of this Helmholtz free energy variation to
the surface chemical potential is
\begin{equation}
f_{s,n}^{(v)}\approx n_{s}m_{C}\left(F_{gr}^{(v)}-F_{dim}^{(v)}\right)\approx-0.5n^{2/3}\,{\rm eV}.\label{eq:f_v_sn}
\end{equation}
The difference of the specific volume for the two phases is
\begin{equation}
V_{gr}-V_{dim}\approx7\times10^{-5}\,{\rm m^{3}/kg},
\end{equation}
which results in the surface energy contribution of
\begin{equation}
pv_{s,n}\approx n_{s}m_{C}P\left(V_{gr}-V_{dim}\right)\approx1.3n^{2/3}\,{\rm eV}.\label{eq:pv_sn}
\end{equation}
The two estimated contributions to $\mu_{s,n}$, Eqs. (\ref{eq:f_v_sn})
and (\ref{eq:pv_sn}), are substantial in that their magnitudes are
comparable to, for example, $3.04n^{2/3}$ and $1.73n^{2/3}\,{\rm eV}$
in Fig. \ref{fig:Ec_vs_n}(b). Furthermore, all the above considerations
do not account for a possible nitrogen- and oxygen-rich ligand layer
on the surface of a carbon particle.\citep{Shaw-2000-235,Mochalin-2012-11,Armstrong-2020-353}
The effect of the ligand layer might become ``extreme'' in the limit
of very small particles, where, for example, one likely needs to consider
${\rm CO}$ and ${\rm CO_{2}}$ molecules instead of an isolated single
carbon atom when calculating rate constants for processes involving
the carbon monomer in Eq. (\ref{cnplusminus}). The just discussed
problems result in the impossibility to accurately estimate the surface
chemical potential $\mu_{s,n}$, even though the values on the order
of $\sim2\text{-}3\,{\rm eV}$ for the surface chemical potential
coefficient $\mu_{s}$ in Eq. (\ref{eq:musn_mus}) do seem reasonable
for large solid carbon particles.\citep{Viecelli-2001-2730} In what follows, we, therefore,
do not rely on specific values of various system parameters, including
$\mu_{s,n}$. Instead, a parametric analysis is performed where the
sensitivity of the carbon condensation kinetics to variations of parameters'
values is studied. Since the dimensionality of the parameter space
is large, we choose a single ``standard'' set of system parameters,
and the sensitivity of the condensation kinetics to variations of
at most two parameters at a time is studied. The choice of the standard
set is somewhat arbitrary, so we choose not necessarily the one reproducing experimental observations the
best, but the one where the different physically distinct steps of the condensation kinetics are represented most clearly. The set of the standard parameters is given in Tab. \ref{tab:std_parms}.
\begin{table*}
\begin{tabular}{|c|c|c|c|c|}
\hline 
Temperature & $\mu_{s,n}$ & $c_{tot}$, Eq. (\ref{eq:ctot}) & $c_{n}(t=0)$ & Viscosity, $\eta$\tabularnewline
\hline 
\hline 
\\[-1em]
$3000\,{\rm K}$ & $\begin{cases}
3.82\,{\rm eV}, & n=1\\
3.04n^{2/3}\,{\rm eV}, & n>1
\end{cases}$ & $4\times10^{28}\,{\rm 1/m^{3}}$ & $c_{tot}\delta_{n,1}$ & $10^{-3}\,{\rm kg/(m\cdot s)}$
\\[-1em]
\tabularnewline
\hline
\end{tabular}\caption{\label{tab:std_parms} Standard set of system parameters. The initial
distribution of carbon particle concentrations is given by $c_{n}(t=0)\propto\delta_{n,1}$,
that is, all the excess carbon initially exists in the form of monomers.
The surface chemical potential is assumed in the capillarity approximation
form, $\mu_{s,n}=3.04n^{2/3}\,{\rm eV}$,\citep{Viecelli-2001-2730}
except for the monomer with a somewhat larger surface chemical potential.
The last column is the dynamic viscosity of the detonation fluid.}
\end{table*}
There is no specific physical reason for the choice of the functional form of standard $\mu_{s,n}$ in the second column of Tab. \ref{tab:std_parms}, except that (i) it combines a simple analytical form for the large particles, which agrees well with our fitting of the energetics due to Mauney \emph{et al.\citep{Mauney-2015-30}}, with (ii) a single adjustable value of $\mu_{s,1}$, which was tuned by us so that the standard system demonstrates the rich kinetic behavior.
The total density of the TATB decomposition products at detonation
conditions can be estimated to be $\rho=2.5\text{-}2.6\,{\rm g/cm^{3}}$.\citep{Wescott-2005-053514,Mader-2007-Explosives,Menikoff-2009-06529}
Using the mass fraction of the excess carbon estimated in the Introduction,
the atomic concentration of excess carbon in detonation products is
$\sim1.5\text{-}3\times10^{28}\,{\rm 1/m^{3}}$. The adopted standard carbon concentration, the third column of Tab. \ref{tab:std_parms}, is on the higher
side of this estimated range. The first two columns of Tab. \ref{tab:std_parms} allow
one to evaluate the total concentration of carbon in the \emph{saturated}
system, i.e., the system where the carbon ``vapor'' - the ``gas'' of small carbon fragments - is in equilibrium
with bulk carbon. Accordingly, the chemical potential is $\mu=\mu_{b}=0$
and the combination of Eqs. (\ref{eq:ctot}), (\ref{eq:cn_eq}) and
(\ref{eq:mu_n_o_mu_s_n}) produces
\begin{equation}
c_{v,tot}^{(sat)}=\sum_{n=1}^{\infty}c^{\circ}e^{-\beta\mu_{n}^{\circ}}\approx5.5\times10^{26}\,{\rm 1/m^{3}}.\label{eq:cvtot_st}
\end{equation}
This value is significantly smaller than the standard concentration of $c_{tot}=4\times10^{28}\,{\rm 1/m^{3}}$ we chose,
and so the standard parameters specify an \emph{over-saturated} system
where the carbon condensation is expected. The dynamic viscosity of
a mixture of small gaseous molecules such as ${\rm H_{2}O}$, ${\rm CO_{2}}$,
${\rm N_{2}}$ etc. can be estimated as $\eta\sim10^{-3}\,{\rm kg/(m\cdot s})$ at near-detonation conditions.\citep{Shaw-1987-2080,Bastea-2002-Transport,Velizhanin-2021-Enskog}

\section{Results of Numerical Modeling\label{sec:Numerical_Results}}

\begin{figure*}
\includegraphics[width=0.4\paperwidth]{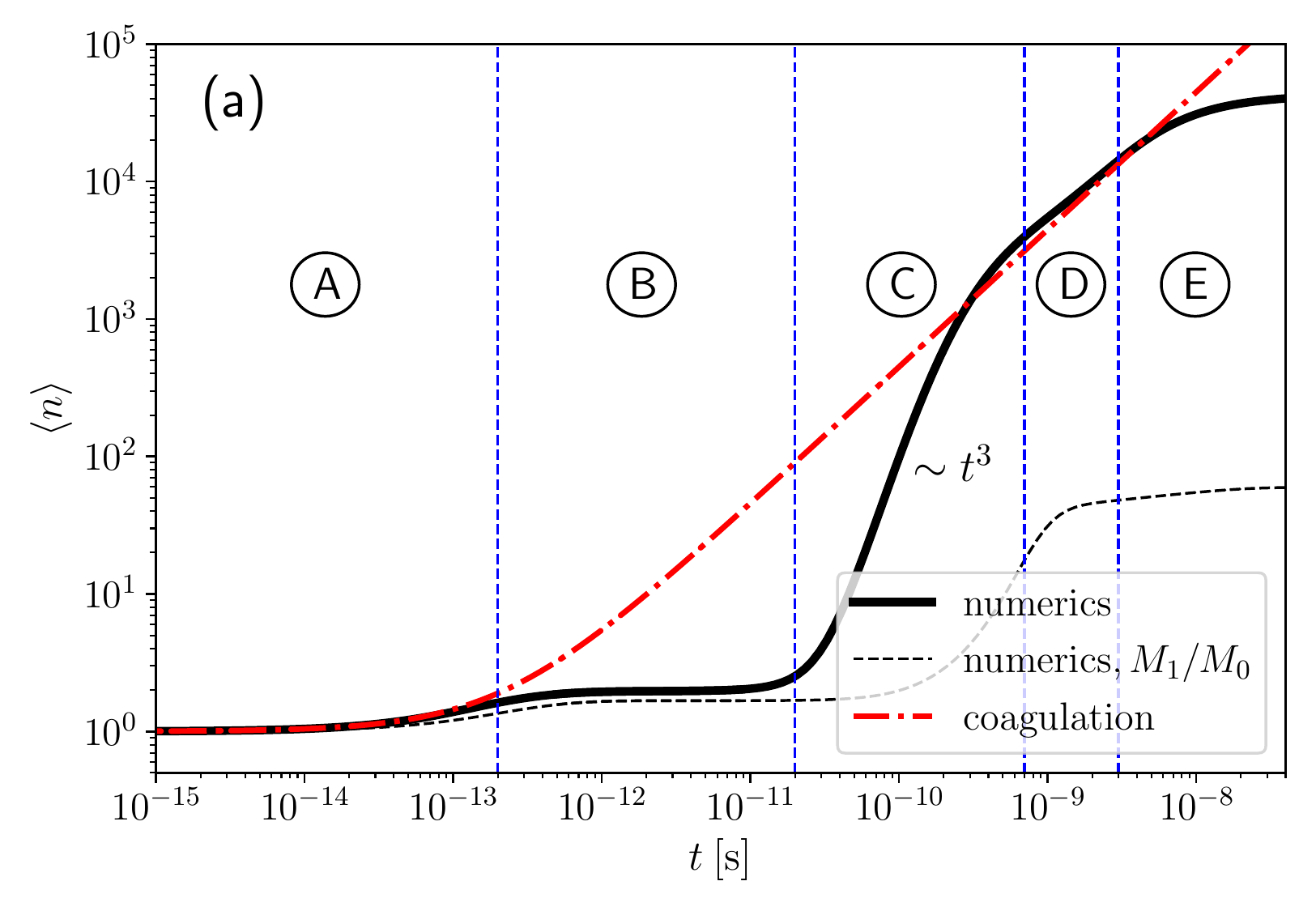}
\includegraphics[width=0.4\paperwidth]{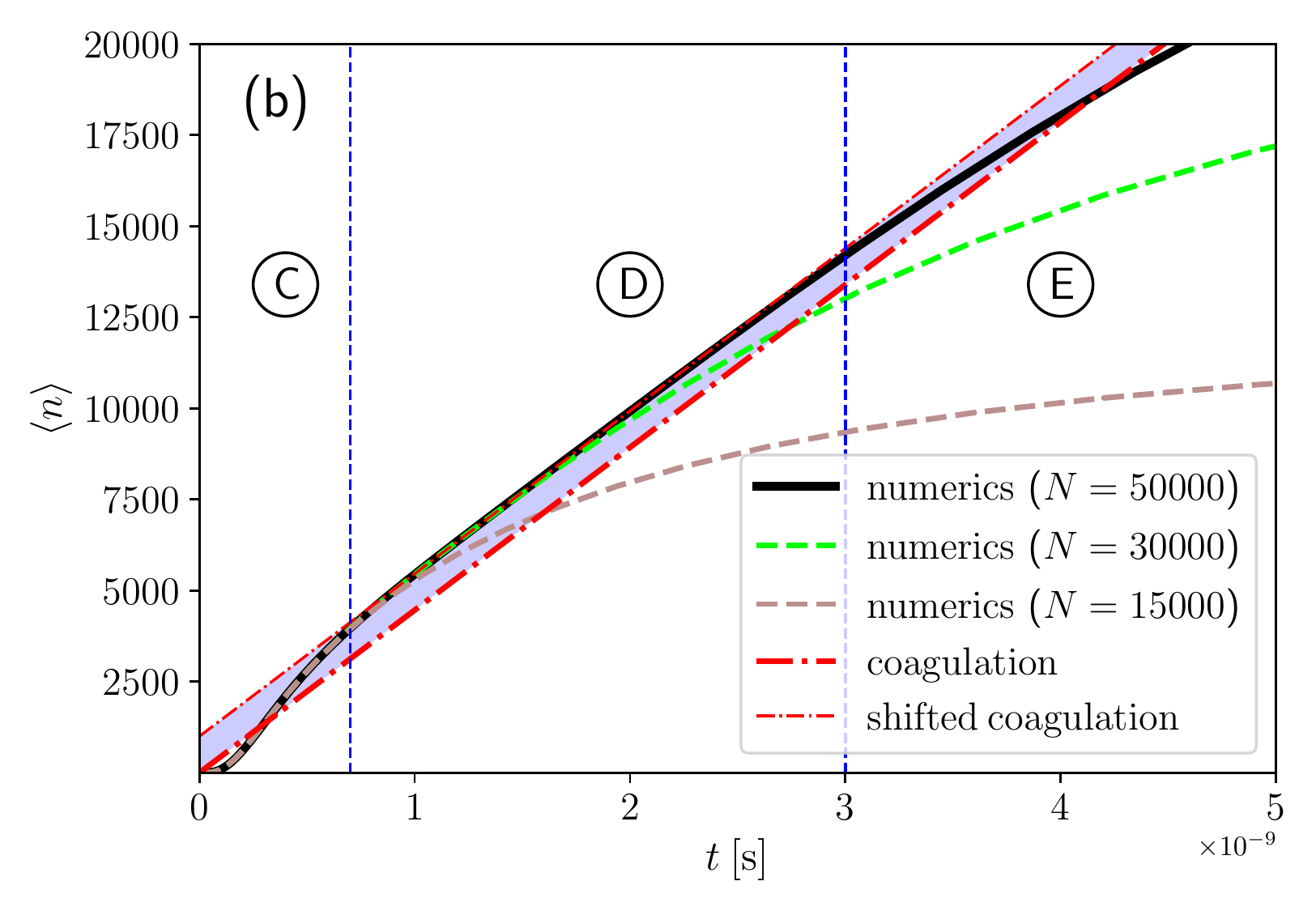}
\caption{\label{fig:nmean_vs_t}Results of numerically solving Eq. (\ref{cnplusminus})
with the standard parameters listed in Tab. \ref{tab:std_parms}. The thick dashed and solid lines in panel (b) show the dependence of the
numerical results on the maximum cluster size $N$. The thick red dash-dotted line is given by Eq. (\ref{eq:nmean_vs_t_coag}), and the thin
red dash-dotted line represents the same expression, plotted with the negative
time shift of $2.24\times10^{-10}\,{\rm s}$. The thin black dashed line in panel (a) shows the effect of the alternative definition of the mean particle size, $M_1/M_0$, as opposed to the ``standard" definition, Eq.~\eqref{eq:nmean_m2_m1}, used throughout this work and plotted by the thick black solid lines in the both panels.}
\end{figure*}

The thick black line in Fig. \ref{fig:nmean_vs_t} shows the time
dependence of the mean carbon particle size
\begin{equation}
\langle n\rangle=M_{2}/M_{1},\label{eq:nmean_m2_m1}
\end{equation}
where $M_{2}$ and $M_{1}$ are the second and first moments, respectively,
of the distribution of particle concentrations, defined by Eq. (\ref{eq:moment_cn}).
This time dependence is obtained by numerically solving Eq. \eqref{cnplusminus} 
with parameters listed in Tab. \ref{tab:std_parms}. More specifically,
the set of ordinary differential equations (ODEs), given by Eqs.~\eqref{cnplusminus} - \eqref{cnminus},
is solved using Python's \texttt{scipy.integrate.solve\_ivp} package.\citep{2020SciPy-NMeth}
The ODEs are stiff, so the backward differentiation formula (BDF) numerical integration method\citep{Shampine-1997-1}
with the analytically calculated Jacobian matrix was found to be efficient
with respect to the computation time. The need to explicitly calculate
the Jacobian matrix is memory demanding, so we could not perform calculations
with the maximum particle size larger than $N\sim5\times10^{4}$,
which corresponds to the particle radius of $\sim4\,{\rm nm}$, assuming
the carbon density of $\rho_{C}=3.5\,{\rm g/cm^{3}}$ (ambient diamond).

Figure \ref{fig:nmean_vs_t}(a) shows $\langle n\rangle$ as a function
of time in the log-log scale to facilitate the observation of different
kinetic stages. The vertical blue dashed lines are guides for the
eye to emphasize the transitions between the observed stages, and
the encircled capital letters, (A) through (E), label the stages.
We briefly discuss the observed features of the stages here. The physics behind the stages will be discussed in detail in the sections
to follow (Secs. \ref{sec:Coagulation} to \ref{sec:Coarsening}). The observed kinetic stages are as follows:
\begin{itemize}
\item Stage (A). The numerics of carbon condensation (black solid line)
is seen to agree with the red dashed line, Eq. (\ref{eq:nmean_vs_t_coag}),
up to $\sim2\times10^{-13}\,{\rm s}$.

\item Stage (B). The mean particle size hardly changes from $t\sim2\times10^{-13}\,{\rm s}$
to $t\sim2\times10^{-11}\,{\rm s}$.

\item Stage (C). The mean particle size increases very rapidly from $t\sim2\times10^{-11}\,{\rm s}$
to $t\sim7\times10^{-10}\,{\rm s}$. More specifically, the mean particle
size appears to increase as a power law $\langle n\rangle\propto t^{a}$, with
the exponent approaching $a\sim3$.

\item Stages (D) and (E). The rate of the carbon particles' aggregation drops
at $t\sim7\times10^{-10}\,{\rm s}$, and, as is seen in Fig. \ref{fig:nmean_vs_t}(b),
the aggregation becomes approximately linear, $d\langle n\rangle/dt={\rm const}$,
in Stage (D), i.e., between $t\sim7\times10^{-10}\,{\rm s}$ and $t\sim3\times10^{-9}\,{\rm s}$
at the largest maximum particle size we can afford computationally,
$N=50000$. The rate of this linear aggregation (slope of the black solid line) looks the same as the one given by Eq. (\ref{eq:nmean_vs_t_coag}).
This becomes even clearer if we shift the thick red dash-dotted line
to the left by $2.24\times10^{-10}\,{\rm s}$ in Eq. (\ref{eq:nmean_vs_t_coag})
to yield the thin red dash-dotted line. At times larger than $t\sim3\times10^{-9}\,{\rm s}$,
the black line deviates from the thin red dash-dotted one because
the reached mean particle sizes become comparable to the maximum particle
size allowed numerically. To further illustrate this, we performed
additional calculations with $N=15000$ and $N=30000$; the resulting brown and green dashed lines, respectively, are seen in Fig.~\ref{fig:nmean_vs_t}(b)
to initially follow the $N=50000$ kinetics, but then deviate. Therefore,
the time dependence of the mean particle size in stages (D) and (E),
together with its dependence on the maximum particle size, suggests
that in the limit of $N\rightarrow\infty$, the kinetics of the aggregation is linear from $t\sim3\times10^{-9}\,{\rm s}$ onward with the slope
given by Eq.~\eqref{eq:dnmean_dt_coag}.
\end{itemize}
At the end of this section, we discuss our choice of the definition of the mean particle size, Eq.~\eqref{eq:nmean_m2_m1}. This definition is not unique, and one can, for example, choose the one given by the ratio of the first and zeroth moments of the distribution of particle concentrations, as it was done in Ref.~\onlinecite{Bastea-2012-214106}. The numerical result for this alternative definition is plotted by the thin black dashed line in Fig.~\ref{fig:nmean_vs_t}(a). This line seems to qualitatively follow all the kinetic stages seen in the thick black solid line, except for stage (D), which is absent in the former, as it transitions directly from stage (C) to stage (E). The definition of the mean particle size, given by Eq.~\eqref{eq:nmean_m2_m1}, does, therefore, seem to better emphasize the transitions between different kinetic stages and, as such, is used exclusively in what follows.

An additional motivation for our choice of the definition comes from Ref.~\onlinecite{Watkins-2017-23129}, where it was suggested that $\langle n\rangle =M_2/M_1$ can be physically relevant when considering the small-angle x-ray scattering (SAXS) by carbon particles. This is because the SAXS intensity from a single particle at vanishing scattering angles is proportional to the volume of the particle squared.\citep{Feigin-Svergun-1987}

\section{Smoluchowski Coagulation\label{sec:Coagulation}}

Smoluchowski coagulation,\citep{Smoluchowski-1916-585,Chandrasekhar-1943-1}
as applied to the kinetics of carbon condensation by Shaw and Johnson,\citep{Shaw-1987-2080}
assumes that the fragmentation of carbon particles can be neglected
and so the evolution of carbon particles is governed only by Eq. (\ref{cnplus}),
whereas Eq. (\ref{cnminus}) is disregarded. The rate constant of aggregation in Eq. (\ref{eq:Knm_k0}) can be written as $K_{n,m}^{+}=k_{0}H_{{\rm FW}}(n,m)$, where
\begin{equation}
H_{{\rm FW}}(n,m)=(n^{1/3}+n^{1/3})(n^{-1/3}+m^{-1/3}).\label{eq:g_ij}
\end{equation}
The subscript ``${\rm FW}$'' stands for the work of Friedlander
and Wang,\citep{Friedlander-1966-126} where they considered the coagulation
with this functional form of the rate constant. An additional approximation
could be made, assuming that particles of only very similar sizes
collide and coalesce, so $H_{{\rm FW}}(n,m)$ is approximately substituted
with $H_{{\rm S}}(n,m)=4$. This approximation dates back to the original
publication by Smoluchowski\citep{Smoluchowski-1916-585} (hence the
subscript ``${\rm S}$''), which was adopted by Shaw and Johnson
to obtain exact analytical results for the kinetics of carbon condensation.\citep{Shaw-1987-2080}
It has been demonstrated that no matter the initial distribution of
carbon in particles of various sizes, the long-time coagulation behavior
becomes self-preserving in a sense that the distribution of particle
concentrations is given by a universal functional form $c_{n}\propto f(n/\tilde{n})$,
where $\tilde{n}$ is a characteristic particle size that depends
on time.\citep{Hidy-1965-123} Typically, this distribution is presented
in the scaled and normalized form in the literature so $f(x)$ has
the zeroth and first moments equal to $1$ exactly. Then, the analytical
result for $H_{{\rm S}}(n,m)=4$ (or any other constant value) is $f_{{\rm S}}(x)=e^{-x}$.\citep{Smoluchowski-1916-585,Shaw-1987-2080}
The self-preserving form $f_{{\rm FW}}(x)$, corresponding to Eq.
(\ref{eq:g_ij}), is more complex and could only be obtained numerically.\citep{Friedlander-1966-126}
Importantly for us, the rescaled distribution $f_{{\rm FW}}(x)$ (actually, our analytical fit to the Friedlander-Wang tabulated numerical results)
has the second moment of $m_{{\rm FW},2}=\int_{0}^{\infty}dx\,x^{2}f_{{\rm FW}}(x)\approx1.888$,
so this distribution is somewhat narrower than $f_{{\rm S}}(x)=e^{-x}$
with the second moment of $m_{{\rm S},2}=2$. Once the self-preserving
form of the distribution is known, it could be substituted into Eq.
(\ref{cnplus}) to obtain the time dependence of the characteristic
particle size $\tilde{n}$. It is done by treating particle sizes
$n$ and $m$ as continuous variables (as opposed to integer ones), substituting summation with integration
in Eq. (\ref{cnplus}), and then integrating the l.h.s and rhs of
the equation over $n$ to obtain the evolution of the zeroth moment
of the distribution. The properly normalized expression for the concentrations
of particles is 
\begin{equation}
c_{n}=\frac{c_{tot}}{\tilde{n}^{2}}f(n/\tilde{n}),\label{eq:cn_ntilde}
\end{equation}
with $c_{tot}$ given by Eq. (\ref{eq:ctot}). Then, the zeroth moment
of Eq. (\ref{cnplus}) is 
\begin{equation}
\int_{0}^{\infty}dn\:\frac{dc_{n}}{dt}=-\frac{k_{0}}{2}\int_{0}^{\infty}dn\int_{0}^{\infty}dm\:H(n,m)c_{n}c_{m}.
\end{equation}
Importantly, $H(n,m)$ is a homogeneous function of degree $0$, i.e.,
$H(n/\tilde{n},m/\tilde{n})=H(n,m)$, and therefore we have\footnote{Note, that Refs.~\onlinecite{Shaw-1987-2080,Bastea-2012-214106} produced $d\tilde{n}/{dt}=wk_{0}c_{tot}$, which is lower than our result by exactly a factor of $2$. The reason for this is the choice for the distance of coalescence, discussed in App.~\ref{app:Rnm}.}
\begin{equation}
\frac{d\tilde{n}}{dt}=2wk_{0}c_{tot},\label{eq:dndt_cnplus}
\end{equation}
where
\begin{equation}
w=\frac{1}{4}\int_{0}^{\infty}dn\int_{0}^{\infty}dm\:H(n,m)f(n)f(m).
\end{equation}
This integral yields $w_{{\rm S}}=1$ for $H(n,m)=H_{{\rm S}}(n,m)=4$
and $f_{{\rm S}}(x)=e^{-x}$. If $H(n,m)=H_{{\rm FW}}(n,m)$ and,
correspondingly, $f(x)=f_{{\rm FW}}(x)$, the double integration can
be performed numerically to yield $w_{{\rm FW}}\approx1.0704$. Then,
the rate of coagulation is defined as
\begin{equation}
R_{c}=\frac{d\langle n\rangle}{dt}=\frac{d}{dt}\left\{ \frac{\int_{0}^{\infty}dn\:n^{2}c_{n}}{\int_{0}^{\infty}dn\:nc_{n}}\right\} =2m_{2}wk_{0}c_{tot}.\label{eq:dnmean_dt_coag}
\end{equation}
In the case of the Smoluchowski approximation, $m_{{\rm S},2}=2$ and
$w_{{\rm S}}=1$. For the Friedlander-Wang case, we have $m_{{\rm FW},2}=1.888$
and $w_{{\rm FW}}=1.0704$. The difference between the two cases,
as far as $R_{c}$ is concerned, is determined by the product of $m_2$
and $w$, and is around $1\%$. The rate of coagulation for the standard
system in the Friedlander-Wang case is
\begin{equation}
R_{c}^{(st)}=4.46\times10^{12}\,{\rm 1/s}.\label{eq:Rc_st_FW}
\end{equation}
The resulting dependence of the mean particle size on time is
\begin{equation}
\langle n\rangle=\langle n\rangle_{0}+R_{c}t=1+R_{c}t,\label{eq:nmean_vs_t_coag}
\end{equation}
where it is assumed that initially all the carbon exists in the form
of monomers, as in Tab. \ref{tab:std_parms}.

Eq. (\ref{eq:nmean_vs_t_coag}) is plotted in Figs. \ref{fig:nmean_vs_t}(a) and \ref{fig:nmean_vs_t}(b) by the thick red dash-dotted lines. As is seen, the pure coagulation agrees well with the numerical results within Stage (A), i.e., up to $t\sim2\times10^{-13}\,{\rm s}$. Coagulation is expected to
dominate the short-time behavior because of the following argument.
The processes of aggregation and fragmentation are treated as reactions
of the second and first order in rate equations (\ref{cnplus}) and (\ref{cnminus}), respectively.
The corresponding rates scale
as $\propto c_{\tilde{n}}^{2}\propto c_{tot}^{2}\tilde{n}^{-2}$ and
$\propto c_{\tilde{n}}\propto c_{tot}$, and, therefore, the aggregation
dominates over fragmentation when the characteristic particle size
$\tilde{n}$ is small, resulting in the coagulation being the dominant
process at short times.\footnote{These considerations will become more quantitative with Eq.~\eqref{eq:dntilde_dt_aggr_fragm} and the corresponding discussion in Sec. \ref{subsec:Coag-Coarsening}. In particular, it will become clear what constitutes the small particle size and how it depends on the energetics of carbon particles via $\mu_s$.}
These considerations break down at low $c_{tot}$,
where the aggregation dominates over fragmentation only at $\tilde{n}\apprle1$.
However, such conditions correspond to an unsaturated system where there is no carbon condensation (see Secs.~\ref{sec:Growth} and \ref{subsec:Kinetic-Regimes} for the detailed discussion).

\section{Nucleation and Growth\label{sec:Growth}}

\begin{figure}
\includegraphics[width=0.4\paperwidth]{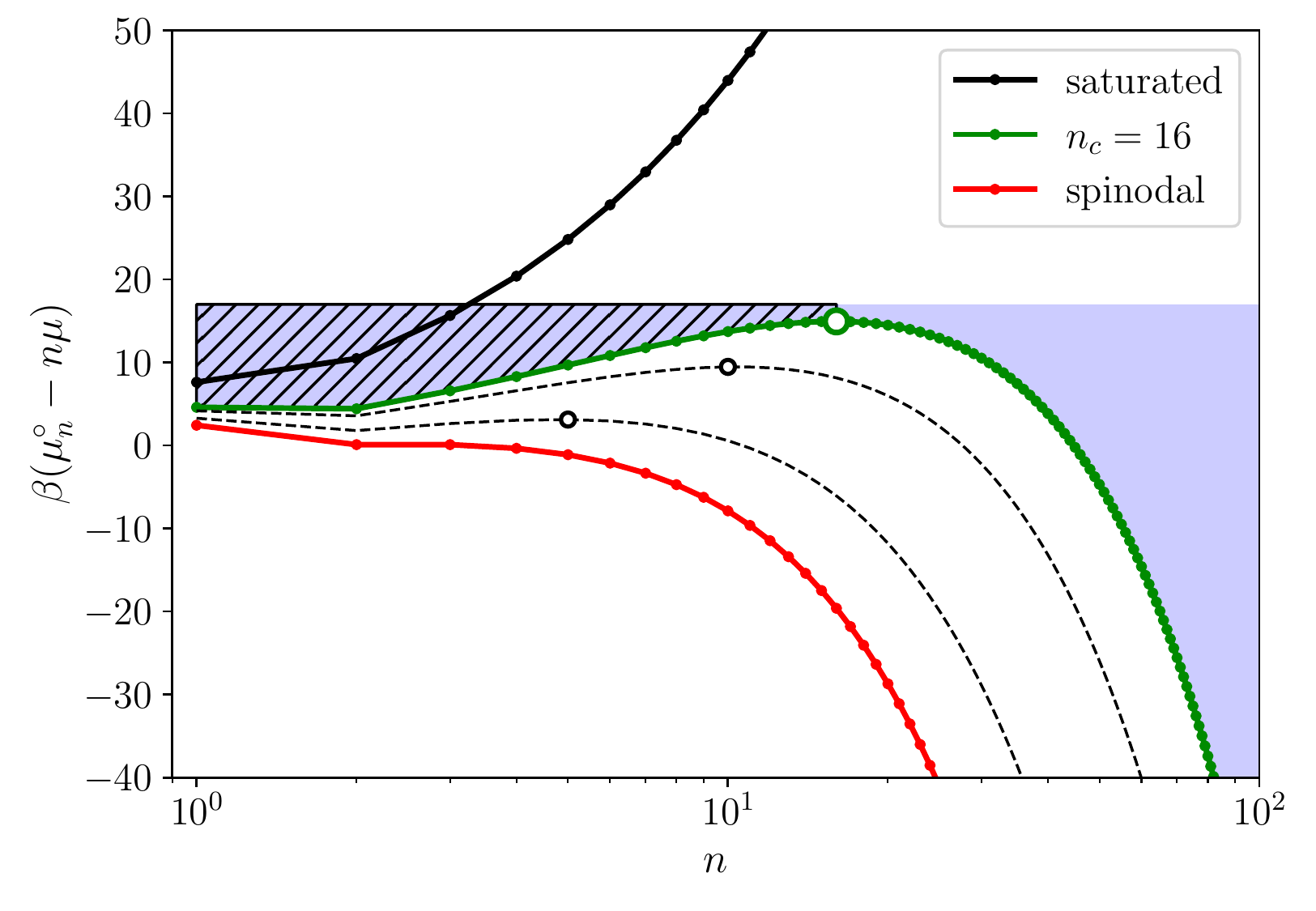}
\caption{\label{fig:mu_vs_n}The negative exponent from Eq. (\ref{eq:cn_eq})
as a function of particle size $n$. The three thick lines (black,
green, and red) correspond to the three different choices for the carbon
chemical potential: $\beta\mu=0$, $\beta\mu\approx3.025$, and $\beta\mu\approx5.185$,
respectively. The empty green circle marks the position of the maximum of the green line. The system parameters are defined by the first
two columns of Tab. \ref{tab:std_parms}. The thin black dashed lines
illustrate how the negative exponent changes when the chemical potential
varies gradually from that of the green line to that of the red line.}
\end{figure}

As just discussed at the very end of Sec. \ref{sec:Coagulation}, the aggregation
is expected to dominate over fragmentation at short times, but the
rates of the two processes become comparable at some point in time,
thus slowing down the rate at which the mean particle size increases. The flattening of the black solid line in Fig. \ref{fig:nmean_vs_t}(a), when getting
from Stage (A) to Stage (B), suggests just that. The similarity of
the aggregation and fragmentation rates suggests close-to-equilibrium
conditions at least for some range of particle sizes, so we first
study the equilibrium behavior of the system. Fig. \ref{fig:mu_vs_n}
shows the negative exponents from Eq. (\ref{eq:cn_eq}) for three
different specifically chosen values of the carbon chemical potential:
$\beta\mu=0$ (black line), $\beta\mu\approx3.025$ (green line),
and $\beta\mu\approx5.185$ (red line).
The first value corresponds to the saturated system - a ``vapor''
of small carbon particles in equilibrium with bulk carbon. The total
carbon concentration in this vapor for the standard set of parameters,
Tab. \ref{tab:std_parms}, is given by Eq. (\ref{eq:cvtot_st}). 

The second value of the chemical potential, $\beta\mu\approx3.025$,
is represented by the green line in Fig. \ref{fig:mu_vs_n}. Since
what is plotted in Fig. \ref{fig:mu_vs_n} is the negative exponent from Eq. (\ref{eq:cn_eq}),
the vertical span of the blue-shaded area gives the logarithm of the
equilibrium particle concentration up to a constant multiplier. One
can see that the particle concentration can be significant at small
and large $n$ but is exponentially suppressed at intermediate particle
sizes. In particular, the concentration is the lowest at $n=n_{c}=16$
- the so-called critical size (hence the subscript ``$c$''), marked
by a green open circle in Fig. \ref{fig:mu_vs_n}. Since the aggregation
and fragmentation rates scale as $c_{n}^{2}$ and $c_{n}$, respectively,
and $c_{n_{c}}$ could be very low, the transfer of particles between
the regions $n\leq n_{c}$ and $n>n_{c}$ by aggregation and fragmentation
is slow. On the other hand, the aggregation and fragmentation of particles
\emph{within} the two regions could be fast. We now switch to a non-equilibrium
situation with the particle concentrations given by
\begin{equation}
c_{n}=\begin{cases}
c^{\circ}e^{-\beta\left(\mu_{n}^{\circ}-n\mu\right)}, & n\leq n_{c}\\
0, & n>n_{c}
\end{cases},\label{eq:cn_mu1}
\end{equation}
where the region of the non-zero concentrations is shown by the diagonal
hatching of the blue-shaded area in Fig. \ref{fig:mu_vs_n}. This
$c_{n}$ does not correspond to equilibrium because the chemical potential
depends on the particle size; it is finite at small $n$ and becomes
abruptly $-\infty$ at $n>n_{c}$. However, since the particle transfer
over the ``potential barrier'' at $n=n_{c}$ is slow relative to
the aggregation and fragmentation at $n<n_{c}$, the quasi-equilibrium
is established for small particles. In other words, one can assume
that there is an equilibrium ensemble of small particles sitting inside
the $n\leq n_{c}$ ``potential well'', and this equilibrium is disturbed
slightly by the slow population transfer over the barrier. This state of the system can then be said to be metastable, as the ensemble
of small particles is long-lived. The value of the chemical potential,
$\beta\mu\approx3.025$, was chosen specifically so that the total
carbon concentration, Eq. (\ref{eq:ctot}), for $c_{n}$ in Eq. (\ref{eq:cn_mu1})
is exactly the standard one ($c_{tot}=4\times10^{28}\,{\rm 1/m^{3}}$), i.e., the one used in the numerical simulations represented by
Fig. \ref{fig:nmean_vs_t}.
One can therefore expect the early-time behavior of the standard system,
Stage (A) in Fig. \ref{fig:nmean_vs_t}, to result in the equilibration
of particles of sizes $n\leq n_{c}=16$. The slow population transfer
over the potential barrier follows this equilibration, resulting in
almost no time dependence of the mean particle size during Stage (B)
in Fig. \ref{fig:nmean_vs_t}. The physical phenomenon when the particle
population reaches the top of the barrier, and then slowly transfers
over it is called \emph{nucleation}.\citep{Friedlander-2000-Smoke,Kalikmanov-2013-Nucleation,Karthika-2016-6663}

Finding the parameters of the quasi-equilibrium ensemble of small
particles, given the total carbon concentration $c_{tot}$, generally
requires a numerical iterative procedure. Specifically, a value for the chemical
potential $\mu$ is guessed and the critical size is determined by
finding where $\mu_{n}^{\circ}-n\mu$ reaches its maximum as a function
of $n$. The combination of Eqs. (\ref{eq:ctot}) and (\ref{eq:cn_mu1})
yields the resulting total carbon concentration. If this concentration
is lower (higher) than the requested $c_{tot}$, the chemical potential
is increased (decreased), and the procedure is repeated until a required
accuracy is reached. This iterative procedure simplifies considerably
if $c_{tot}$ is dominated by monomers, e.g., because $\mu_{n}^{\circ}$
increases rapidly with $n$. Then, $c_{tot}\approx c_{1}=c^{\circ}e^{-\beta\left(\mu_{1}^{\circ}-\mu\right)}$,
and the chemical potential $\mu$ is found directly. We do not use
this approximation in this work, since, for example, it is not applicable
for the standard system; see the discussion pertaining to Fig. \ref{fig:cn_vs_n_t} below. Fig. \ref{fig:nc_dens_plot} shows the critical size $n_{c}$ as a function of total carbon concentration and temperature for the otherwise standard system.
\begin{figure}
\includegraphics[width=0.4\paperwidth]{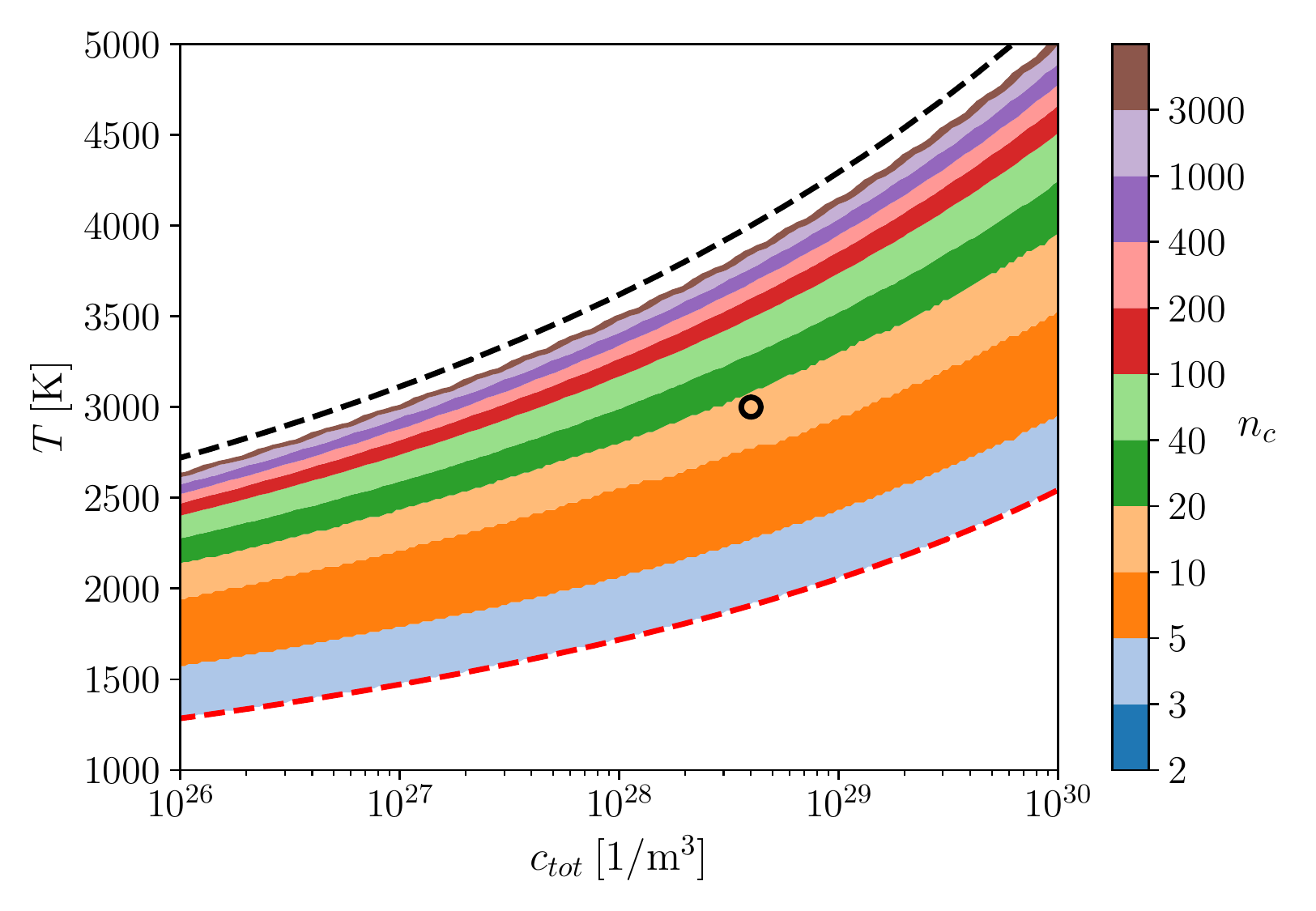}
\caption{\label{fig:nc_dens_plot}The density plot of the critical size, $n_{c}$,
as a function of $c_{tot}$ and temperature (other parameters are
standard). The circle corresponds to all the parameters being standard.}
\end{figure}
The kinetic behavior of the system depends of course on all the parameters
in Tab. \ref{tab:std_parms}, but the total carbon concentration and
temperature are deemed the most important ones because (i) they are
directly controlled when going through a range of HEs in the experiment\citep{Bastea-2017-42151,Watkins-2018-821}
and (ii) variations in some other parameters affect the kinetics less
critically. For example, the viscosity affects the kinetics by simply
rescaling the time.

As is seen in Fig. \ref{fig:nc_dens_plot}, the critical size decreases
when $c_{tot}$ increases at a fixed temperature, and the described
iterative procedure does not have a solution when $c_{tot}$ is outside
a certain finite range. First, if $c_{tot}$ is lower than $c_{v,tot}$
in Eq. (\ref{eq:cvtot_st}), the system is unsaturated, i.e., there
is no carbon condensation, and, in particular, no nucleation. The
resulting final equilibrium state of the system is unsaturated carbon
vapor with concentrations given by Eq. (\ref{eq:cn_eq}), where $\beta(\mu_{n}^{\circ}-n\mu)$
looks qualitatively similar to the black line in Fig. \ref{fig:mu_vs_n}.
The saturated carbon concentration, $c_{v,tot}$, as a function of
temperature is plotted by a black dashed line in Fig. \ref{fig:nc_dens_plot}.

The position of the maximum of $\beta(\mu_{n}^{\circ}-n\mu)$, i.e.,
the critical size, is seen to decrease when $c_{tot}$ is increased
at a fixed temperature, and the maximum disappears when $c_{tot}$
becomes larger than a certain threshold. At this threshold, the expression
$\beta(\mu_{n}^{\circ}-n\mu)$ is plotted by the red line in Fig.
\ref{fig:mu_vs_n}, and the two thin black dashed lines show the evolution
of $\beta(\mu_{n}^{\circ}-n\mu)$ when $c_{tot}$ is increased from
$4\times10^{28}\,{\rm 1/m^{3}}$(green line) to $\approx4.6\times10^{30}\,{\rm 1/m^{3}}$
(red line). Specifically, these two dashed lines are obtained by solving
the iterative problem for $c_{tot}=10^{29}\,{\rm 1/m^{3}}$ and $c_{tot}=10^{30}\,{\rm 1/m^{3}}$,
resulting in $n_{c}=10$ and $n_{c}=5$, respectively. The red dashed
line in Fig. \ref{fig:nc_dens_plot} corresponds to this high-$c_{tot}$
threshold where the maximum disappears in $\beta(\mu_{n}^{\circ}-n\mu)$.
At even larger concentrations, the nucleation barrier is absent, and
so the initial state of the system - carbon ``vapor'' - is unstable,
as opposed to metastable when the nucleation barrier is present. The
condensation without nucleation is typically referred to as the spinodal
decomposition or spinodal transformation in the literature.\citep{Binder-1987-783,Puri-2004-407,Clouet-2009-203}
For the standard system with varied $c_{tot}$, the spinodal threshold corresponds to the
total carbon concentration of
\begin{equation}
c_{tot}^{(spin)}\approx 4.6\times10^{30}\,{\rm 1/m^{3}}.\label{eq:cspin_tot_st}
\end{equation}

\subsection{Quasi-Equilibrium\label{subsec:Quasi-Equilibrium}}

Above, we considered a situation where an ensemble of carbon particles
becomes effectively trapped by a nucleation barrier so the distribution
of carbon particle concentrations is very approximately given by Eq.
(\ref{eq:cn_mu1}), and there is a non-vanishing flux of the carbon
particles over the barrier. It is clear, however, that in the established
quasi-equilibrium, the concentration $c_{n}$ cannot vanish abruptly
at the top of the nucleation barrier, since otherwise the flux of
carbon particles vanishes immediately beyond the barrier. It is the
goal of this subsection to find a more accurate expression for $c_{n}$
in the quasi-equilibrium, and the carbon flux over the barrier corresponding
to it. To this end, we introduce the concentration flux as 
\begin{equation}
J_{n}(t)=\frac{d}{dt}\left[\sum_{m=n+1}^{\infty}c_{m}(t)\right],\label{eq:Jn_ddt_M0}
\end{equation}
 that is $J_{n}$ is the number of carbon particles, per unit volume
per unit time, that cross an imaginary boundary, put between particle
sizes $n$ and $n+1$, from left to right. It follows from Eqs. (\ref{cnplus})
and (\ref{cnminus}) that $J_{n}$ can be written as 
\begin{equation}
J_{n}=\sum_{i}\sum_{m=n-i+1}^{n}K_{m,i}^{+}c_{m}c_{i}-\sum_{i}\sum_{m=n-i+1}^{n}K_{m,i}^{-}c_{m+i}.\label{eq:J_n}
\end{equation}
According to Eq. (\ref{eq:cn_mu1}), most of the carbon population
sits in particles of sizes $n\leq n_{c}$. We can, therefore, neglect
aggregation processes where the two colliding particles are both larger
than $n_{c}$. The respective reverse processes of fragmentation are
neglected as well for consistency. This implies that the index $i$
in Eq. (\ref{eq:J_n}) runs only from $i=1$ to $i=i_{v}<n_{c}$ Here, $i_v$ is some effective largest particle size in the metastable ensemble of
small carbon particles below the nucleation barrier. As mentioned above, this
ensemble is referred to as ``vapor'', hence the subscript ``$v$''
of $i_{v}$. The concentration of particles in the vapor is given
by Eq. (\ref{eq:cn_eq}), specifically
\begin{equation}
c_{v,i}=c^{\circ}e^{\beta(i\mu-\mu_{i}^{\circ})},\label{eq:c_v_i}
\end{equation}
where $\mu$ is the chemical potential of the quasi-equilibrated vapor, generally found iteratively, as described above.
Substituting this expression into Eq. (\ref{eq:J_n}) results in
\begin{align}
J_{n} & =\sum_{i}\sum_{m=n-i+1}^{n}K_{m,i}^{+}c_{m}c^{\circ}e^{\beta(i\mu-\mu_{i}^{\circ})}\nonumber \\
 & -\sum_{i}\sum_{m=n-i+1}^{n}K_{m,i}^{+}c^{\circ}e^{\beta(\mu_{m+i}^{\circ}-\mu_{m}^{\circ}-\mu_{i}^{\circ})}c_{m+i}.
\end{align}
Assuming that the concentration $c_{n}$ does not change too rapidly
as a function of $n$ allows one to treat it as a continuous function
of $n$ and use its Taylor expansion, $c_{m}\approx c_{n}+(m-n)c_{n}^{\prime}$,
where $c_{n}^{\prime}=\frac{\partial c_{n}}{\partial n}$, to produce
\begin{equation}
J_{n}=a(n)c_{n}-b(n)c_{n}^{\prime}\label{eq:Jn_a_b}
\end{equation}
where 
\begin{equation}
a(n)=\sum_{i}\sum_{m=n-i+1}^{n}K_{m,i}^{+}c^{\circ}\left[e^{\beta(i\mu-\mu_{i}^{\circ})}-e^{\beta(\mu_{m+i}^{\circ}-\mu_{m}^{\circ}-\mu_{i}^{\circ})}\right]\label{eq:an}
\end{equation}
and
\begin{align}
b(n) & =\sum_{i}\sum_{m=n-i+1}^{n}K_{m,i}^{+}c^{\circ}\left[(n-m)e^{\beta(i\mu-\mu_{i}^{\circ})}\right.\nonumber \\
 & \left.+e^{\beta(\mu_{m+i}^{\circ}-\mu_{m}^{\circ}-\mu_{i}^{\circ})}(m+i-n)\right]\label{eq:bn}
\end{align}
This result is essentially identical to that previously obtained by Binder and
Stauffer.\citep{Binder-1976-343} Equation \eqref{eq:Jn_a_b} can be thought
of as a one-dimensional drift-diffusion equation, where $c_{n}$ is
interpreted as the probability to find some effective Brownian particle at
coordinate $n$. The first rhs term is then the drift term with
the drift velocity $a(n)$, and the second rhs term describes diffusion
with the diffusion coefficient $b(n)$. Importantly, $b(n)$ is always
positive, which actually allows for its interpretation as the diffusion
coefficient, and $a(n)$ is negative at small $n$ and positive at
large $n$ for the over-saturated vapor ($\mu>0$). If $n$, at which
$a(n)$ crosses from negative to positive, is sufficiently larger
than $i_{v}$, the condition for this crossing is $i\mu=\mu_{n+i}^{\circ}-\mu_{n}^{\circ}$.
This condition is essentially equivalent to that for the maximum of
$\beta(\mu_{n}^{\circ}-n\mu)$ discussed above, and so $a(n)$ crosses
from negative to positive at the critical size, $n=n_{c}$. Another
important observation for derivations to follow is that the combination
of Eqs. (\ref{eq:mu_n_o_mu_s_n}) and (\ref{eq:musn_mus}) suggests
that $\lim_{n\rightarrow\infty}\left(\mu_{n+i}^{\circ}-\mu_{n}^{\circ}\right)=0$,
resulting in $a(n),b(n)\propto K_{n,i}^{+}$ at large $n$. This,
in turn, produces $a(n)/b(n)={\rm const}>0$ at $n\rightarrow\infty$.

Solving Eq. \eqref{eq:Jn_a_b} can, in principle, be approached by supplementing
it with $\frac{\partial c_{n}}{\partial t}=-J_{n}^{\prime}$ and then
solving the resulting second-order diffusion-drift partial differential
equation for $c_{n}$ as a function of both $n$ and $t$. The problem
is simplified, however, by the quasi-equilibrium assumption with the
time-independent $c_{v,i}$. We further introduce an ``absorbing''
boundary at some $n=n_{a}\gg n_{c}$ so that $c_{n}=0$ for $n\geq n_{a}$.
Under these conditions, we have an infinite source ($c_{v,i}$) and
sink (absorbing boundary) of carbon particles, so the system is expected
to reach its steady state with $n$-independent concentration flux
$J_{n}=J={\rm const}$, and Eq. (\ref{eq:Jn_a_b}) transforms into
\begin{equation}
c_{n}^{\prime}-\frac{a(n)}{b(n)}c_{n}=-\frac{J}{b(n)}.\label{eq:cnprime_cn_J}
\end{equation}
This steady-state problem is a standard way to approach the kinetics of nucleation in the literature.\citep{Ford-1997-5615,Clouet-2009-203}
The obtained ordinary differential equation ($J$ is still an unknown
constant) can be solved by first solving the homogeneous one ($J=0$),
which produces the Green's function for the lhs of Eq. (\ref{eq:cnprime_cn_J}),
and then using the Green's function to obtain a particular solution
of the full inhomogeneous differential equation. The result is
\begin{equation}
c_{n}=Ae^{\int_{0}^{n}dm\,\frac{a(m)}{b(m)}}-\int_{0}^{n}dn'\,\frac{J}{b(n')}e^{\int_{n'}^{n}dm\,\frac{a(m)}{b(m)}},\label{eq:cn_A_J}
\end{equation}
where the first rhs term is the general homogeneous solution, and
the second rhs term is a particular inhomogeneous solution. The
two constants of integration, $J$ and $A$, are to be found from
the conditions of the known concentrations at very small sizes, $c_{n}=c_{v,n}$,
and that $c_{n}$ vanishes at the absorbing boundary condition. Substituting
the latter condition ($c_{n}=0$ at $n=n_{a}$) into Eq. (\ref{eq:cn_A_J})
results in 
\begin{equation}
J=\frac{A}{\int_{0}^{n_{a}}dn'\,\frac{1}{b(n')}e^{-\int_{0}^{n'}dm\,\frac{a(m)}{b(m)}}}.
\end{equation}
The integral $-\int_{0}^{n'}dm\,\frac{a(m)}{b(m)}$ is a concave function
of $n'$, and reaches its maximum where $a(n')$ vanishes, i.e., at
$n'=n_{c}$. This integral decays linearly at large $n'$, and therefore
\begin{equation}
F(n_{a})=\int_{0}^{n_{a}}dn'\,\frac{1}{b(n')}e^{-\int_{0}^{n'}dm\,\frac{a(m)}{b(m)}}
\end{equation}
converges to a finite value at $n_{a}\rightarrow\infty$. We can therefore push the
absorbing boundary to $n_a=\infty$ and evaluate the flux as
\begin{equation}
J=\frac{A}{F(\infty)}.
\end{equation}
Substituting this expression into Eq. (\ref{eq:cn_A_J}) yields
\begin{equation}
c_{n}=Ae^{\int_{0}^{n}dm\,\frac{a(m)}{b(m)}}\left[1-\frac{F(n)}{F(\infty)}\right].\label{eq:cn_Fn_Finfty}
\end{equation}
Functions $a(n)$ and $b(n)$ do have a complicated dependence on
$n$ so the integrals above are not easy to evaluate. We first find
$c_{n}$ and $n$ near the critical size $n_{c}$. At around the critical
size, Eqs.~(\ref{eq:an}) and (\ref{eq:bn}) can be approximated
by Taylor expansions of $\mu_{m}^{\circ}$ with respect to $m$ as
\begin{align}
a(n) & =\sum_{i}\sum_{m=n-i+1}^{n}K_{m,i}^{+}c_{v,i}\left[1-e^{\beta(\mu_{m+i}^{\circ}-\mu_{m}^{\circ}-i\mu)}\right]\nonumber \\
 & \approx\left[\beta\mu-\beta\mu_{n}^{\circ\prime}\right]\sum_{i}\sum_{m=n-i+1}^{n}iK_{m,i}^{+}c_{v,i}.
\end{align}
and, similarly
\begin{equation}
b(n)\approx\sum_{i}\sum_{m=n-i+1}^{n}iK_{m,i}^{+}c_{v,i},\label{eq:bn_approx}
\end{equation}
and, therefore, approximately we have $a(n)/b(n)\approx\beta\left[\mu-\mu_{n}^{\circ\prime}\right]$.
This function is negative at small $n<n_{c}$, and reaches a positive
plateau at large $n$, as mentioned above. We evaluate
\begin{equation}
g(n)=-\int_{0}^{n}dm\,\frac{a(m)}{b(m)}=-\beta(n\mu-\mu_{n}^{\circ})+B,
\end{equation}
at $n\sim n_{c}$, where $B$ is the integration constant originating
from (i) the incorrect lower integration limit of $0$ (lowest particle size
is $n=1$) and also (ii) the fact that at very low $m$, the ratio $a(m)/b(m)$ cannot be accurately represented by $\beta\left[\mu-\mu_{m}^{\circ\prime}\right]$.
Function $F(n)$ then becomes
\begin{equation}
F(n)=\int_{0}^{n}dm\,\frac{1}{b(m)}e^{-\beta(m\mu-\mu_{m}^{\circ})+B}.
\end{equation}
In this expression, the integrand is seen to have a pronounced maximum
at $m\sim n_{c}$, and therefore, $F(n)\ll F(\infty)$ if $n<n_{c}$.
At $n$ below the critical size, we, therefore, obtain from Eq.~(\ref{eq:cn_Fn_Finfty})
\begin{equation}
c_{n}\approx Ae^{\int_{0}^{n}dm\,\frac{a(m)}{b(m)}}=Ae^{\beta(n\mu-\mu_{n}^{\circ})-B}.
\end{equation}
Since sufficiently below the critical size one must have $c_{n}=c_{v,n}$,
the integration constants are determined by $c^{\circ}=Ae^{-B}$.
The concentration flux then becomes
\begin{equation}
J=\frac{c^{\circ}}{\int_{0}^{\infty}dn'\,\frac{1}{b(n')}e^{-\beta(n'\mu-\mu_{n'}^{\circ})}}=\frac{c^{\circ}}{\tilde{F}(\infty)},\label{eq:J_co_Ftilde}
\end{equation}
and the concentration is 
\begin{equation}
c_{n}=c_{v,n}\left[1-\frac{\tilde{F}(n)}{\tilde{F}(\infty)}\right],\label{eq:cn_Ftilde}
\end{equation}
where
\begin{equation}
\tilde{F}(n)=F(n)e^{-B}=\int_{0}^{n}dm\,\frac{1}{b(m)}e^{-\beta(m\mu-\mu_{m}^{\circ})}.\label{eq:Ftilde}
\end{equation}
Eq. (\ref{eq:cn_Ftilde}) gives a more accurate dependence of the
quasi-equilibrium particle concentration on size than Eq. (\ref{eq:cn_mu1}).
An additional very useful approximation for $b(n)$ in Eq.~(\ref{eq:bn_approx})
is obtained by substituting $K_{m,i}^{+}$ in there with $K_{n,i}^{+}$
to yield
\begin{equation}
b(n)\approx\sum_{i}\sum_{m=n-i+1}^{n}iK_{n,i}^{+}c_{v,i}=\sum_{i}i^{2}K_{n,i}^{+}c_{v,i}.\label{eq:bn_approx1}
\end{equation}
In function $\tilde{F}(n)$, we still have the integration, which is somewhat ambiguous since it starts from $0$, whereas the minimum
cluster size is $n=1$. We correct it empirically by comparing Eq.~(\ref{eq:cn_Ftilde})
for concentration, supplemented with Eqs.~(\ref{eq:Ftilde}) and
(\ref{eq:bn_approx1}), with that for the case of ideal (i.e., monomer-only)
vapor where instead of the approximate differential equation, Eq.~(\ref{eq:cnprime_cn_J}),
one has a recurrence relation between $c_{n+1}$ and $c_{n}$.\citep{Ford-1997-5615,Clouet-2009-203}
More specifically, a comparison with Eq.~(84) in Ref.~\onlinecite{Clouet-2009-203}
yields a corrected expression for $\tilde{F}(n)$ as
\begin{equation}
\tilde{F}(n)=\sum_{m=1}^{n-1}\frac{1}{b(m)}e^{-\beta\left[m\mu-\mu_{m}^{\circ}\right]}.
\end{equation}
Equation \eqref{eq:cn_Ftilde} then becomes
\begin{equation}
c_{n}=c_{v,n}\frac{J}{c^{\circ}}\sum_{m=n}^{\infty}\frac{1}{b(m)}e^{-\beta\left[m\mu-\mu_{m}^{\circ}\right]}.
\end{equation}
At large $n$, we approximate $b(m)\approx b(n)$ and $\mu_{m}^{\circ}\approx\mu_{n}^{\circ}+(m-n)\mu_{n}^{\circ\prime}$
inside the summation to obtain
\begin{equation}
c_{n}\approx\frac{J}{b(n)}\frac{1}{1-e^{-\beta\left[\mu-\mu_{n}^{\circ'}\right]}},
\end{equation}
where Eq.~(\ref{eq:bn_approx1}) at large $n$ becomes
\begin{equation}
b(n)\approx\sum_{i}i^{2}k_{0}n^{1/3}i^{-1/3}c_{v,i}=b_{\infty}n^{1/3},\label{eq:b_b_infty}
\end{equation}
where $b_{\infty}=\sum_{i}i^{5/3}k_{0}c_{v,i}$. The concentration then becomes ($\mu_{n}^{\circ\prime}$ is neglected at large $n$)
\begin{equation}
c_{n}=c_{\infty}n^{-1/3},\label{eq:cn_c_infty}
\end{equation}
where $c_{\infty}=\frac{J}{b_{\infty}\left(1-e^{-\beta\mu}\right)}$.
We now estimate $J$ analytically. To this end, we first calculate
$\tilde{F}(\infty)$ using the Laplace approximation at $m\sim n_{c}$
as
\begin{align}
\tilde{F}(\infty) & \approx\frac{1}{b(n_{c})}e^{-\beta(n_{c}\mu-\mu_{n_{c}}^{\circ})}\int_{-\infty}^{\infty}dm\,e^{-\beta\left|\mu_{m}^{\circ\prime\prime}\right|(m-n_{c})^{2}/2}\nonumber \\
 & =\frac{1}{b(n_{c})}e^{-\beta(n_{c}\mu-\mu_{n_{c}}^{\circ})}Z^{-1},\label{eq:Ftilde_Z}
\end{align}
where $b(n_{c})$ cannot be approximated by Eq.~(\ref{eq:b_b_infty}),
since $n_{c}$ is not necessarily very large. The so-called Zeldovich
factor is denoted by $Z=\left(\frac{\beta\left|\mu_{n_{c}}^{\circ\prime\prime}\right|}{2\pi}\right)^{1/2}=\left(\frac{\beta\mu_{s}}{9\pi n_{c}^{4/3}}\right)^{1/2}$.

\begin{figure}
\includegraphics[width=0.4\paperwidth]{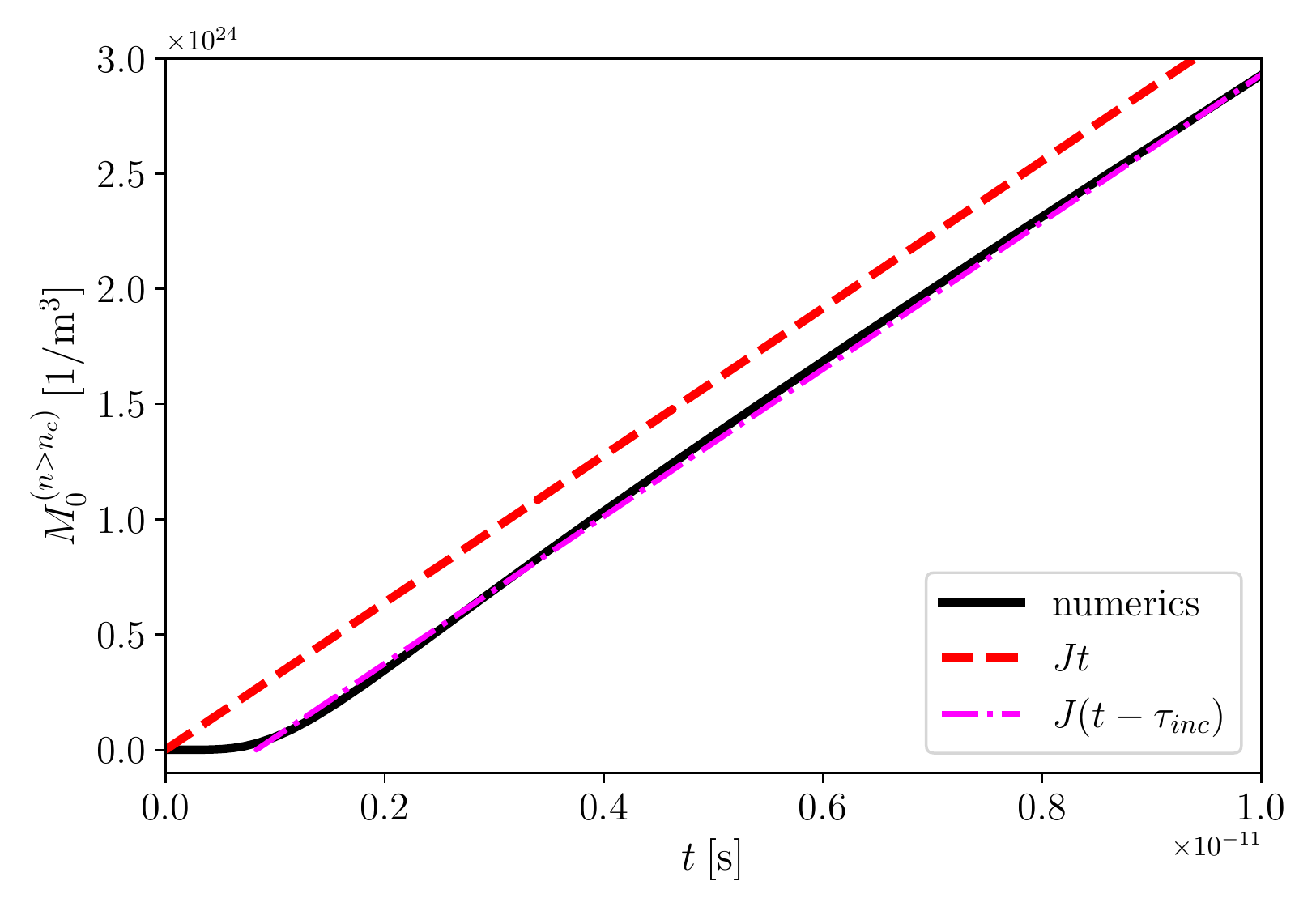}
\caption{\label{fig:M0_vs_t}The numerical result for $M_{0}^{(n>n_{c})}=\sum_{n=n_{c}+1}^{N}c_{n}$
for the standard system is plotted by the black solid line. The red dashed line represents $Jt$, where the steady-state concentration flux
$J$ is given by Eq. (\ref{eq:J_final}). The magenta dash-dotted line is $J(t-\tau_{inc})$ with $\tau_{inc}=8.3\times10^{-13}\,{\rm s}$.}
\end{figure}

To summarize our analytical results, the final expression for the steady-state concentration
flux is obtained by combining Eqs. (\ref{eq:Ftilde_Z}) and (\ref{eq:J_co_Ftilde})
to produce
\begin{equation}
J=b(n_{c})Zc^{\circ}e^{-\beta(\mu_{n_{c}}^{\circ}-n_{c}\mu)},\label{eq:J_final}
\end{equation}
where $b(n_{c})$ is calculated using Eq. (\ref{eq:bn_approx1}),
not Eq. (\ref{eq:b_b_infty}). Once the flux is evaluated, the steady-state
concentration of large particles is obtained from Eq. (\ref{eq:cn_c_infty}).
Equation \eqref{eq:Jn_ddt_M0} for the concentration flux can be interpreted
as the time derivative of the zeroth moment of the distribution of
particle concentrations, Eq. (\ref{eq:moment_cn}) where, however,
the summation runs not from $n=1$ but from $n>n_{c}$. The numerical
result (standard parameters) for such a moment, $M_{0}^{(n>n_{c})}$, is plotted by the black line in Fig. \ref{fig:M0_vs_t}.
As is seen, this moment does not increase initially since there are no
carbon particles at the top of the nucleation barrier, so the quasi-equilibrium
is not established yet. After $t\sim 10^{-12}\,{\rm s}$, $M_{0}^{(n>n_{c})}$
starts increasing linearly, and its slope is seen to be very similar to
that of $Jt$, plotted by the red dashed line, where $J$ is evaluated
from Eq. (\ref{eq:J_final}), thus supporting the above derivations
and the entire quasi-equilibrium concept. In particular, that the
numerical slope does not seem to change appreciably within the time range in Fig. \ref{fig:M0_vs_t} means that the quasi-equilibration of the carbon vapor below the nucleation barrier is indeed established
before the noticeable fraction of the carbon is transferred over the nucleation barrier.
To find the time required to establish the quasi-equilibrium - \emph{incubation
time}\citep{Clouet-2009-203} - we shift the $Jt$ line by adding a time delay. The resulting magenta dash-dotted line is observed to agree well with the numerical results for the incubation time of $\tau_{inc}=8.3\times10^{-13}\,{\rm s}$.

\begin{figure}
\includegraphics[width=0.4\paperwidth]{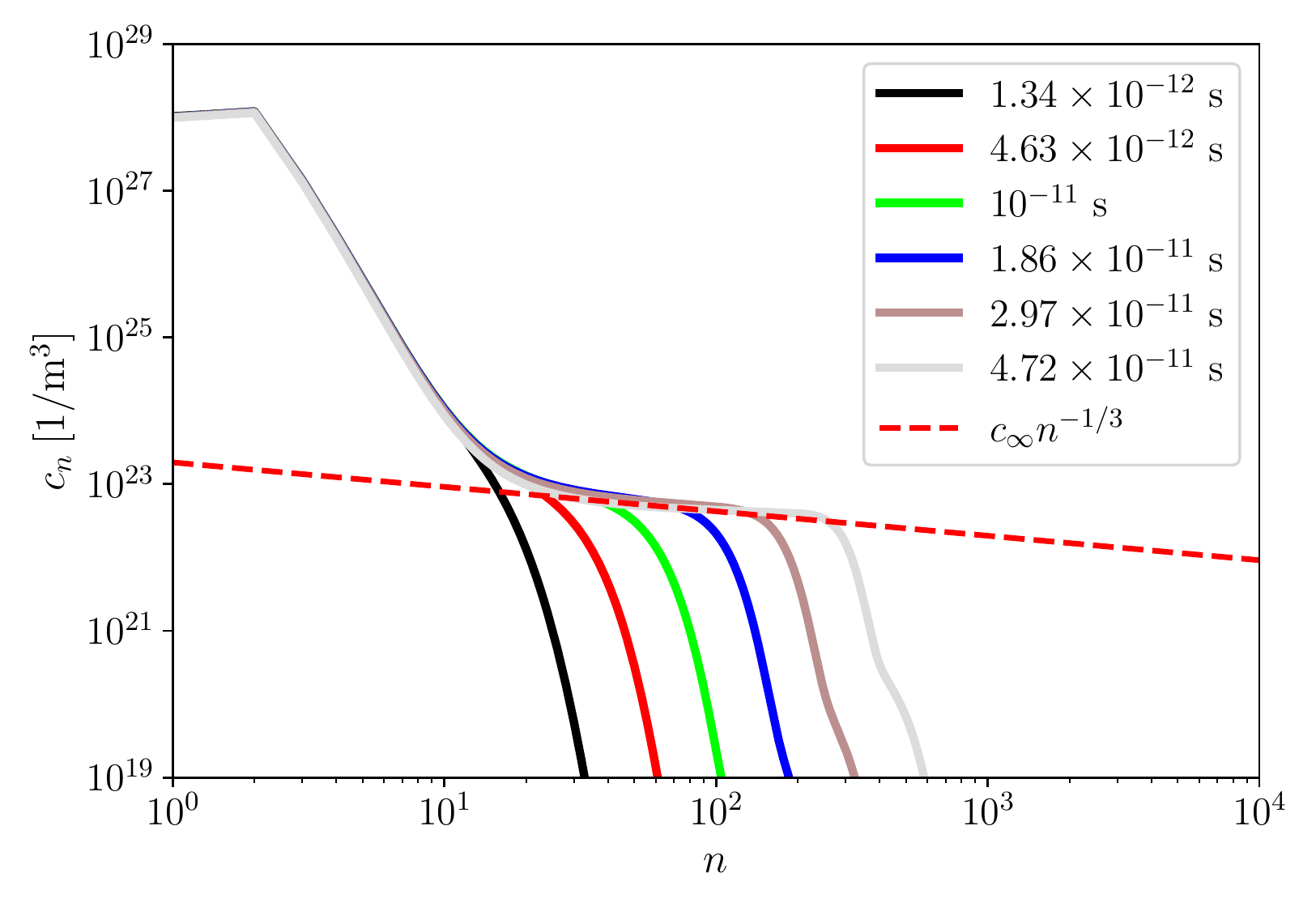}
\caption{\label{fig:cn_vs_n_t}Time evolution of the distribution of carbon
particle concentrations (solid lines), obtained numerically for the
standard system, Tab. \ref{tab:std_parms}. The analytical result for the asymptotic (large
times, large particles) distribution of concentrations, Eq. (\ref{eq:cn_c_infty}),
is plotted by the red dashed line.}
\end{figure}

The evolution of the distribution of particle concentrations in the
same exact numerical simulation is shown in Fig. \ref{fig:cn_vs_n_t}
by the solid lines.
As is seen, the vapor concentrations ($n<n_{c}=16$) do not change
within the time range from $t\sim10^{-12}\,{\rm s}$ to at least $\sim5\times10^{-11}\,{\rm s}$,
again supporting the quasi-equilibrium approximation. Concentrations
of larger particles can be observed to gradually converge to the asymptotic
dependence, Eq. (\ref{eq:cn_c_infty}), plotted by the red dashed line. Another important observation here is that even though the concentration
of particles in the vapor decreases very rapidly with the size, so
that $i_{v}$ can be safely adopted to be e.g., $5$, the metastable vapor is
not dominated by monomers, since the concentration of dimers ($n=2$)
is actually higher, albeit not by much. In this situation, one cannot
use the approximation that the vapor is dominated by monomers, which
is by far the prevalent one in the literature. In particular, it means
that (i) the actual summation has to be performed in Eq. (\ref{eq:bn_approx1})
instead of just taking the first $i=1$ term, and (ii) the iterative
procedure to find the quasi-equilibrium chemical potential $\mu$,
described above, cannot be simplified by assuming $c_{tot}\approx c_{v,1}$.

\subsection{From Quasi-Equilibrium to Growth}

\begin{figure*}
\includegraphics[width=0.4\paperwidth]{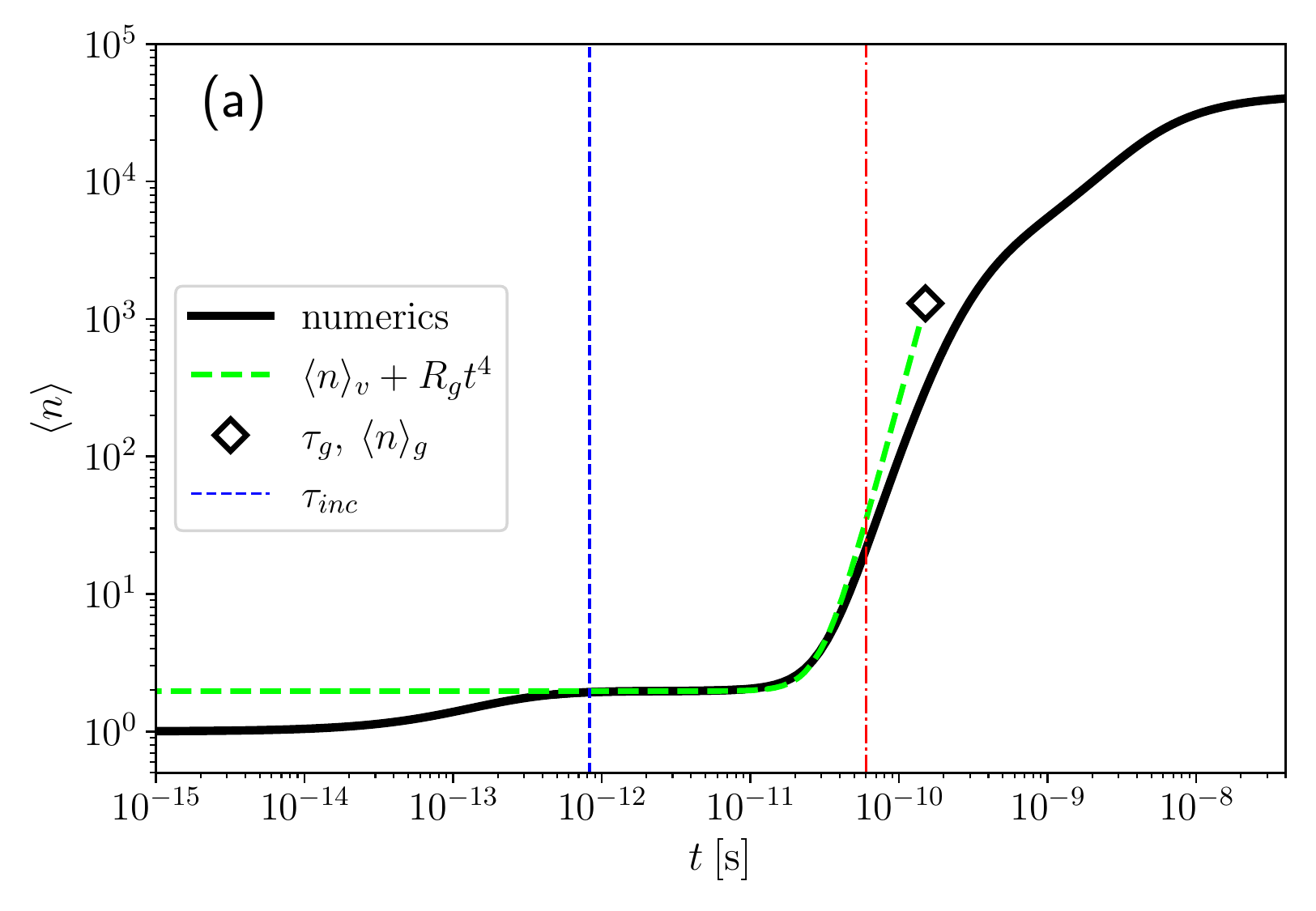}
\includegraphics[width=0.4\paperwidth]{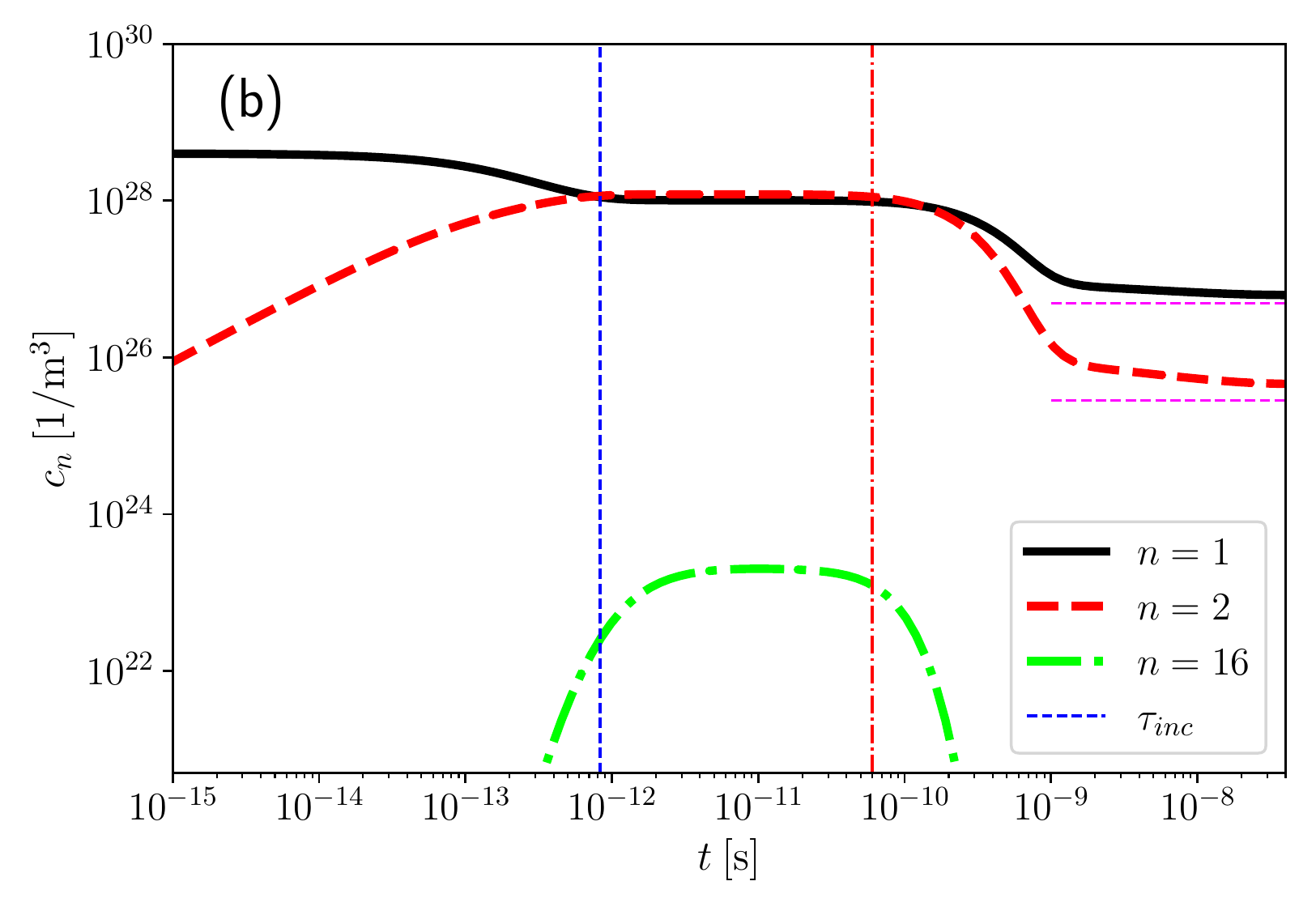}
\caption{\label{fig:nmean_t_growth}(a) Comparison of the numerical result
for the mean particle size (black solid line) and the analytical result
for the growth step (green dashed line). The black solid line is re-plotted
from Fig. \ref{fig:nmean_vs_t}(a). The green dashed line represents Eq.
(\ref{eq:nmean_Rgt4}) with the mean particle size, corresponding
to the quasi-equilibrium carbon vapor, added to the rhs. The thin red dash-dotted line is re-plotted from Fig. \ref{fig:nmean_vs_t}(b). The diamond
shape marks the point where the growth ends, according to Eqs. (\ref{eq:taug})
and (\ref{eq:nmeang}). (b) The time dependence of concentrations
of carbon particles of sizes $n=1$, $2$ and $16$, obtained in the
same simulations as in panel (a). The thin magenta dashed lines give the
saturated monomer and dimer concentrations, $c_{v,1}^{(sat)}$ and
$c_{v,2}^{(sat)}$, respectively.}
\end{figure*}

The key observation from the numerical results in Fig. \ref{fig:cn_vs_n_t}
is that the dependence of $c_{n}$ on $n$ seems to follow the steady-state prescription of Eq.~(\ref{eq:cn_c_infty}), but only up to
a certain $n=n_{g}=n_{g}(t)$, beyond
which the concentration drops very rapidly. This should allow us,
in principle, to evaluate this $n_{g}$ as a function of time, and
therefore to describe the evolution of the concentration distribution.
One way to obtain this dependence is to notice that the discussed
observation allows one to find $n_{g}$ and $\Delta n$, so that $0<\Delta n\ll n_{g}$
and $c_{n_{g}-\Delta n}$ still follows Eq.~(\ref{eq:cn_c_infty}),
whereas $c_{n_{g}+\Delta n}$ is already much smaller. That the inequality
$\Delta n\ll n_{g}$, obvious for the specific example in Fig. \ref{fig:cn_vs_n_t},
can be satisfied in general is demonstrated in App. \ref{sec:Drift-Diffusion}.
The ``hydrodynamic cell'' limited by the cluster sizes $n_{g}-\Delta n$
and $n_{g}+\Delta n$ can be introduced so that the concentration
in-fluxes are $J$ and $\sim0$ from the left and right, respectively.
The characteristic time $\Delta t$ of filling this cell in up to
the steady-state concentration is determined by $J\Delta t=2\Delta nc_{n_{g}}$,
upon which $n_{g}$ shifts to the right as $n_{g}\rightarrow n_{g}+2\Delta n$.
Substituting finite differences with differentials, and using Eq.~(\ref{eq:cn_c_infty})
for $c_{n_{g}}$ at large $n_{g}$, we obtain
\begin{equation}
\frac{dn_{g}}{dt}=\frac{J}{c_{n_{g}}}=\alpha n_{g}^{1/3},\label{eq:dng_dt}
\end{equation}
where $\alpha=b_{\infty}\left(1-e^{-\beta\mu}\right)$. Physically,
$n_{g}$ increases with time because whenever a large particle ($n\gg n_{c}$)
is formed, small carbon clusters ($n<n_{c})$ from the over-saturated
vapor bind to its surface almost irreversibly, further increasing
its size. This step of the kinetics of the first-order phase transition
is typically referred to as the \emph{growth} in the literature,\citep{Penrose-1978-Kinetics,Binder-1987-783,Krapivsky-2010-Kinetic} hence the subscript ``$g$'' of $n_g$. The ODE in Eq. (\ref{eq:dng_dt}) is straightforwardly solved to produce
$n_{g}=\left(\frac{2}{3}\alpha t\right)^{3/2}$. Assuming that the
second moment of the concentration distribution is dominated by large
particles, one obtains
\begin{align}
M_{2} & =\sum_{n=1}^{\infty}n^{2}c_{n}\approx\int_{0}^{n_{g}}dn\,n^{2}c_{\infty}n^{-1/3}=\frac{3}{8}c_{\infty}n_g^{8/3} \nonumber\\
& =\frac{2}{27}c_{\infty}\alpha^{4}t^{4}.
\end{align}
The time dependence of the average particle size during the growth is then expressed by
\begin{equation}
\langle n\rangle=M_{2}/M_{1}=R_{g}t^{4},\label{eq:nmean_Rgt4}
\end{equation}
where $R_{g}=\frac{2}{27}\frac{c_{\infty}}{c_{tot}}\alpha^{4}$. We
can also estimate the time when the growth ends, $\tau_{g}$, by assuming
that all carbon is transferred from the metastable vapor to large particles by the
end of the growth and so $c_{tot}\approx\int_{0}^{n_{g}}dn\:nc_{n}$,
where $c_{n}$ is given by Eq.~(\ref{eq:cn_c_infty}). Straightforward algebra produces
\begin{equation}
\tau_{g}=\frac{3}{2\alpha}\left(\frac{5}{3}\frac{c_{tot}}{c_{\infty}}\right)^{2/5}.\label{eq:taug}
\end{equation}
The mean particle size reached by the end of the growth step is obtained
from Eq. (\ref{eq:nmean_Rgt4}) as
\begin{equation}
\langle n\rangle_{g}=R_{g}\tau_{g}^{4}.\label{eq:nmeang}
\end{equation}
Fig. \ref{fig:nmean_t_growth}(a) re-plots the numerical result (black solid line) from Fig. \ref{fig:nmean_vs_t}(a) to compare it to the
analytical prediction for the kinetics of the growth, obtained in this subsection.
In particular, the green line in Fig. \ref{fig:nmean_t_growth}(a) represents Eq. (\ref{eq:nmean_Rgt4})
with the only modification that we add $\langle n\rangle_{v}=c_{tot}^{-1}\sum_{n=1}^{N}c_{v,n}$,
where $c_{v,n}$ is given by Eq. (\ref{eq:c_v_i}), to the rhs of Eq. (\ref{eq:nmean_Rgt4}). This is done because the growth step
effectively starts from the quasi-equilibrium where the mean particle
size is $\langle n\rangle_{v}$. The diamond shape marks the point
of the green curve where the growth is expected to end, according
to Eqs. (\ref{eq:taug}) and (\ref{eq:nmeang}).
The vertical lines in Fig. \ref{fig:nmean_t_growth}(a) correspond to $\tau_{inc}=8.3\times10^{-13}\,{\rm s}$
(blue dashed), introduced when discussing Fig. \ref{fig:M0_vs_t},
and $\tau=6\times10^{-11}\,{\rm s}$ (red dash-dotted), chosen to
mark where the green line starts visibly deviating from the black
one. The analytical result for the kinetics of the growth is thus
seen to (i) describe correctly the time when $\langle n\rangle$ starts
increasing noticeably ($\sim2\times10^{-11}\,{\rm s}$), and (ii) agree
with the numerics for another order of magnitude in the mean particle
size. We consider this a very good result given the level of approximations
made in the analytical derivations in this section.
The main reason for the deviation of the analytical result from the numerics after
the vertical red dash-dotted line in Fig. \ref{fig:nmean_t_growth}(a)
is that the derivations in this section assumed $c_{n}=c_{v,n}$ to
be time-independent. However, when a noticeable fraction of carbon is
transferred over the nucleation barrier, this assumption becomes incorrect.
This is illustrated by Fig. \ref{fig:nmean_t_growth}(b) where the
black solid and red dashed lines show the time dependence of $c_{1}$
and $c_{2}$, respectively, in the numerical simulation.
At very short times, monomers dominate since the initial state of the system is monomers
only. Then, at $t\sim\tau_{inc}$ (blue dashed line), the quasi-equilibrium
is established and the critical nucleus ($n=n_c=16$) is formed, as is
seen from a plateau appearing on the thick green dash-dotted line.
The growth proceeds with approximately time-independent $c_{v,1}$,
$c_{v,2}$, and $c_{v,16}$ from $t\sim\tau_{inc}$ until these
concentrations of particles in vapor start dropping visibly at $\sim6\times10^{-11}\,{\rm s}$,
which, in turn, results in the decreasing rate of the growth. This decrease
is not accounted for in the analytics, resulting in the green line
increasing faster than the black one in Fig. \ref{fig:nmean_t_growth}(a).
Specifically, the numerics yields $\frac{d\ln\langle n\rangle}{d\ln t}$
of at most $\sim3$ during the growth step, hence the ``$\sim t^{3}$'' label in Fig. \ref{fig:nmean_vs_t}(a). Analytically, however, this log-log
derivative should reach the value of $4$, as is clear from Eq. (\ref{eq:nmean_Rgt4}),
but the decreasing concentration of carbon in the vapor does not allow
the standard system to quite reach this value numerically.

\begin{figure*}
\includegraphics[width=0.4\paperwidth]{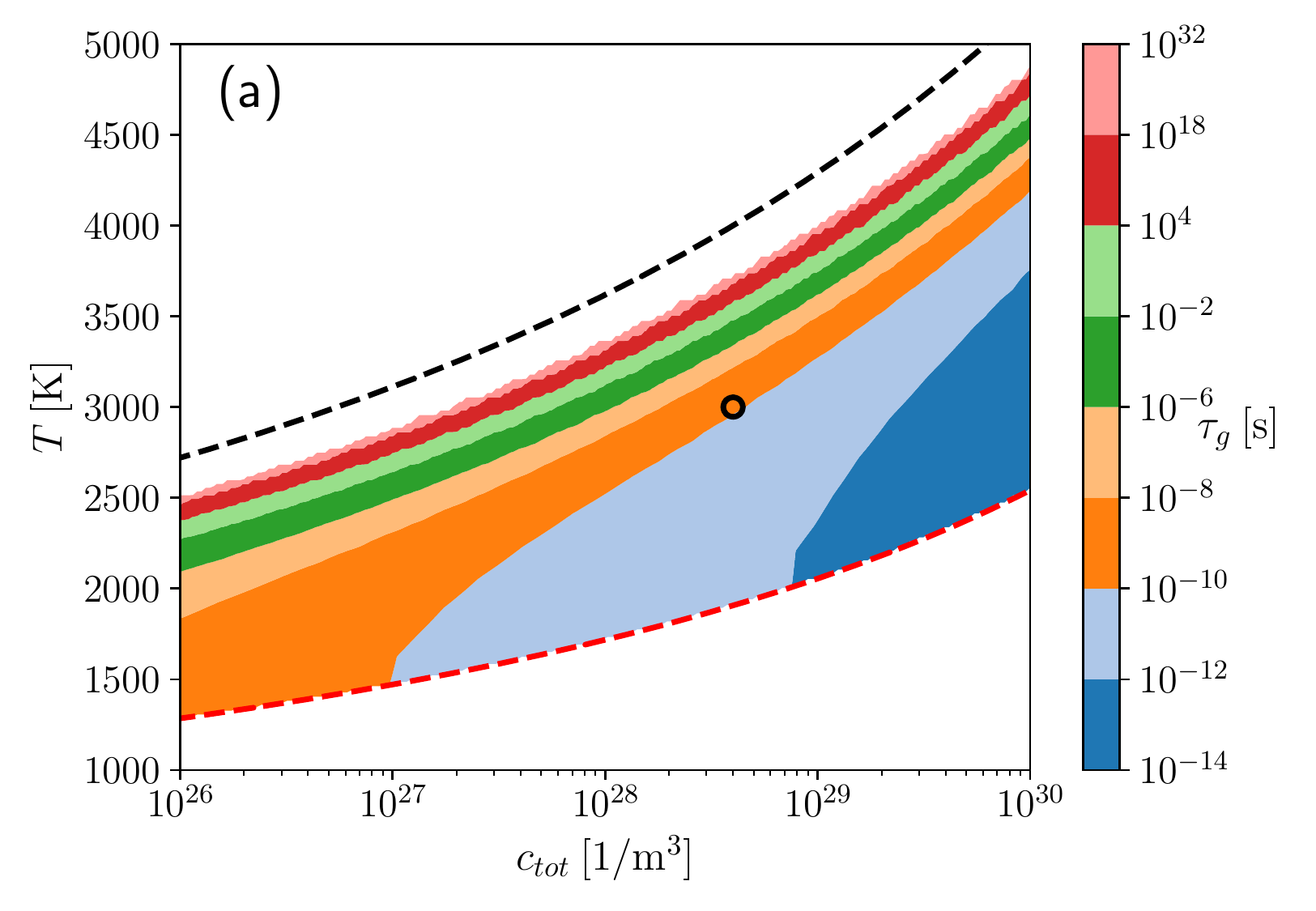}
\includegraphics[width=0.4\paperwidth]{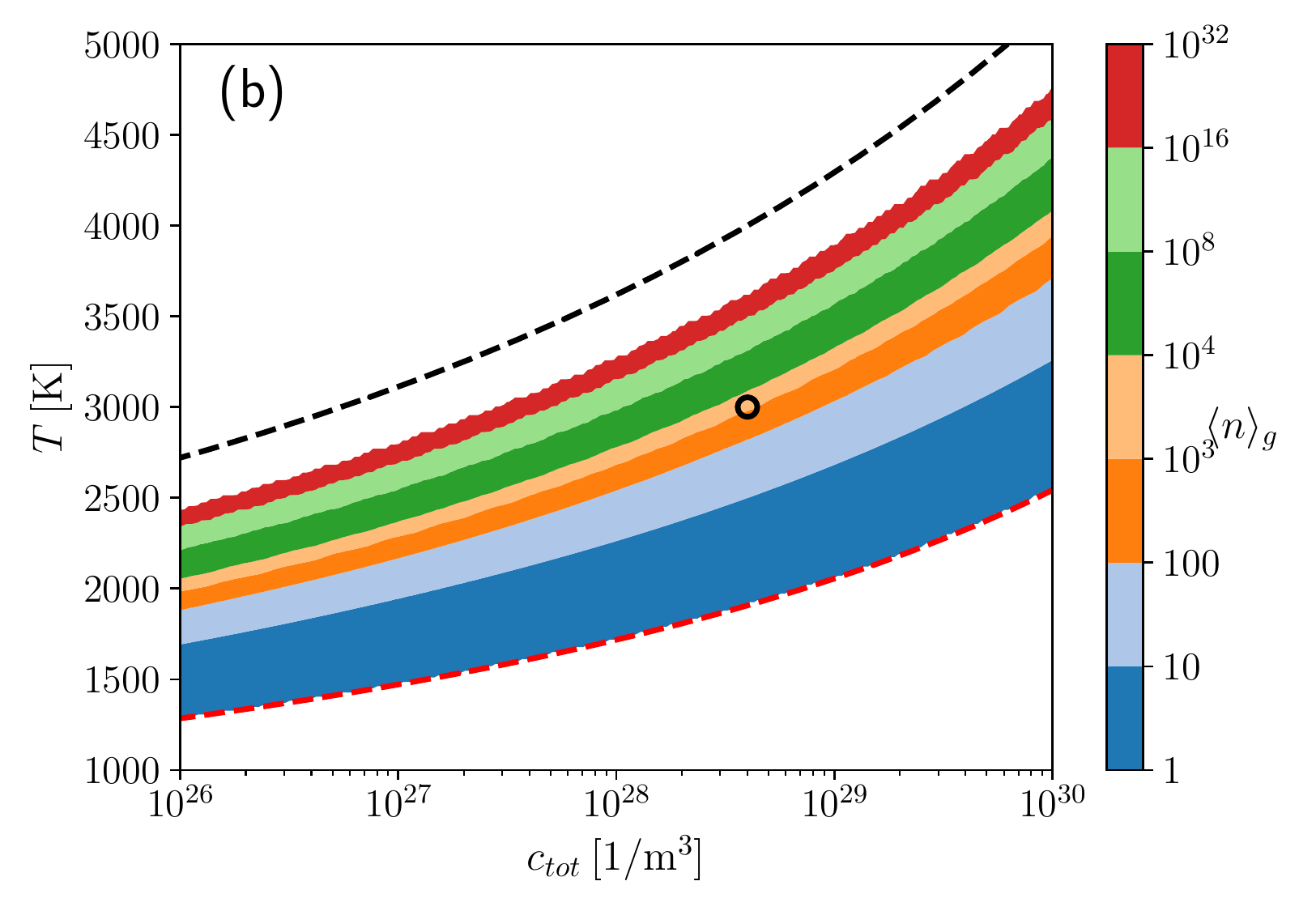}
\caption{\label{fig:taug_nmeang_dens_plots}(a) Growth times and (b) mean particle
size at the end of the growth as functions of the total carbon concentration
$c_{tot}$ and temperature $T$ for the otherwise standard system.
The black circle and the black and red dashed lines are reproduced from Fig. \ref{fig:nc_dens_plot}.}
\end{figure*}

Fig. \ref{fig:taug_nmeang_dens_plots} plots the growth time, Eq.
(\ref{eq:taug}), and the resulting mean particle size, Eq. (\ref{eq:nmeang}),
as functions of the total carbon concentration and temperature for
the otherwise standard system.

Table \ref{tab:growth} characterizes the growth step for several representative
sets of system parameters. 
\begin{table*}

\begin{tabular}{|c|c|c|c|c|c|}
\hline 
 & $n_{c}$ & $\tau_{g}\,[{\rm s}]$ & $\langle n\rangle_{g}$ & $\langle n\rangle$ at $t=10^{-6}\,{\rm s}$ & $\langle n\rangle_{v}$\tabularnewline
\hline 
\hline 
Standard system & $16$ & $1.5\times10^{-10}$ & $1.3\times10^{3}$ & - & $1.97$\tabularnewline
\hline 
$c_{tot}=1.5\times10^{28}\,{\rm 1/m^{3}}$ & $30$ & $4.1\times10^{-8}$ & $1.1\times10^{6}$ & - & $1.68$\tabularnewline
\hline 
$\mu_{s,n}=3.04n^{2/3}\,{\rm eV}$ & $189$ & $1.3\times10^{8}$ & $3.7\times10^{29}$ & $1.4\times10^{-27}$ & $1.02$\tabularnewline
\hline 
$\mu_{s,n}=3.04n^{2/3}\,{\rm eV}$, $T=2500\,{\rm K}$ & $13$ & $4.3\times10^{-10}$ & $2.8\times10^{3}$ & - & $1.10$\tabularnewline
\hline 
Equation \eqref{eq:Esn} & Spinodal &  &  &  & \tabularnewline
\hline 
Equation \eqref{eq:Esn}, $c_{tot}=2\times10^{20}\,{\rm 1/m^{3}}$ & 72 & $7.7\times10^{2}$ & $2.4\times10^{9}$ & $6.9\times10^{-27}$ & $1.01$\tabularnewline
\hline 
\end{tabular}\caption{\label{tab:growth}The growth parameters for several representative parameterizations of the system. The first and second numerical columns give
the critical size and the growth time, respectively. The third numerical
column gives the mean particle size at the end of the growth step,
Eq.~\eqref{eq:nmeang}. If the growth time is longer than $1\,{\rm \mu s}$,
the fourth numerical column gives the mean particle size reached in
the growth by this $1\,{\rm \mu s}$, evaluated using Eq.~\eqref{eq:nmean_Rgt4}.
The last column is the mean particle size in the quasi-equilibrium
vapor, evaluated as $\langle n\rangle_{v}=c_{tot}^{-1}\sum_{i=1}^{n_{c}}c_{v,i}$,
where $c_{v,i}$ is given by Eq. (\ref{eq:c_v_i}), and the chemical
potential in there is found using the iterative procedure described
above.}

\end{table*}
The first numerical row corresponds to the standard system, and each
subsequent row characterizes the growth for a system that differs
from the standard one by parameters specified in the first column.
In particular, the second numerical row corresponds to a lowered
total carbon concentration, which results in significantly longer
growth times. The third numerical row corresponds to a system where
the surface chemical potential for particles of any size (including
monomers) is given by $\mu_{s,n}=3.04n^{2/3}\,{\rm eV}$, which is
the parameterization used for diamond particles in Ref.~\onlinecite{Bastea-2017-42151}.
We obtained the much larger critical size of $n_{c}=189$, than $n_{c}\sim10$
in Ref.~\onlinecite{Bastea-2017-42151}, and this discrepancy is briefly
addressed in App. \ref{sec:nc_mono_dominated}. This large critical
size, $n_{c}=189$, corresponds to long growth times, and the time
of $\tau_{g}=1.3\times10^{8}\,{\rm s}$ is obviously much longer than
the duration of any realistic detonation experiment.
The second to last column of Tab. \ref{tab:growth} gives the mean particle size, Eq. (\ref{eq:nmean_Rgt4}),
reached in the growth assuming that the growth is only allowed to
run for $1\,\mu{\rm s}$ - an order of magnitude estimate for carbon
clustering times observed experimentally.\citep{Ten-2014-369,Bagge-Hansen-2015-245902,Watkins-2017-23129,Watkins-2018-821}
This results in $\langle n\rangle\sim10^{-27}$, which implies that
the nucleation barrier is so high that the growth is not observed
at realistic times. However, assuming the same $\mu_{s,n}=3.04n^{2/3}\,{\rm eV}$
but with a lower temperature of $T=2500\,{\rm K}$ (fourth numerical
row), results in much smaller $n_{c}$ and growth times, demonstrating
the high sensitivity of the growth parameters to temperature. The
last two rows of Tab. \ref{tab:growth} characterize the growth in the case where
$E_{s,n}$ in Eq. (\ref{eq:Esn}) is used as $\mu_{s,n}$ for two different
total carbon concentrations: the standard one, $c_{tot}=4\times10^{28}\,{\rm 1/m^{3}}$, and a much lower one, $c_{tot}=2\times10^{20}\,{\rm 1/m^{3}}$. The former concentration results in a spinodal transformation, i.e., the concentration
is too high to have a nucleation barrier. This is because $E_{s,n}$,
plotted by the thick black line in Fig. \ref{fig:Ec_vs_n}(b), renders
small particles significantly more energetic than $\mu_{s,n}$ in
Tab. \ref{tab:std_parms}, resulting in a much less stable carbon
vapor phase. The latter, perhaps unrealistically low, concentration
results in the metastable vapor with a finite nucleation barrier.

\section{Coarsening\label{sec:Coarsening}}

The carbon vapor is depleted at the end of the growth step, so the
large carbon particles can increase in size any further only by redistributing carbon
between themselves. This redistribution of carbon is typically referred
to as \emph{coarsening} in the literature.\citep{Lifshitz-1961-35,Penrose-1978-Kinetics,Krapivsky-2010-Kinetic,Alexandrov-2018-035102}
The two physical processes that occur concurrently during the coarsening
step are the coagulation and Ostwald ripening. These two process are
not independent, since both affect the shape of the self-preserving
distribution of concentrations,\citep{Alexandrov-2018-035102} but
we consider them separately, thus assuming that either one of them
strongly dominates in each specific situation. In the following subsections (Secs. \ref{subsec:Coag-Coarsening} and \ref{subsec:Ostwald-Ripening}),
we discuss the coagulation and Ostwald ripening in coarsening.

\subsection{Coagulation in Coarsening\label{subsec:Coag-Coarsening}}

\begin{figure}
\includegraphics[width=0.4\paperwidth]{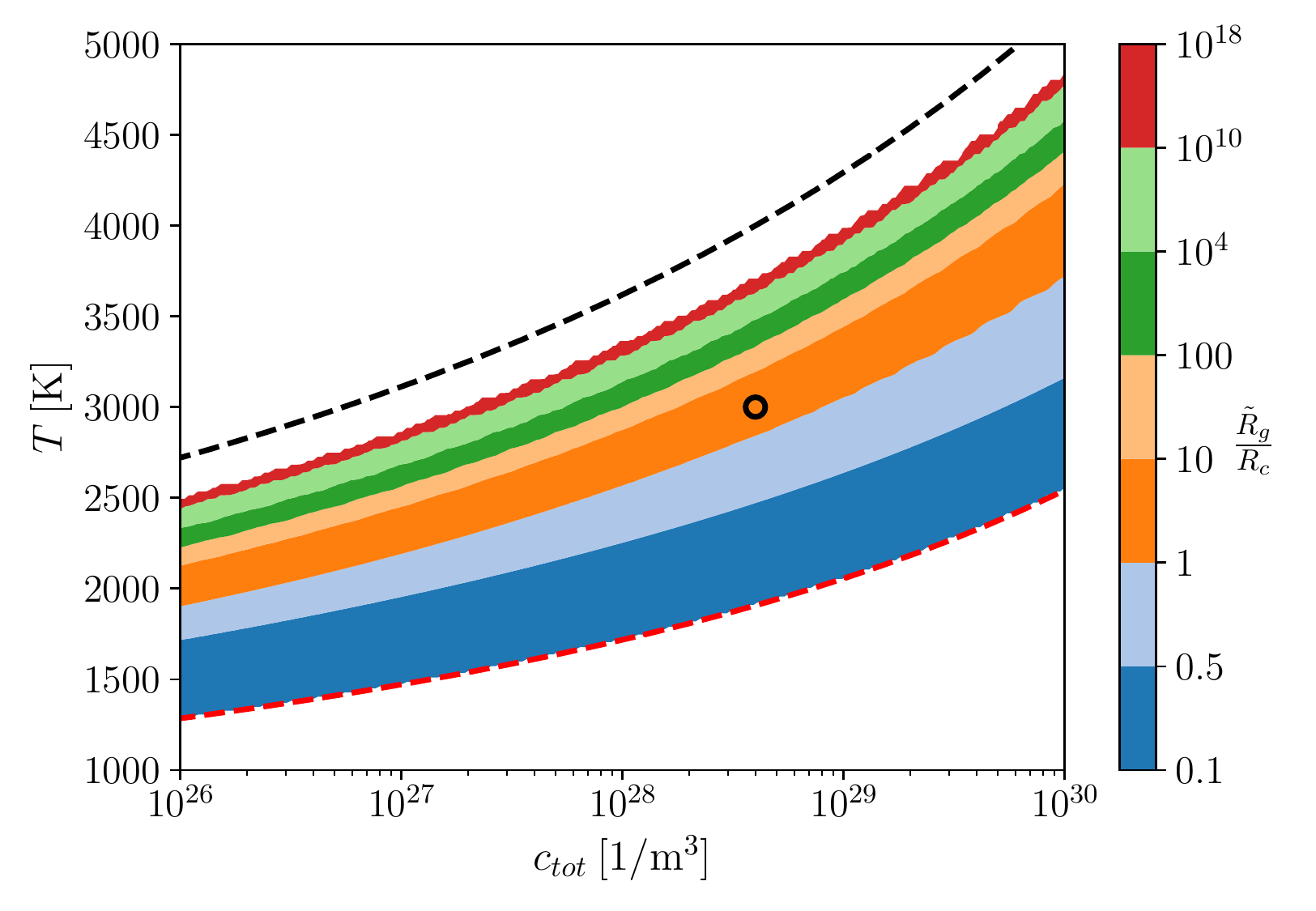}
\caption{\label{fig:rgrc_dens_plot}The ratio of the ``linearized'' growth
rate, Eq. (\ref{eq:Rg_linear}), and the rate of coagulation, Eq.
(\ref{eq:Rr_id}), as a function of the total carbon concentration
$c_{tot}$ and temperature $T$, for the otherwise standard system.
The black circle and the black and the red dashed lines are reproduced from Fig. \ref{fig:nc_dens_plot}.}
\end{figure}

Once the nucleation barrier is fully overcome at the end of the growth
step, the aggregation is expected to dominate over fragmentation,
so the carbon particles undergo the Smoluchowski coagulation. To make
this argument more quantitative, we reproduce the procedure used above
to derive Eq. (\ref{eq:dndt_cnplus}), but this time also accounting
for fragmentation. Adopting the Smoluchowski approximation, $c_{n}\propto e^{-n/\tilde{n}}$,
and assuming that the coagulation is dominated by collisions of particles
of similar size, one obtains

\begin{equation}
\frac{d\tilde{n}}{dt}\approx2k_{0}c_{tot}\left[1-\frac{c^{\circ}}{c_{tot}}\tilde{n}^{2}e^{\beta\left(\mu_{\tilde{n}}^{\circ}-2\mu_{\tilde{n}/2}^{\circ}\right)}\right].
\end{equation}
Since $\mu_{n}^{\circ}=\mu_{s}n^{2/3}$ for large particles, we obtain
\begin{equation}
\frac{d\tilde{n}}{dt}\approx2k_{0}c_{tot}\left[1-\frac{c^{\circ}}{c_{tot}}\tilde{n}^{2}e^{-\beta\mu_{s}\tilde{n}^{2/3}\left(2^{1/3}-1\right)}\right].\label{eq:dntilde_dt_aggr_fragm}
\end{equation}
The second term in the brackets corresponds to fragmentation, and
it vanishes at $\tilde{n}\rightarrow\infty$, so the aggregation dominates
over fragmentation, resulting in the Smoluchowski coagulation. That
at some smaller $\tilde{n}$ aggregation does not dominate is consistent
with the results of the previous section (Sec. \ref{sec:Growth}), where the quasi-equilibrium
state - carbon vapor - was discussed. As it was already observed
and briefly discussed in Sec. \ref{sec:Numerical_Results}, the long-time
asymptotic behavior of the condensation kinetics does, indeed, seem
to be dominated by coagulation. More specifically, Eq. (\ref{eq:nmean_vs_t_coag}),
augmented by the time shift of $-2.24\times10^{-10}\,{\rm s}$, provides
a very accurate description of the numerical condensation kinetics
in Fig. \ref{fig:nmean_vs_t}(b). It is interesting to note that this
shift, needed to have an agreement between the red dash-dotted and black solid lines, is negative, implying that the growth is ``faster''
than the coagulation, i.e., if the completed growth is described by
a ``linearized'' rate $\tilde{R}_{g}$ so that 
\begin{equation}
\langle n\rangle_{g}=\tilde{R}_{g}\tau_{g},\label{eq:Rg_linear}
\end{equation}
then $\tilde{R}_{g}$ is larger than $R_{c}$ for the standard system.
Fig. \ref{fig:rgrc_dens_plot} plots the ratio of $\tilde{R}_{g}$ to $R_{c}$
as a function of the total carbon concentration and temperature.
Apparently, $\tilde{R}_{g}$ is not always larger
than $R_{c}$ - the growth becomes slower when the spinodal (red dashed line) is approached. The sign of $\tilde{R}_{g}/R_{c}-1$ gives the
sign of the intercept when the line with the coagulation slope is
fitted to the numerical data. In particular, the negative $\tilde{R}_{g}/R_{c}-1$
produces a negative intercept, which looks like the growth step ``delays''
the final coagulation. Accordingly, the positive $\tilde{R}_{g}/R_{c}-1$
results in effectively giving the coagulation a ``head start''.

\subsection{Ostwald Ripening\label{subsec:Ostwald-Ripening}}

The other important coarsening mechanism is the competitive growth,
or \emph{Ostwald ripening}, where a large particle increases in size not
because it collides with one of the comparable sizes (coagulation), but
because it absorbs small clusters previously evaporated by smaller
(but still large) ones. The direct result of $\mu_{s,n}\propto n^{2/3}$
scaling for large particles is the Kelvin equation - the concentration
of vapor, which is in equilibrium with a large carbon particle of
size $n$, decreases with $n$.\citep{Atkins-2006-Physical} In the
Ostwald ripening, larger particles predominantly absorb vapor clusters
at the price of smaller particles that mostly evaporate, and a self-preserving
balance is maintained between the distribution of concentrations of large carbon particles and the concentration of the vapor. This mechanism has traditionally not been discussed in the context of carbon condensation
in detonation,\citep{Shaw-1987-2080,Chevrot-2012-084506,Bastea-2012-214106}
so we consider it here to compare its rate with that of coagulation.
The theory of the Ostwald ripening has been well developed in the literature,\citep{Lifshitz-1961-35,Kahlweit-1975-1,Krapivsky-2010-Kinetic}
but it mostly concerns a monomer-dominated vapor. In what follows,
we outline some of the derivation steps needed to generalize the
previous results to the case of the non-ideal vapor, i.e., the vapor where the carbon particles of sizes $n=2,3,...$ have to be accounted for in addition to the monomers. To this end, we consider
a large particle of mass $n$ in the vapor atmosphere. The vapor is
in equilibrium with itself, but not with large clusters. Then, the size-increase rate for the large particle of size $n$ can be written as
\begin{equation}
\frac{dn}{dt}=\sum_{i}iK_{n,i}^{+}c_{v,i}-\sum_{i}iK_{n,i}^{-},
\end{equation}
where, as in Eq. (\ref{eq:J_n}), $i$ runs only over vapor and $c_{v,i}$
is given by Eq. (\ref{eq:c_v_i}). Assuming that $n\gg i$, Eq. (\ref{eq:Knm_k0})
produces $K_{n,i}^{+}=k_{0}n^{1/3}i^{-1/3}$ and the rate constant of fragmentation is given by Eq. (\ref{eq:Kminus_Kplus}). Combining the above expressions, we obtain for the size-increase rate
\begin{equation}
\frac{dn}{dt}=k_{0}c^{\circ}n^{1/3}\sum_{i}i^{2/3}\left\{ e^{\beta(i\mu-\mu_{i}^{\circ})}-e^{\beta(i\mu_{n}^{\circ\prime}-\mu_{i}^{\circ})}\right\} ,\label{eq:dn_dt_mu}
\end{equation}
where, as above, a finite difference is substituted with differentiation
as $\mu_{n}^{\circ}-\mu_{n-i}^{\circ}\approx i\frac{d\mu_{n}^{\circ}}{dn}=i\mu_{n}^{\circ\prime}$.
To proceed further, we note that the vapor is almost in equilibrium
with large clusters in the Ostwald ripening, and thermodynamically,
the large clusters are almost bulk and so $\beta\mu\ll1$. Equation \eqref{eq:musn_mus},
accurate for large particles, produces $\mu_{n}^{\circ\prime}=\frac{2}{3}\mu_{s}n^{-1/3}$,
and so Eq. (\ref{eq:dn_dt_mu}) becomes
\begin{equation}
\frac{dn}{dt}=k_{0}c_{v,1}^{(sat)}S_{5/3}\left\{ \beta\mu n^{1/3}-\frac{2}{3}\beta\mu_{s}\right\} ,\label{eq:dn_dt_dmu}
\end{equation}
where we define $S_{p}=\sum_{i=1}^{i_{v}}i^{p}e^{-\beta(\mu_{i}^{\circ}-\mu_{1}^{\circ})}$
and $c_{v,i}^{(sat)}=c^{\circ}e^{-\beta\mu_{i}^{\circ}}$. Parameters
$S_{p}$ are defined in such a way that $S_{p}=1$ if the vapor is
ideal. The total mass of vapor per unit volume is evaluated as (using
again that $\mu$ is small)
\begin{equation}
c_{v,tot}=\sum_{i}ic^{\circ}e^{\beta(i\mu-\mu_{i}^{\circ})}=c_{v,tot}^{(sat)}+c_{v,1}^{(sat)}S_{2}\beta\mu.
\end{equation}
The last two expressions combine into
\begin{equation}
\frac{dn}{dt}=k_{0}\frac{S_{5/3}}{S_{2}}n^{1/3}\left\{ c_{v,tot}-c_{v,tot}^{(sat)}-\frac{2}{3}c_{s,1}S_{2}\beta\mu_{s}n^{-1/3}\right\} \label{eq:dn_dt_c_totv}
\end{equation}
This equation needs to be complemented with the mass conservation
per unit volume
\begin{equation}
c_{tot}=c_{v,tot}+\sum_{n}nc_{n},\label{eq:mass_cons}
\end{equation}
where the summation over $n$ is performed over large clusters only,
i.e., the vapor phase excluded. Equations \eqref{eq:dn_dt_c_totv} and \eqref{eq:mass_cons} constitute the closed system of equations since
the condensation/evaporation of large clusters affects $c_{v,tot}$ due
to mass conservation, Eq. (\ref{eq:mass_cons}), which, in turn, affects
the rate of the size increase in large clusters via Eq. (\ref{eq:dn_dt_c_totv}).

\begin{figure*}
\includegraphics[width=0.4\paperwidth]{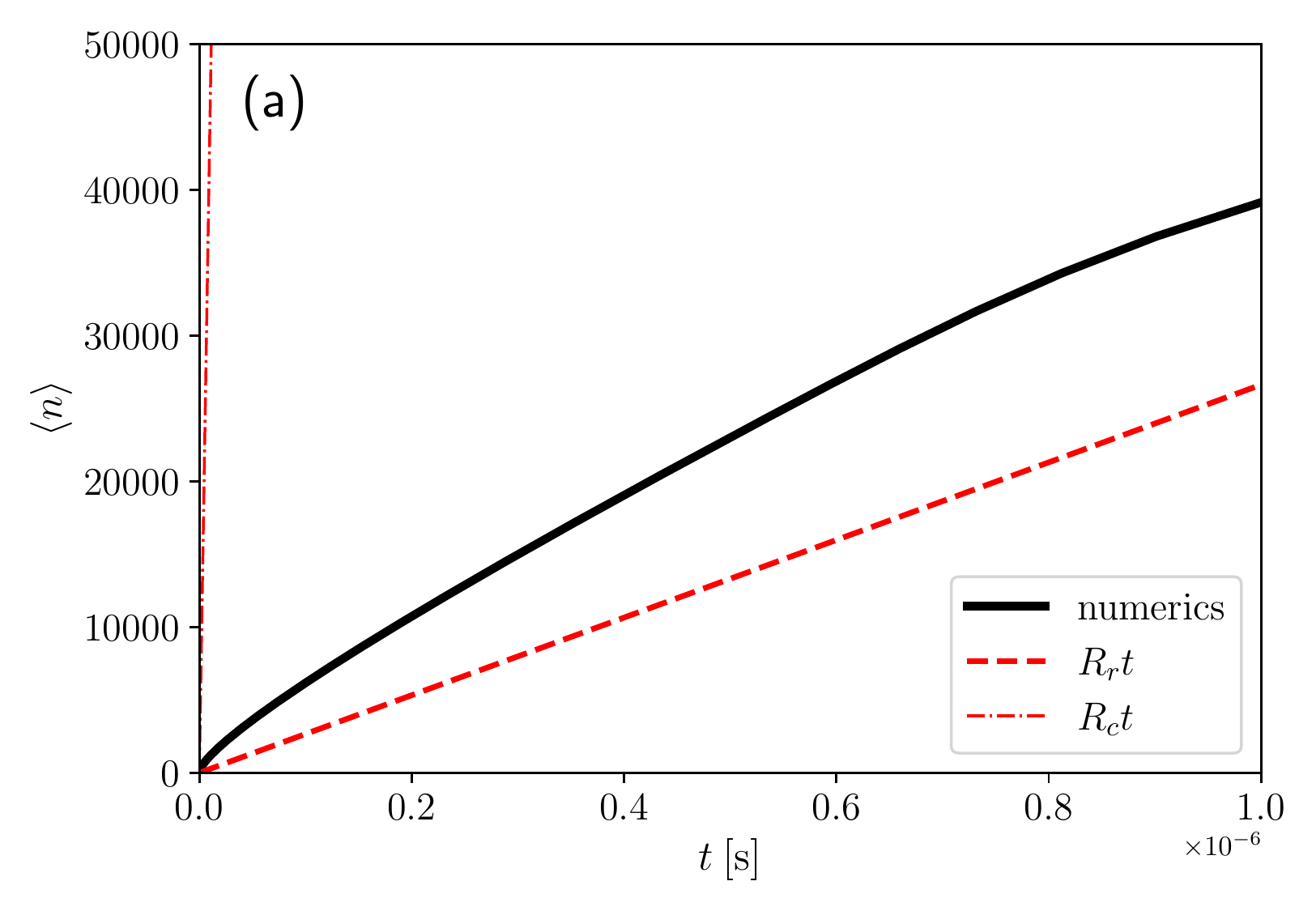}
\includegraphics[width=0.4\paperwidth]{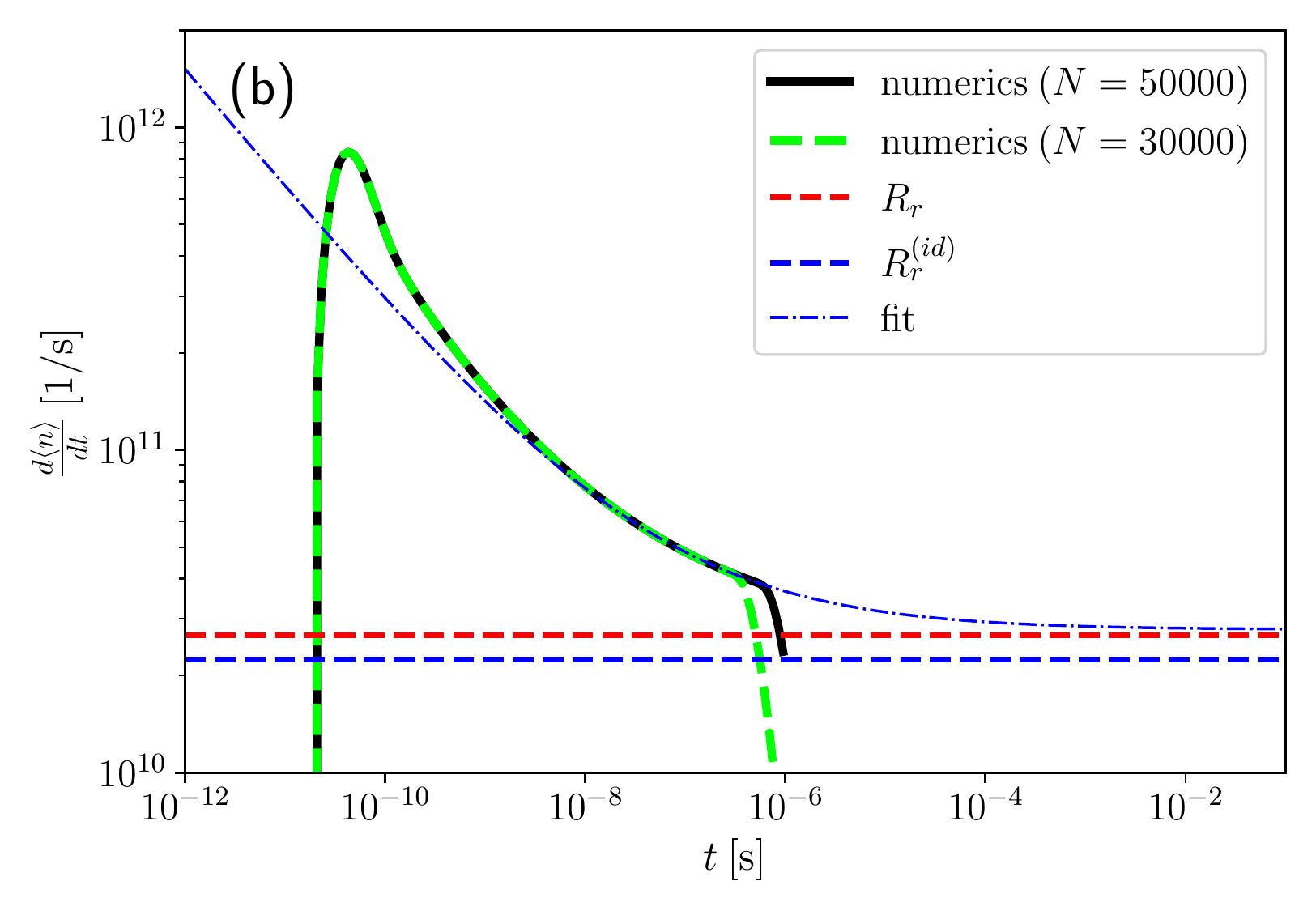}
\caption{\label{fig:nmean_vs_t_ripe}Kinetics of coarsening for the standard
system, obtained by numerically turning off the coagulation. The figure
shows the mean particle size in panel (a) and its time derivative
in panel (b). The thick black lines in the both panels represent the
numerical result. The thick red dashed line represents the analytical
prediction for the Ostwald ripening rate, Eq. (\ref{eq:Rr_nid}). The thick blue dashed line in panel
(b) is the monomer-only analytical result for the Ostwald ripening, Eq.
(\ref{eq:Rr_id}). The thin blue dash-dotted line in panel (b) is the fitting result,
Eq. (\ref{eq:fit_expr}). The thin red dash-dotted line in panel (a) represents the coagulation, Eq.~\eqref{eq:dnmean_dt_coag}.}
\end{figure*}

An important observation about Eqs. (\ref{eq:dn_dt_c_totv}) and (\ref{eq:mass_cons})
is that they include non-ideality of the vapor phase only through
parameters $S_{5/3}$ and $S_{2}$, which could be ``absorbed''
into definitions of $k_{0}$ and $\mu_{s}$. This implies that if
we had an expression for the rate of the Ostwald ripening, as a function
of $k_{0}$ and $\mu_{s}$, for the ideal vapor (i.e., monomers only),
we would be able to use it for the non-ideal vapor by simply re-defining
$k_{0}$ and $\mu_{s}$ in there. The Lifshitz-Slyozov (LS) theory\citep{Lifshitz-1961-35}
for the monomer-only Ostwald ripening gives the following rate of
size increase in a particle 
\begin{equation}
\frac{d\tilde{n}}{dt}=\frac{8}{81}\beta\mu_{s}k_{0}c_{v,1}^{(sat)},
\end{equation}
where, similar to Sec. \ref{sec:Coagulation}, the concentration of
large particles is given by a universal self-preserving distribution
$c_{n}\propto f(n/\tilde{n})$, with $\tilde{n}$ being the characteristic
particle size . The moments of this distribution are denoted by $m_{p}=\int dn\:n^{p}c_{n}$,
and using the self-preserving distribution obtained in Ref.~\onlinecite{Lifshitz-1961-35},
one can calculate numerically $m_{1}/m_{0}=1.1296\tilde{n}$ and $m_{2}/m_{1}=1.4241\tilde{n}$.\footnote{Note that unlike Sec. \ref{sec:Coagulation}, the LS distribution is not normalized to have $m_{0}=m_{1}=1$} This gives the rate of ripening in the monomer-only (i.e., ideal) case as
\begin{align}
R_{r}^{(id)} &= \frac{d\langle n\rangle}{dt}=\frac{1}{M_1}\frac{dM_{2}}{dt}=1.4241\frac{d\tilde{n}}{dt}\nonumber\\
&=0.14065\beta\mu_{s}k_{0}c_{v,1}^{(sat)}.\label{eq:Rr_id}
\end{align}
Eq. (\ref{eq:dn_dt_c_totv}) suggests that to convert $R_{r}^{(id)}$
to the non-ideal one, $R_{r}$, we need to substitute $k_{0}\rightarrow S_{5/3}k_{0}/S_{2}$
and $\mu_{s}\rightarrow S_{2}\mu_{s}$. Accordingly, Eq. (\ref{eq:Rr_id})
is converted from the ideal to non-deal form as
\begin{equation}
R_{r}=S_{5/3}R_{r}^{(id)}=0.14065\beta\mu_{s}k_{0}M_{v,5/3}^{(sat)},\label{eq:Rr_nid}
\end{equation}
where $M_{v,p}^{(sat)}=\sum_{i=1}^{i_{v}}i^{p}c_{v,i}^{(sat)}=S_{p}c_{v,1}^{(sat)}$.
The resulting ideal and non-ideal Ostwald ripening rates for the standard system are
\begin{align}
R_{r}^{(id)}&=2.246\times10^{10}\,{\rm 1/s},\nonumber \\
R_{r}&=2.662\times10^{10}\,{\rm 1/s},\label{eq:Rr_st}
\end{align}
showing $\sim 20\%$ correction due to the non-ideality. These values
are more than two orders of magnitude lower than the corresponding
coagulation rate, Eq. (\ref{eq:Rc_st_FW}), resulting in an excellent
agreement of the ``shifted'' coagulation (thin dash-dotted line)
with the numerical results (thick black solid line) in Fig. \ref{fig:nmean_vs_t}(b).
In order to be able to compare the analytical results for ripening
with numerics, we performed new numerical simulations where (i) all
the rate constants $K_{n,m}^{\pm}$ were set to zero if $\min(n,m)>i_{v}$,
thus effectively turning off the coagulation, and (ii) all the carbon
initially sat in carbon particles of size $n_{0}=50$ so that the
system does not need to slowly overcome the nucleation barrier. The
numerical results were not sensitive to the specific value of $i_{v}$
- effective largest particle size in the vapor - when it was significantly
larger than $1$. The thick black line in Fig. \ref{fig:nmean_vs_t_ripe}(a)
represents the results of such numerical simulations for the standard
system.
The thick dashed and thin dashed dotted lines represent the evolutions of the mean particle size according to the Ostwald ripening and coagulation, respectively. Clearly, coagulation analytics overestimates the rate of coarsening by far.
The slopes of the black solid and red dashed lines look similar, especially
at longer times, so one could claim a tentative agreement. The more
accurate test of agreement is provided in Fig. \ref{fig:nmean_vs_t_ripe}(b),
where the time derivative of the mean particle size is plotted. The
thick black line is seen to increase rapidly at first, establishing an
approximately self-preserving distribution, and then it is seen to
drop slowly, until it starts decreasing very quickly at $\sim6\times10^{-7}\,{\rm s}$
due to the finite maximum particle size in simulations (compare black solid and green dashed lines). This observed convergence to the constant
coarsening rate is slow; it would require much larger maximum cluster
sizes and simulation times to reach a few-percent agreement. Instead,
we fit the numerical results for $\frac{d\langle n\rangle}{dt}$ with
an empirical formula
\begin{equation}
A_{0}\left[1+(A_{1}t)^{-A_{2}}\right],\label{eq:fit_expr}
\end{equation}
and then interpret $A_{0}$ as the rate of coarsening at $t\rightarrow\infty$.
Such fitting of the thick black line in Fig. \ref{fig:nmean_vs_t_ripe}(b)
in the time range from $10^{-7}$ to $5\times10^{-7}\,{\rm s}$
produces $A_{0}=2.777\times10^{10}\,{\rm 1/s}$, $A_{1}=2.240\times10^{7}\,{\rm 1/s}$,
and $A_{2}=0.3719$. Increasing the fitting range by extending it down to $5\times10^{-8}\,{\rm s}$ changes $A_0$ by less than $1\%$.
The resulting coarsening rate, $A_{0}$, is different
from the analytical prediction, $R_{r}$ in Eq. (\ref{eq:Rr_st}),
by only about $4\%$, whereas the deviation from $R_{r}^{(id)}$ is
$\sim24\%$. This is an excellent agreement that also demonstrates
that the vapor non-ideality can be important when evaluating the rate
of the Ostwald ripening.

\begin{figure}
\includegraphics[width=0.4\paperwidth]{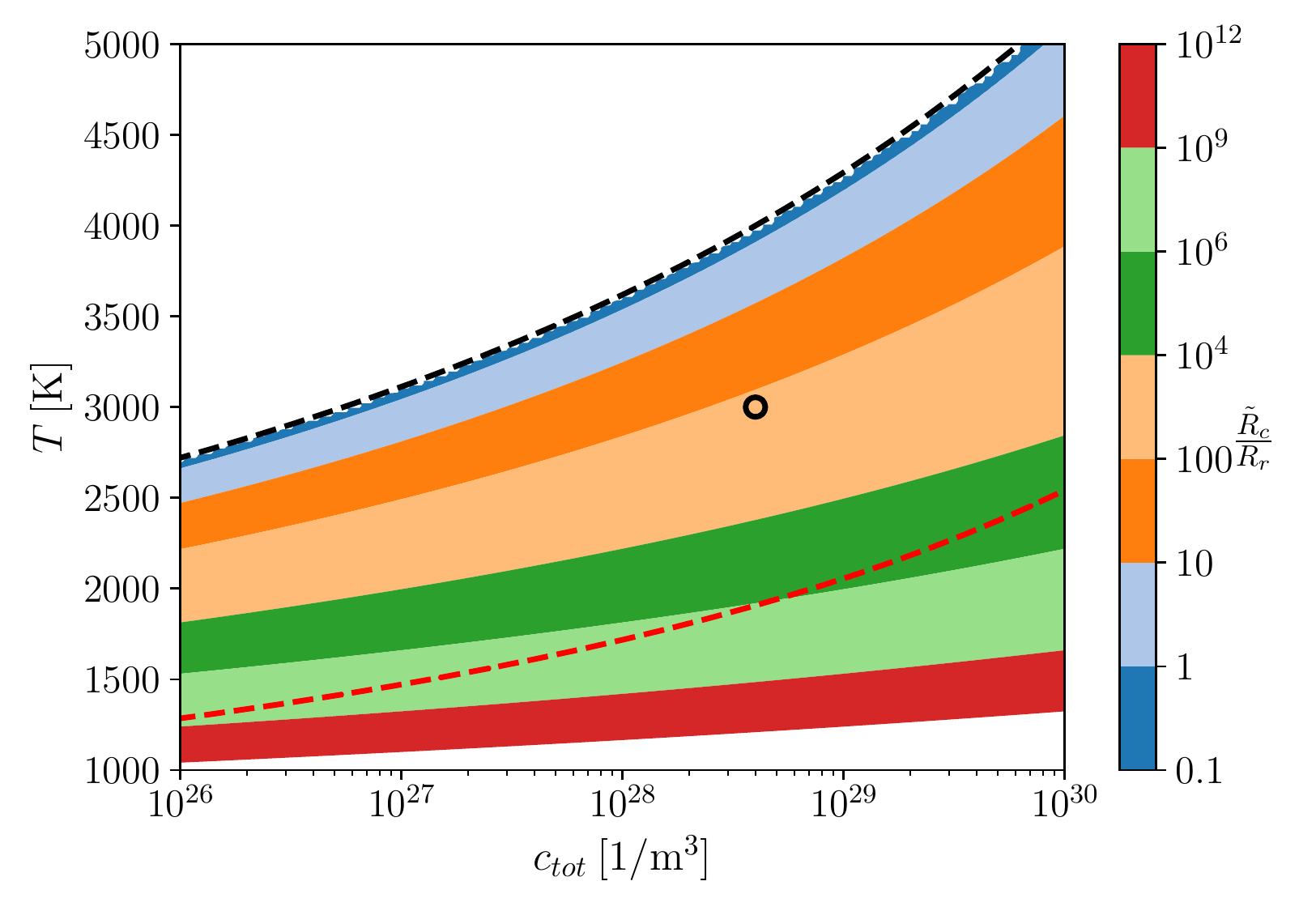}
\caption{\label{fig:rc_rr_dens_plot}The ratio of the corrected coagulation,
Eq. (\ref{eq:Rc_corr}), and Ostwald ripening rates, Eq. (\ref{eq:Rr_nid}),
as a function of the total carbon concentration $c_{tot}$ and temperature
$T$ for the otherwise standard system. The black circle and the black and red dashed lines are reproduced from Fig. \ref{fig:nc_dens_plot}.}
\end{figure}

It is instructive to discuss the sensitivity of the Ostwald ripening
and coagulation rates on various system parameters. The Ostwald ripening
rate, Eqs. (\ref{eq:Rr_id}) and (\ref{eq:Rr_nid}), is most (i.e.,
exponentially) sensitive to temperature since the concentration of
the saturated vapor increases exponentially with $T$. The rate of coagulation,
Eq. (\ref{eq:dnmean_dt_coag}), seems to depend on all the parameters
only relatively weakly (e.g., non-exponentially).
It is important to note, however, that Eq. (\ref{eq:dnmean_dt_coag}) is directly
applicable to the coarsening step only when the degree of over-saturation
of the system is high, $c_{tot}\gg c_{v,tot}^{(sat)}$, such as the
case for the standard system. This is because at very long times,
the probability for a large particle to collide with another large
particle is proportional to the total concentration of carbon sitting
in large particles, which is not $c_{tot}$, but $c_{tot}-c_{v,tot}^{(sat)}$
instead. The coagulation rate in coarsening is then
\begin{equation}
\tilde{R}_{c}=\left[1-\frac{c_{v,tot}^{(sat)}}{c_{tot}}\right]R_{c},\label{eq:Rc_corr}
\end{equation}
where the original coagulation rate $R_c$ is given by Eq. (\ref{eq:dnmean_dt_coag}).
This new corrected rate becomes arbitrarily small when approaching
the saturation line (e.g., the black dashed line in Fig. \ref{fig:taug_nmeang_dens_plots})
moving from low to high temperatures, since $c_{tot}=c_{v,tot}^{(sat)}$
is the definition of the saturation line. Therefore, we expect that
the dominating mechanism of coarsening is determined by temperature:
the Ostwald ripening is prevalent at higher temperature, and the coagulation
dominates the lower temperature regime. Figure \ref{fig:rc_rr_dens_plot}
plots the ratio of the corrected coagulation rate, Eq. (\ref{eq:Rc_corr}),
and the Ostwald ripening rate, Eq. (\ref{eq:Rr_nid}), as a function
of $c_{tot}$ and $T$.
This ratio is seen to decrease when the temperature increases, and
when the system is sufficiently close to saturation (black dashed line), the ratio becomes less than one, signifying the onset of the
regime where the Ostwald ripening dominates the coarsening. Unlike
the density plots above, this one extends to temperatures below those
given by the spinodal (red dashed line). This is because all the above
density plots concerned the growth step, which is only present when
there is a nucleation barrier, i.e., when the system is in between
the saturation and spinodal lines. Coarsening, however, can occur
when the carbon vapor is unstable, i.e., at temperatures lower or the total carbon concentrations higher than those corresponding to the onset of the spinodal transformation.

\section{Discussion\label{sec:Discussion}}

In the following subsections (Secs. \ref{subsec:Kinetic-Regimes} to \ref{subsec:critical_size}), we discuss several general topics which
would be out of place in the previous more specific sections. In particular,
Sec. \ref{subsec:Kinetic-Regimes} discusses the general trends of
the carbon condensation kinetics as the system parameters vary. Ostwald
ripening in a realistic system, where carbon-oxygen reactions are
treated explicitly, is discussed in Sec. \ref{subsec:Ripening_realistic}.
Thermal equilibration of carbon particles surround by the detonation
fluid is considered in Sec. \ref{subsec:thermal_eq}. Capillarity
approximation, introduced in Sec. \ref{sec:Energetics}, is briefly
revisited in Sec. \ref{subsec:capillarity_revisited}. The physical
meaning and importance of the particle of critical size in condensation,
introduced in Sec. \ref{sec:Growth}, are revisited in Sec. \ref{subsec:critical_size}. 

\subsection{Kinetic Regimes\label{subsec:Kinetic-Regimes}}

\begin{figure}
\includegraphics[width=0.4\paperwidth]{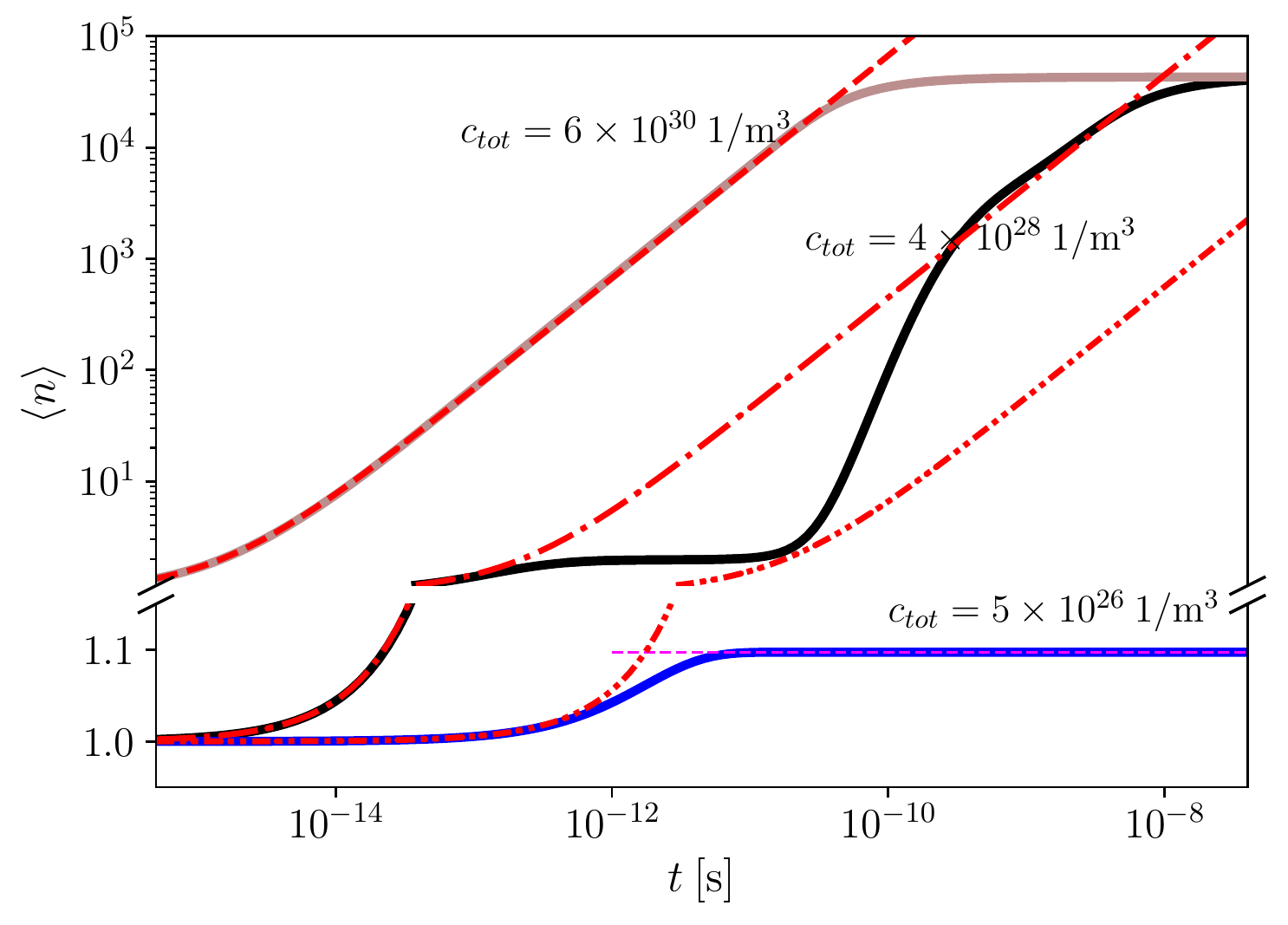}
\caption{\label{fig:nmean_vs_t_ctot}Numerical results for the evolution of
the mean particle size in three distinct kinetic regimes. The black solid line and the red dash-dotted line correspond to the standard
system and are reproduced from Fig. \ref{fig:nmean_vs_t}(a). The thick brown and blue solid lines correspond to numerical results at
substantially higher and lower carbon concentrations, respectively.
The red dashed and dot-dot-dashed lines represent the corresponding analytical
coagulation results, Eq. (\ref{eq:nmean_vs_t_coag}). The horizontal thin magenta dashed line is the equilibrium mean particle size, $\langle n\rangle_{v}\approx1.01$,
for the unsaturated vapor corresponding to the total carbon concentration
of $c_{tot}=5\times10^{26}\,{\rm 1/m^{3}}$.}
\end{figure}

In Secs. \ref{sec:Numerical_Results}-\ref{sec:Coarsening}, the discussion
was focused on the carbon condensation in the standard system, Tab.
\ref{tab:std_parms}. The parameters of the standard system were intentionally
chosen to demonstrate a rich kinetic behavior that consisted of the
coagulation step, Sec. \ref{sec:Coagulation}, the growth step, preceded
by the nucleation event when particles of the critical size are first
formed, Sec. \ref{sec:Growth}, and the coarsening step where the
two pathways (coagulation and Ostwald ripening) of varying efficiency
are observed, Sec. \ref{sec:Coarsening}. This succession of kinetic
steps is characteristic of a metastable system, which, from the perspective of
e.g., Fig. \ref{fig:nc_dens_plot}, is parametrically in between the saturation and
spinodal lines. In this section, we discuss the carbon condensation
(or the absence thereof) in situations where the system parameters are
sufficiently different from the standard ones to produce qualitatively
different behavior. The thick solid lines in Fig. \ref{fig:nmean_vs_t_ctot}
represent the numerical results for the time evolution of the mean
particle size for the standard system (black line), reproduced from
Fig. \ref{fig:nmean_vs_t}(a), and the two systems which are standard
except for the much higher (brown line) and much lower (blue line)
total carbon concentration.
In particular, the brown solid line corresponds to the total carbon
concentration of $c_{tot}=6\times10^{30}\,{\rm 1/m^{3}}$, which is
chosen to be somewhat higher than the concentration of the threshold of spinodal transformation for the
standard system, Eq. (\ref{eq:cspin_tot_st}). The system is over-saturated
at this high concentration resulting in carbon condensation, so the
mean particle size increases monotonically. The red dashed line is
calculated from Eq. (\ref{eq:nmean_vs_t_coag}), and its agreement
with the numerical result is excellent at all times until the maximum
particle size allowed in simulations is reached at $t\sim10^{-10}\,{\rm s}$.
It is thus expected that these two lines would agree at all times
if the maximum numerically allowed particle size approaches infinity. This level
of agreement is actually not trivial since Eq. (\ref{eq:nmean_vs_t_coag})
accounts only for aggregation of particles, whereas the fragmentation
is present in numerical simulations. The ratio of the effective fragmentation
and aggregation rates in Eq. (\ref{eq:dntilde_dt_aggr_fragm}) is
$\frac{c^{\circ}}{c_{tot}}\tilde{n}^{2}e^{-\beta\mu_{s}\tilde{n}^{2/3}\left(2^{1/3}-1\right)}$,
and this expression does have a single maximum as a function of the
characteristic particle size $\tilde{n}$. The location of this maximum
is denoted by $\tilde{n}_{*}$, and this is where the fragmentation
is the strongest relative to aggregation. It is clear from the functional
form of this expression that the fragmentation can be neglected at
(i) $\tilde{n}\ll\tilde{n}_{*}$ or $\tilde{n}\gg\tilde{n}_{*}$ if
$c_{tot}$ is not too large, or (ii) any $\tilde{n}$ if $c_{tot}$
is large. The former scenario corresponds to the standard system where
the analytical coagulation (red dash-dotted line in Fig. \ref{fig:nmean_vs_t_ctot})
is in agreement with numerics (thick black line) at very small and
very large particle sizes, but not in between where the fragmentation
is strong, resulting in the nucleation and growth. The latter scenario,
represented by the brown line in Fig. \ref{fig:nmean_vs_t_ctot},
is where the fragmentation is weak everywhere, so the coagulation
describes the condensation everywhere because $c_{tot}$ is large.
Different thermodynamics of carbon particles generally results in
a different total concentration threshold between the spinodal and
metastable behaviors, $c_{tot}^{(spin)}$. For instance, the parameterization
from the fifth numerical row of Tab. \ref{tab:growth} results in
the spinodal behavior already at the standard concentration, so the
numerical result for this parameterization (not shown) is in excellent
agreement with the red dash-dotted line in Fig. \ref{fig:nmean_vs_t_ctot}
until the maximum particle size of $N=50000$ is reached numerically.

It is important to mention that the fragmentation term in Eq. (\ref{eq:dntilde_dt_aggr_fragm})
concerns the fragmentation only into particles of similar size. The
fragmentation into very dissimilar sizes, which gives rise to the
Ostwald ripening, has to be accounted for separately. We checked that
for the range of the system parameters considered in this paper, the
coagulation always dominates over Ostwald ripening in coarsening when
the system is unstable with respect to condensation, i.e., $c_{tot}>c_{tot}^{(spin)}$.

The blue line in Fig. \ref{fig:nmean_vs_t_ctot} corresponds to $c_{tot}=5\times10^{26}\,{\rm 1/m^{3}}$,
which is slightly less than the total concentration of saturation
for the standard system, Eq. (\ref{eq:cvtot_st}). Under these conditions,
the system is stable so no carbon condensation is expected. Indeed,
the kinetics is in agreement with the coagulation (red dash-dot-dotted line) at very short times, but then the carbon condensation stops. Finding the equilibrium distribution of particle concentrations,
corresponding to the adopted $c_{tot}$, results in a heavily monomer-dominated
``vapor". The mean particle size in this vapor is shown by the
horizontal magenta line which agrees with numerics very well at $t\apprge10^{-11}\,{\rm s}$.

\subsection{Ostwald Ripening in Realistic Detonation Products\label{subsec:Ripening_realistic}}

In Sec. \ref{subsec:Ostwald-Ripening}, we discussed the mechanism
of the Ostwald ripening and its contribution to the coarsening step.
This discussion was based on kinetic equations \eqref{cnplus} and \eqref{cnminus} that disregarded all the other species present
in detonation products but the excess carbon. Even though the obtained
qualitative trends (e.g., the dependence on temperature and total
carbon concentration) are likely reliable, it is doubtful that the
results are quantitatively accurate. The problem is that, unlike coagulation,
the rate of Ostwald ripening is very sensitive to the energetics and
reactivity of very small carbon particles. Contrary to what is implicitly
assumed by the adopted kinetic equations, Eqs. (\ref{cnplus}) and
(\ref{cnminus}), these small particles are not likely to be pure-carbon
monomers, dimers, trimers etc. Instead, the ensemble of small ``carbon
particles'' with thermodynamically significant concentrations is
expected to comprise chemically ``stabilized'' carbon in the form
of molecules such as ${\rm CO}$ and ${\rm CO_{2}}$. In addition to
this, large carbon particles themselves are expected to have a layer
of surface ligands, e.g., oxygen-bearing chemical groups, that saturate
dangling carbon bonds.\citep{Mochalin-2012-11,Armstrong-2020-353,Lindsey-2020-054103}
We thus expect the following reversible reactions to contribute to
the Ostwald ripening
\begin{align}
{\rm CO}+{\rm C}_{n}{\rm O}_{m} & \rightleftarrows{\rm C}_{n+1}{\rm O}_{m+1},\nonumber \\
{\rm CO_{2}}+{\rm C}_{n}{\rm O}_{m} & \rightleftarrows{\rm C}_{n+1}{\rm O}_{m+2},\label{eq:CO_CO2_reacs}
\end{align}
where large carbon particles ${\rm C}_{n}{\rm O}_{m}$ would typically
have $m\propto n^{2/3}$ because of the oxygen on the surface.\citep{Armstrong-2020-353,Lindsey-2020-054103}
These reactions require the generalized theory of Ostwald ripening
that can account for alloying\citep{Philippe-2013-4237}, as well as inhomogeneous distribution of
different elements within the carbon particle (e.g., oxygen on the surface, carbon in the core). To the best of our knowledge,
such a theory is yet to be developed. However, some qualitative speculations,
based on Eq. (\ref{eq:CO_CO2_reacs}) and results of Sec. \ref{subsec:Ostwald-Ripening},
can already be put forward. Because of the presence of oxygen, the
significant mass fraction of all the carbon in HE ($\sim40\%$ for
TATB\citep{Ornellas-1982-52821}) does not condense and end up trapped
within carbon mono- and dioxide. Equation \eqref{eq:HE_decomp} suggests
that this mass fraction only increases with temperature. Very approximately,
we treat this gaseous form of carbon as the vapor of small carbon
clusters in Sec. \ref{subsec:Ostwald-Ripening}, whose concentration
enters the expression for the Ostwald ripening rate, Eq. (\ref{eq:Rr_id}),
as $c_{v,1}^{(sat)}$. We then take $c_{v,1}^{(sat)}=c_{tot}/2$, which produces
\begin{equation}
R_{r}^{(id)}\approx0.8k_{0}c_{tot}
\end{equation}
from Eq. (\ref{eq:Rr_id}), assuming that $\mu_{s}=3.04\,{\rm eV}$ and
$T=3000\,{\rm K}$. The corresponding corrected coagulation rate,
Eq. (\ref{eq:Rc_corr}), gives
\begin{equation}
\tilde{R}_{c}=\left[1-\frac{c_{v,1}^{(sat)}}{c_{tot}}\right]R_{c}\approx2k_{0}c_{tot}.
\end{equation}
These rates are comparable, and it is feasible that at higher temperatures, $c_{v,1}^{(sat)}$ becomes even larger, thus further increasing $R_{r}^{(id)}$
and respectively decreasing the corrected coagulation rate $\tilde{R}_{c}$.
This very crude prediction is consistent with the numerical observation\citep{Armstrong-2020-353}
that the carbon condensation proceeds via Ostwald ripening at $T=6500\,{\rm K}$,
yet it does not contradict the typical assumption that it is the Smoluchowski coagulation
that dominates the carbon condensation at lower temperatures ($T\sim2500\text{-}3000\,{\rm K}$).\citep{Shaw-1987-2080,Chevrot-2012-084506,Bastea-2012-214106}
Nevertheless, the approximate considerations here are of course no
substitution for an accurate treatment of Eqs. (\ref{eq:CO_CO2_reacs})
within the theoretical framework of the Ostwald ripening.

\subsection{Thermal Equilibration of Carbon Particles\label{subsec:thermal_eq}}

All the derivations in Sec. \ref{subsec:Products_Thermo}, where we
obtained the relation between the aggregation and fragmentation rate
constants, Eq. (\ref{eq:Kminus_Kplus}), tacitly assumed that the
carbon particles are always in the thermal equilibrium (i.e., have
the same temperature) with the surrounding fluid. This might not be
exactly the case since the carbon condensation releases heat, so the
carbon particles can have some excess temperature. This phenomenon and its influence on the kinetics of condensation are discussed in general
literature.\citep{Ratke-2002-Growth,Seinfeld-2016-Atmospheric} In
this subsection, we assess whether this excess temperature is substantial
in the carbon condensation in detonation products. To this end, we
first calculate the heat flux from the carbon particle to the surround
fluid. Solving a simple spherically symmetric problem of steady-state
thermal diffusion, one obtains for the total heat flux from the surface of the particle as
\begin{equation}
J=4\pi R\kappa\Delta T,
\end{equation}
where $\kappa$ is the heat conductivity of the fluid, $R$ is the
particle radius, and $\Delta T$ is the excess temperature. The excess
thermal energy of the particle is estimated as
\begin{equation}
\Delta Q=\frac{4}{3}\pi R^{3}C_{P}\rho\Delta T,
\end{equation}
where $C_{P}$ is the heat capacity of carbon per unit mass, and $\rho$
is the density of carbon in the particle. Then, the characteristic
thermal equilibration time is
\begin{equation}
\tau_{T}=\frac{\Delta Q}{J}=\frac{C_{P}R^{2}\rho}{3\kappa}.
\end{equation}
This time needs to be compared with the coagulation time, $\tau_{c}$,
required to obtain a particle of radius $R$. This coagulation time
is obtained from Eq. (\ref{eq:nmean_vs_t_coag}) as
\begin{equation}
\tau_{c}=\langle n\rangle/R_{c}=\frac{4}{3}\pi R^{3}\frac{\rho}{m_{C}R_{c}}.
\end{equation}
The ratio of these two times is
\begin{equation}
\frac{\tau_{T}}{\tau_{c}}=\frac{R_{*}}{R},\label{eq:tau_T_tau_c}
\end{equation}
where the characteristic particle radius is introduced as
\begin{equation}
R_{*}=\frac{C_{P}m_{C}R_{c}}{4\pi\kappa}.
\end{equation}
The coagulation rate $R_{c}=4.46\times10^{12}\,{\rm 1/s}$ is adopted
from Eq. (\ref{eq:Rc_st_FW}). Heat capacity is estimated as $C_{P}\approx1000\,{\rm \frac{J}{kg\cdot K}}$.\citep{Velizhanin-2020-070051}
Thermal conductivity of the detonation fluid is $\kappa\approx1\,{\rm \frac{W}{m\cdot K}}$.\citep{Bastea-2002-Transport,Velizhanin-2021-Enskog}
The result is
\begin{equation}
R_{*}\sim10^{-11}\,{\rm m}.
\end{equation}
This is an unphysically small size, and so one has $\tau_{T}/\tau_{c}\ll1$
in Eq. (\ref{eq:tau_T_tau_c}) for all realistic carbon particles,
which, in turn, results in a very fast thermal equilibration of carbon
particles if the coagulation is assumed to be the dominant process
in carbon condensation. For this thermal equilibration to be not fast
enough for nanometer-sized particles, one needs to assume that, for
example, the coarsening is dominated by the Ostwald ripening with
the rate higher than that of coagulation by at least two orders of
magnitude. However, this assumption would not agree with the experimental
observations that at least several hundred nanoseconds are required
for the formation of nanometer-sized particles in detonation.\citep{Ten-2010-387,Ten-2014-369,Bagge-Hansen-2015-245902,Watkins-2017-23129,Watkins-2018-821,Hammons-2021-5286}
We thus conclude that the thermal equilibration is very fast at detonation-like
conditions and the carbon particles can be assumed to be in thermal
equilibrium with the detonation fluid.

\subsection{Capillarity Approximation Revisited\label{subsec:capillarity_revisited}}

The capillarity approximation, Eq. (\ref{eq:musn_mus}), was assumed
for large carbon particles in this work. This approximation seems
to be physically sound\citep{Machlin-2007-Aspects,Kelton-2010-Nucleation}
and is also supported by the results of the fitting of energetics
of carbon particles, illustrated in Fig. \ref{fig:Ec_vs_n}. Nevertheless,
it is worth mentioning here that there has been a long and hot debate in the
literature on the validity and possible generalizations, modifications
and corrections to the capillarity approximation, starting, to the best of our knowledge, with the 1962 paper by Lothe and Pound.\citep{Lothe-1962-2080}
The multitude of papers ensued.\citep{Lothe-1966-630,Reiss-1967-2496,Lothe-1968-1849,Reiss-1968-5553,Reiss-1977-1,Weakliem-1994-6408,Reiss-1997-4506,Reiss-1998-8548,Kusaka-2006-031607,Vosel-2009-204508}
The debate seemed to focus on how to correctly re-interpret and modify
the Gibbs theory of capillarity\citep{Gibbs-1948-Thermodynamics}
to be quantum- and statistical-mechanically consistent. In particular,
the possibly large contributions of the translational motion and
rotation of a particle as the whole to the entropy of this particle were not originally accounted for,\citep{Gibbs-1948-Thermodynamics,Frenkel-1939-538}
but accounting for them warrants ``subtracting'' the corresponding
degrees of freedom from the vibration motions in the particle.\citep{Lothe-1962-2080,Lothe-1968-1849}
It is important to note, however, that the debate concentrated not
so much on the validity of the capillarity approximation, Eq. (\ref{eq:musn_mus}),
but more on whether one can use the macroscopic value of surface tension
for microscopic particles, and if or how it should be modified to
have a quantum- and statistical-mechanically consistent theory. As
was pointed out in Ref.~\onlinecite{Kusaka-2006-031607}, the debate could
be avoided altogether if the thermodynamic properties of particles
are obtained directly from, for instance, atomistic calculations.
As was mentioned in the Introduction, these thermodynamic properties are not too accurately
known at present.
Under these conditions, the debate is moot, as far as this work is concerned, so we
resorted to not relying on specific values of the standard chemical
potential of carbon particles. Instead, we show how general kinetic
trends and rates of specific kinetic steps depend on various parameters
of the system.

\subsection{The Physical Significance of the Critical Size\label{subsec:critical_size}}

Finally, we would like to revisit the physical significance and importance
of the critical size. The notion of the critical size appeared naturally
in Sec. \ref{sec:Growth} when discussing the thermodynamics of the
metastable vapor phase, and, in particular, the nucleation barrier.
The original kinetic problem, Eqs. (\ref{cnplus}) and (\ref{cnminus}),
is complex and non-linear in a sense that there are quadratic (with
respect to particle concentrations) terms in there corresponding to
aggregation via binary collisions. Simplification of that, via the
quasi-equilibrium approximation, Sec. \ref{subsec:Quasi-Equilibrium},
resulted in the linear kinetic equation, Eq. (\ref{eq:Jn_a_b}). This
equation describes an effective Brownian particle with the diffusion coefficient given by Eq.~\eqref{eq:bn}, propagating in a one-dimensional
landscape defined by the potential energy $U(n)=-\int^{n}dn'\,a(n')$,
where $a(n)$ is given by Eq. (\ref{eq:an}). This potential energy
has a maximum at $n=n_{c}$, and, therefore, the problem becomes identical
to the famous Kramers problem of a Brownian particle escaping from
a potential well over a potential barrier.\citep{Haenggi1990-251}
This latter problem is ubiquitous in physics and chemistry, and the
parameters of the potential barrier are of paramount importance when
analyzing the \emph{kinetics} of the escape. However, the potential
barrier is not significant from the \emph{purely thermodynamic} standpoint,
since the probability to find the particle on top of the potential barrier is typically too small to contribute noticeably to the moments of the probability distribution of the particle position $n$. For example, in a problem of escaping from one potential well
to the other in a double-well problem, the states on top of the potential
barrier are the \emph{least} populated ones. In the transition state
theory of chemical reactions,\citep{Atkins-2006-Physical} the parameters
of the transition state are of course of fundamental importance when calculating
reaction rate constants, but it is the least populated state (relative
to reactants/products) that is almost never observed experimentally.
In the analysis in Sec. \ref{sec:Growth}, the notion and the value
of the critical size are important in estimations of the growth rate
that agree well with the numerical simulations, Fig. \ref{fig:nmean_t_growth}(a).
However, the critical size of $n_{c}=16$ does not look any special
when inspecting the numerically obtained evolution of the mean particle
size, black line in Fig. \ref{fig:nmean_t_growth}(a). In Fig. \ref{fig:cn_vs_n_t},
the critical size corresponds to the imaginary border between the
metastable vapor phase and the large particles with concentrations
given by Eq. (\ref{eq:cn_c_infty}). However, there are many particles
either smaller or much larger than $n_{c}$ during the growth. Therefore,
unlike Ref.~\onlinecite{Bastea-2017-42151}, we see no physical reason to
compare the mean carbon particle size, observed at any point during
the condensation kinetics, to the critical size.

\section{Conclusion\label{sec:Conclusion}}

In this work, we considered the carbon condensation,
or carbon clustering, in detonation from the perspective of the theory of the kinetics
of first-order phase transitions. The kinetics of carbon condensation
was described on the level of rate equations, Eqs. (\ref{cnplus})
and (\ref{cnminus}), where, unlike in previous efforts, the thermodynamically
consistent fragmentation of carbon particles was included. The spectrum
of the kinetic behaviors was observed to be rich, with the system (i) undergoing
the barrierless (spinodal) condensation at high concentrations of
excess carbon; (ii) going through the nucleation, growth, and coarsening
steps of the phase-transition kinetics at intermediate concentrations;
and (iii) forming a stable carbon ``vapor'' without condensation
at low concentrations.

The treatment of carbon condensation in this
paper is only an approximation to reality. The first reason for this
is, as it was touched upon in Sec. \ref{subsec:Ripening_realistic},
there is a complex chemistry involved in the HE decomposition and
carbon condensation,\citep{Zhang-2009-10619} which is not accounted
for by the rate equations in this work. In particular, the rhs of Eq. (\ref{eq:HE_decomp}) and Refs.~\onlinecite{Zhang-2009-10619,Mochalin-2012-11,Armstrong-2020-353,Lindsey-2020-054103}
suggest that the carbon-oxygen chemistry has to be important when
considering carbon condensation. The effect of this chemistry on the
coarsening step is very approximately discussed in Sec. \ref{subsec:Ripening_realistic}.

The second reason is that the thermodynamic parameters of the detonation
products are assumed time-independent everywhere in this work. In
reality, the temperature and total carbon concentration decrease with
time because of the expansion of the detonation products, and the
dynamics of this expansion depends on the geometry (e.g., size and shape)
of the HE charge, details of the detonation ignition, and the possible
confinement of the charge.\citep{Bdzil-2007-263,Watkins-2017-23129}
It is thus feasible that a very slow condensation takes place immediately
after the fast reaction zone, but then it speeds up because of the
temperature drop in expansion; compare the third and fourth numerical
rows in Tab. \ref{tab:growth}.

An additional complication from the expansion of the detonation products is that at lower temperatures, the assumption of the diffusion-limited
reaction of aggregation, Eq. (\ref{eq:Kplus_diff_lim}), likely breaks down. The diffusion-influenced rate constant of coagulation, $K^{+}_*$, can be expressed as $1/K^+_{*}=1/K^+ + 1/K^+_c$,\citep{Atkins-2006-Physical} where $K^+$ is the diffusion-limited rate constant, Eq. (\ref{eq:Kplus_diff_lim}), and $K^+_c$ is the rate constant of coalescence of two carbon particles in physical contact.  
At high temperatures, one likely has $K^+\ll K^+_c$, resulting in the diffusion-limited aggregation reaction, assumed throughout this work.
However, the coalescence is expected to be a thermally activated process with a substantial activation barrier,\citep{Bastea-2012-214106,Watkins-2017-23129} and so it should become the rate-limiting step at lower temperatures.
This is expected to result in the effective ``freeze-out'', or quenching, of the condensation
process when the temperature of detonation products becomes lower than a certain threshold.\citep{Bastea-2012-214106,Watkins-2017-23129} In particular, the freeze-out temperature of $2500\:\rm{K}$ was introduced by the authors of Ref.~\onlinecite{Watkins-2017-23129} to rationalize the experimental observation that the condensation of carbon particles stopped within $\sim 200-300$ nanoseconds after the passing of the detonation front, resulting in $\approx 8\,{\rm nm}$ diameter carbon particles. We note here in the passing that the finite ultimate size of carbon particles and the likely disordered nature of carbon inside those particles resulted in the need to offset carbon by $+8.75\,{\rm kcal/mol}$ relative to its bulk phase in thermochemical calculations.\citep{Ornellas-1982-52821}

Finally, the carbon phase inside the carbon particles can undergo transformations as a result of the decreasing temperature and pressure.\citep{Bagge-Hansen-2019-3819}
Nevertheless, the theory of the kinetics of first-order phase transitions seems to be
very robust in a sense that very chemically and physically distinct
systems demonstrate the same general trends and kinetic steps of coagulation, nucleation, growth and coarsening.\citep{Binder-1976-343,Penrose-1978-Kinetics,Binder-1987-783,Slezov-2009-Kinetics,Krapivsky-2010-Kinetic}
We, therefore, strongly believe that this work will be useful when
analyzing the results of carbon condensation coming from experiments
and atomistic simulations. On the other hand, insights from atomistic simulations into the energetics of carbon particles, as well as detailed condensation kinetics and underlying mechanisms, would be very helpful in further developing and refining the first-order phase transition theory of the carbon condensation in detonation.

\begin{acknowledgments}
K.A.V. is grateful to Joshua Coe for multiple inspiring discussions
on carbon condensation. This work was supported by the US Department of Energy through the ASC-PEM-HE Program, Los Alamos National Laboratory.
Los Alamos National Laboratory is operated by Triad National Security,
LLC, for the National Nuclear Security Administration of the U.S. Department
of Energy (Contract No. 89233218NCA000001).
\end{acknowledgments}

\section*{Author Declarations}
\subsection*{Conflict of Interest}
The authors have no conflicts to disclose.

\section*{Data Availability}

The data that support the findings of this study are available from the corresponding author upon reasonable request.

\appendix

\section{Treatment of Aggregation and Fragmentation of Identical Particles}

\label{app:over_count}

The factors of $1/2$ in the first rhs terms of Eqs.~(\ref{cnplus})
and (\ref{cnminus}) compensate for over-counting when physically
the same reactions, e.g., $(n)+(m)\rightarrow(n+m)$ and $(m)+(n)\rightarrow(n+m)$,
are counted twice. There is, however, no such over-counting for the
reaction $(m)+(m)\rightarrow(2m)$ so instead of the factors of $1/2$,
it has to be $(1+\delta_{n-m,m})/2$ in the first rhs terms of Eqs.~(\ref{cnplus}) and
(\ref{cnminus}). Furthermore, since the processes $(m)+(m)\rightarrow(2m)$
and $(2m)\rightarrow(m)+(m)$ lead to the loss and gain, respectively,
of two particles of size $m$, there have to be extra factors of $1+\delta_{nm}$
in the second rhs terms of Eqs.~(\ref{cnplus}) and (\ref{cnminus}), as was adopted in Eq. (1) of Ref.~\onlinecite{Chevrot-2012-084506}.
These extra factors do thus simply renormalize the ``diagonal''
rate constants, so that Eqs.~(\ref{cnplus}) and (\ref{cnminus})
can be used as is if the diagonal rate constants are substituted with
\begin{equation}
K_{m,m}^{\pm}\rightarrow2K_{m,m}^{\pm}.\label{eq:Kp_2Kp}
\end{equation}
Eq.~(\ref{eq:Kplus_diff_lim}) gives the rate constant of diffusion-limited
aggregation as $K_{n,m}^{+}=4\pi(D_{n}+D_{m})(R_{n}+R_{m})$. However,
it can be demonstrated that this rate constant has to be effectively
multiplied by an extra factor of $1/2$ for the identical particles, i.e.,
when $n=m$. This extra factor is discussed in detail between Eqs.~(13.57)
and (13.59) in the work of Seinfeld and Pandis.\citep{Seinfeld-2016-Atmospheric}
The overall result is that this extra factor of $1/2$ cancels out
exactly the extra factor of $2$ in Eq.~(\ref{eq:Kp_2Kp}), and therefore
Eqs.~(\ref{cnplus}) and (\ref{cnminus}), supplemented by Eq.~(\ref{eq:Kplus_diff_lim}),
are to be used as is. Accordingly, we consider the extra factor of $1+\delta_{j,k}$ in Eq. (1) of Ref.~\onlinecite{Chevrot-2012-084506} to be erroneous. It is worth noting, however, that the effect of the presence or absence of these factors on the numerical results is non-negligible only when the mean particle size is very small. 

\section{Distance of Coalescence}

\label{app:Rnm}

There does not seem to be a clear agreement in the literature on what
the distance of coalescence $R_{nm}$ has to be. The two prevalent
choices assumed in the literature are $R_{nm}=R_{n}+R_{m}$,\citep{Friedlander-2000-Smoke,Jacobson-2005-Atmospheric,Seinfeld-2016-Atmospheric} and $R_{nm}=\left(R_{n}+R_{m}\right)/2$,\citep{Chandrasekhar-1943-1,Shaw-1987-2080,Bastea-2012-214106,Bastea-2017-42151,Chevrot-2012-084506}
where $R_{n}$ is the radius of the particle of size $n$. The choice
of $R_{nm}=\left(R_{n}+R_{m}\right)/2$ was given by Chandrasekhar,\citep{Chandrasekhar-1943-1} who seemed to obtain it from Smoluchowski
himself, since Smoluchowski \citep{Smoluchowski-1916-585} did have
an equation which suggested the choice $R_{nm}=\left(R_{n}+R_{m}\right)/2$.
However, the exact meaning of $R_{n}$ was not specified by Smoluchowski
until much later in his paper, and Chandrasekhar (or at least those
citing Chandrasekhar) seemed to interpret $R_{n}$ as the particle
radius. Contrary to that, when discussing, later in his paper, the effective
reaction radius for two identical particles, Smoluchowski chose $R=2a$,
where $a$ was explicitly said to be the particle radius. We therefore
assume the choice $R_{nm}=R_{n}+R_{m}$, where $R_{n}$ is the particle
radius, which can be evaluated from Eq.~(\ref{eq:Rn_nC}).

\section{Drift-Diffusion Equation in the Large-Time Limit\label{sec:Drift-Diffusion}}

As discussed in the main text, Eq. (\ref{eq:Jn_a_b}) is the drift-diffusion
equation with the effective drift and diffusion coefficients given
by Eqs. (\ref{eq:an}) and (\ref{eq:bn}), respectively. At large
times, $J_{n}=J$ up to $n=n_{g}\gg1$ and the drift and diffusion
coefficients vary slowly with the particle size $n$. Assuming that
they are in fact size-independent, we obtain the drift-diffusion equation
$J_{n}=ac_{n}-bc_{n}^{\prime}$, which can be solved exactly treating
$n$ as a continuous variable. The Green's function for this equation is
\begin{equation}
c_{n}=\frac{1}{\left(4\pi bt\right)^{1/2}}e^{-\frac{\left(n-n_{0}-at\right)^{2}}{4bt}}.
\end{equation}
What this solution means is that the population of particles, sitting
initially in particles of the same size $n_{0}$, will drift toward large particles with ``velocity'' $a$, broadening as $\Delta n\sim\sqrt{bt}$.
This Green's function can be used to obtain the distribution of particle
concentrations at time $t^{\prime}$ if at some earlier time $t$, the distribution is $c_{n}=c_{0}\theta(n-n_{0})$. Obviously, this
entire stepwise distribution will propagate toward larger $n$ at the velocity
$a$, and the width of the transition from $c_{n}=c_{0}$ to $c_{n}\sim0$
will be on the order of $\Delta n\sim\sqrt{bt}$. Applying these considerations
to numerical results in Fig. \ref{fig:cn_vs_n_t}, we obtained $n_{g}\sim at$
and $\Delta n\propto\sqrt{bt}$, and therefore the assumption of $\Delta n\ll n_{g}$
is satisfied at large times.

\section{Critical Size in a Monomer-Dominated System\label{sec:nc_mono_dominated}}

We assume here that when the quasi-equilibrium is established, the
carbon vapor is dominated by monomers, i.e., carbon vapor is almost
ideal. Then, the chemical potential of the vapor is determined by
\begin{equation}
c_{tot}=c_{1}=c^{\circ}e^{\beta\left(\mu-\mu_{1}^{\circ}\right)},
\end{equation}
and the concentration of in-vapor particles of arbitrary size is given
by
\begin{equation}
c_{n}=c^{\circ}e^{\beta(n\mu-\mu_{n}^{\circ})}=c^{\circ}\left(\frac{c_{tot}}{c^{\circ}}\right)^{n}e^{\beta\left(n\mu_{1}^{\circ}-\mu_{n}^{\circ}\right)}.
\end{equation}
The critical size is defined by where this population has the minimum
with respect to $n$, so the derivative of the exponent with respect
to the particle size has to vanish
\begin{equation}
\beta\mu_{1}^{\circ}+\ln\frac{c_{tot}}{c^{\circ}}-\beta\mu_{n}^{\circ\prime}=0.
\end{equation}
Using Eq. (\ref{eq:mu_n_o_mu_s_n}), i.e., $\mu_{n}^{\circ}=\mu_{s,n}+\beta^{-1}\ln\left(c^{\circ}\Lambda_{C}^{3}n^{-3/2}\right)$,
we obtain
\begin{equation}
\beta\left(\mu_{s,1}-\mu_{s,n}^{\prime}\right)+\frac{3}{2n}+\ln\left(c_{tot}\Lambda_{C}^{3}\right)=0,
\end{equation}
which is almost identical to Eqs. (28) and (29) for the critical size
in Ref.~\onlinecite{Abraham-1968-732}. The only difference originates
from the absence of the chemically neutral buffer gas/fluid in the
latter formulation, which, assuming the gas of carbon particles to be perfect,
would result in vanishing of the last rhs term of Eq. (\ref{eq:mun_fntr})
in the present work. Further assuming that $\mu_{s,n}=\mu_{s}n^{2/3}$ for
all the particles including monomers, one obtains
\begin{equation}
\beta\mu_{s}\left(1-\frac{2}{3}n_{c}^{-1/3}\right)+\frac{3}{2n_{c}}+\ln\left(c_{tot}\Lambda_{C}^{3}\right)=0.\label{eq:nc_analyt_mus}
\end{equation}
This equation, combined with the standard set of parameters from Tab.
\ref{tab:std_parms}, produces $n_{c}\approx180$. This compares well
with $n_{c}=189$ in the second numerical row of Tab. \ref{tab:growth}.
That the just obtained critical size is slightly lower than the exact
one in Tab. \ref{tab:std_parms} is because of the assumption of ideality, $c_{1}=c_{tot}$, here. 

Eq. (\ref{eq:nc_analyt_mus}) could be compared to Eq. (7) in Ref.~\onlinecite{Bastea-2017-42151},
which, using the notation of the present paper, is 
\begin{equation}
\frac{1}{3}\beta\mu_{s}n_{c}^{2/3}-\frac{5}{2}\ln n_{c}+\ln\left(c_{tot}\Lambda_{C}^{3}\right)=0.\label{eq:nc_analyt_Bastea}
\end{equation}
This equation is rather different from the one we obtained, and using
it lowers the critical size from $n_{c}\approx180$ to $n_{c}\sim5$.
The physical origin of Eq. (\ref{eq:nc_analyt_Bastea}) is not clear.

\bibliography{references}

\end{document}